\documentclass[12pt]{article}
\pdfoutput=1
\usepackage[nosort]{cite}
\usepackage[normalem]{ulem}

\usepackage[dvipsnames]{xcolor}
\usepackage{epsfig}
\usepackage{amsfonts}
\usepackage{amscd}
\usepackage{latexsym}
\usepackage{amsmath, amssymb}
\usepackage{verbatim}
\usepackage{color}
\usepackage{fancyhdr}
\usepackage{hyperref}
\usepackage{tikz}
\usepackage{slashed}
\usepackage{multirow}
\usetikzlibrary{calc}
\usetikzlibrary{topaths}
\usetikzlibrary{decorations}
\usetikzlibrary{decorations.pathmorphing}
\usetikzlibrary{arrows, decorations.markings, cd}
\usetikzlibrary{calc,arrows,cd,decorations.markings}
\tikzset{
->-/.style args={#1rotate#2}{decoration={markings, mark=at position #1 with {\arrow[scale=1.5,rotate = #2 ]{stealth}}}, postaction={decorate}}
}
\usetikzlibrary{shapes.geometric}
\usetikzlibrary{knots}
\usepackage{tikz-3dplot}

\usepackage{float}
\usepackage{draft}
\usepackage{graphicx, subfig}
\usepackage{mciteplus}
\usepackage{bbm}

\usepackage{amsthm}
\usepackage{cleveref}
\usepackage{algorithm}
\usepackage[letterpaper,left=1.1in,top=1in,right=1.1in,bottom=1.3in]{geometry}
\usepackage{makecell}
\usepackage{enumerate}
\usepackage{tablefootnote}
\usepackage{longtable}

\usepackage[us,hhmmss]{datetime}

\usepackage{pifont}

\usepackage[scr=euler]{mathalfa}

\usepackage[english]{babel}
\usepackage{soul}
\usepackage{upgreek}
\usepackage[scr]{rsfso}

\DeclareMathAlphabet{\zapfcal}{OT1}{pzc}{m}{it}

\providecolor{darkblue}{rgb}{0,0,0.6}

\numberwithin{equation}{section}

\def\B{\mathcal{B}}

\colorlet{mylinkcolor}{NavyBlue}
\colorlet{mycitecolor}{Aquamarine}
\colorlet{myurlcolor}{Aquamarine}
\newcommand\myshade{90}

\hypersetup{
  linkcolor  = mylinkcolor!\myshade!black,
  citecolor  = mycitecolor!\myshade!black,
  urlcolor   = myurlcolor!\myshade!black,
  colorlinks = true,
}

\begin{document}
\begin{titlepage}

\title{Topological Aspects of the $c=2$ Model in the Continuum and on the Lattice}

\author{Yichul Choi$^1$, Nathan Seiberg$^1$, Sahand Seifnashri$^1$, and Wucheng Zhang$^2$}

\address{${}^1$ School of Natural Sciences, Institute for Advanced Study, 1 Einstein Drive, Princeton, NJ 08540, U.S.A.}
\address{${}^2$ Department of Physics, Princeton University, Jadwin Hall, Washington Road, Princeton, NJ 08544, U.S.A.}
\begin{abstract}
\noindent
We discuss the continuum two-dimensional conformal field theory, whose target space is a two-torus. In addition to its standard zero-form internal symmetries, the model also has a $-1$-form symmetry associated with its $\theta$-parameter, which is also known as a background $B$-field. These symmetries combine into a higher group with anomalies. A spatially winding $B$-field generically breaks the spatial translation symmetry completely, and reduces the continuous target-space (internal) momentum symmetry to a (projectively realized) discrete subgroup. When $B$ varies linearly, a discrete translation subgroup survives, and it combines with the internal symmetry into a non-Abelian group, while the other translations survive as non-invertible symmetries. Even though the model lacks the standard spatial translation symmetry, the non-invertible translation symmetry explains why it is solvable via a Fourier decomposition. We then consider a lattice version of this system, using its modified Villain formulation.  Surprisingly, all these subtleties in the symmetries of the continuum model exist on a finite lattice.  Finally, we discuss various generalizations of this model, demonstrating that the phenomena we uncovered are more broadly valid.
\end{abstract}

\end{titlepage}
\tableofcontents

\section{Introduction}

Many modern developments in high-energy physics and condensed matter physics hinge on a detailed understanding of subtle topological aspects of continuum quantum field theory (QFT).  Examples include periodic $\theta$-parameters, Chern-Simons terms, chiral symmetries, anomalies, certain exact dualities, topologically nontrivial configurations of fields, and so on. 

A common lore in the past was that most of these topological features are absent in discrete lattice systems. Defying this myth, there has been significant progress in recent years toward realizing many of these topological properties in finite lattice models. For instance, a modified version of the Villain model \cite{Villain:1974ir}, which first appeared in \cite{Gross:1990ub}, and more recently in \cite{Sulejmanpasic:2019ytl, Gorantla:2021svj} as well as in \cite{Cheng:2022sgb, Fazza:2022fss} in its Hamiltonian formulation, provides us with a discrete lattice realization of the 1+1d $c=1$ free compact boson conformal field theory (CFT) with its winding symmetry, 't Hooft anomalies, and dualities realized exactly on a finite lattice. Other examples are discussed in  \cite{Choi:2021kmx, Choi:2022zal, Jacobson:2023cmr, Berkowitz:2023pnz, Jacobson:2024hov, Chen:2024ddr,  Jacobson:2024muj, Xu:2024hyo, Katayama:2025pmz, Fidkowski:2025rsq, Thorngren:2026ydw, Seifnashri:2026ema, Ikeda:2026lyl, Lu:2026jnq, Aoki:2026pvq, Argurio:2026txf, Bacq:2026ega, Aoki:2026krx,Wei:2026msr}. These developments are particularly important in the context of (generalized) global symmetries, their anomalies, and related topics \cite{Gaiotto:2014kfa}. 

In this paper, we discuss the 1+1d $c=2$ CFT of two compact bosons (describing a sigma model whose target space is a 2-torus). We uncover a rich and subtle symmetry structure that stems from the topology of the dynamical and classical background fields.  We then study a modified Villain lattice model that flows to this continuum theory. Surprisingly, the lattice model captures all the richness of the continuum theory's symmetry structure that we mentioned above.

\subsection{The continuum theory}

The Lagrangian density of the continuum theory is
\ie \label{eq:lagint}
\frac{1}{4\pi} G_{IJ} d\varphi^I \wedge \star d\varphi^J + \frac{B}{2\pi}  d\varphi^1 \wedge d\varphi^2 \,.
\fe
Here, $\varphi^I \sim \varphi^I + 2\pi$, $I=1,2$, are compact scalar fields,\footnote{Repeated flavor indices $I, J, \cdots$ are always summed over throughout the paper.} and $G_{IJ}$ is the metric of the target space torus.
The inverse metric is denoted by $G^{IJ}$.
An important feature of the theory \eqref{eq:lagint} is that it admits a periodic coupling constant $B \sim B+1$, which is analogous to $\theta$-parameters in gauge theories.\footnote{The normalization of $B$ in \eqref{eq:lagint} is common in the string theory literature. Another natural convention would be to write the coupling in terms of a $2\pi$-periodic $\theta$-parameter, $\theta = 2\pi B$. \label{fn:theta}}
The invariance of the theory under $B \to B+1$ is a part of a much larger T-duality group of the theory.\footnote{The transformation $B \to B+1$, in a different duality frame, becomes a large diffeomorphism of the target space torus that corresponds to $T \in SL(2,\mathbb{Z})$.}  (See \cite{Giveon:1994fu} for a review.)

The theory based on \eqref{eq:lagint} has a $U(1)\times U(1)$ target-space momentum symmetry and a $U(1)\times U(1)$ winding symmetry.\footnote{In addition, it has some discrete symmetries, and for special values of $G_{IJ}$ and $B$, it has a larger symmetry.  We will ignore these symmetries.} Therefore, we will couple it to four classical background $U(1)$ gauge fields, $A^I$ for the momentum symmetries and $\widetilde A_I$ (with $I=1,2$) for the winding symmetries.

Also, we will promote $B$ to a spacetime-dependent background field, and regard it as a background gauge field for a $-1$-form global symmetry \cite{Cordova:2019jnf, Cordova:2019uob}. In particular, we will be interested in $B$ that winds around space.\footnote{Analogous interfaces with varying $\theta$-parameters  have been discussed by many authors in different systems.  Examples include \cite{Gaiotto:2008sd, Estes:2012nx, Gaiotto:2014kfa, Gaiotto:2017yup, Gaiotto:2017tne, Hsin:2018vcg, Aitken:2018cvh, Cordova:2019jnf, Cordova:2019uob, Hsin:2020cgg, Gorantla:2020fao, Choi:2022jqy, Nguyen:2022lie, Choi:2022rfe}. And there are also many discussions about it in the context of topological insulators.}

\subsection{Defining the continuum action}

The first challenge is to define the action of \eqref{eq:lagint} coupled to all these background gauge fields.  One approach uses differential cohomology. (See the nice lectures \cite{Moore:2025tmt}.)  Another approach, which we will follow below, extends both the dynamical fields $\varphi^I$ and the classical background fields $A^I$, $\widetilde A_I$, and $B$ to a 2+1d bulk and defines the action in that bulk.  Here, we should make sure that the action is independent of the extension of the dynamical fields to the bulk. Dependence on the extension of the classical background fields to the bulk is associated with 't Hooft anomalies in the corresponding symmetries.

A more pedestrian approach uses a trivialization of all the bundles.  We cover our spacetime (or the bulk attached to it) with patches.  In each patch, all the fields $\varphi^I$, $A^I$, $\widetilde A_I$, and $B$ are real valued.  And they change from patch to patch using transition functions in the overlap regions.  Then, the action is defined in terms of all this data.

The expression for the Lagrangian density in each patch depends on real-valued fields and is the naive one.  Then, the data in the transition functions yields ``correction terms'' localized in the overlap regions.  These correction terms depend on the background fields $A^I$, $\widetilde A_I$, and $B$, and their transition functions $\Lambda^I$, $\widetilde \Lambda_I$, and $K\in \bZ$, as well as the dynamical fields $\varphi^I$ and their windings $\widetilde{Q}^I\in \bZ$.  (This will be discussed in detail below.)

In the quantum theory, we integrate over $\varphi^I$ and sum over the winding numbers $\widetilde{Q}^I$.  We do not integrate or sum over the background fields and their transition functions.  Then, it is common to refer to the overlap regions and the correction terms for background fields as ``topological defects.''\footnote{The fact that the defects are topological is related to the independence of the choice of trivialization \cite{Gaiotto:2014kfa}.}

In the path-integral formulation of the theory with Lorentzian time $t\in {\mathbb R}_t$, there is no need for correction terms at fixed time, whereas with compact Euclidean time $\tau \sim \tau +2\pi$, there is no essential difference between the correction terms in space and in time.
In contrast, in the Hamiltonian formulation, the correction terms in space are defects in space, but the correction terms in Euclidean time are related to constraints on a larger Hilbert space. These issues will be discussed in detail below.  
In particular, in Appendix \ref{particleontorus}, we will use the standard problem of a quantum-mechanical particle on a torus in a constant magnetic field as a pedagogical example illustrating some of the associated subtleties.

\subsection{The symmetry}

The careful definition of the action and its properties yields a rich set of phenomena associated with the system's symmetries. 
\begin{itemize}
    \item The $-1$-form symmetry mixes with other zero-form global symmetries of the theory to form a higher-group symmetry \cite{Kapustin:2013uxa, Cordova:2018cvg, Benini:2018reh, Heidenreich:2020pkc}.\footnote{An $n$-group symmetry involves $m$-form symmetries with $m \leq n-1$. Therefore, the symmetry structure we find may be referred to as a ``$1$-group symmetry'' or as a ``lower-group symmetry.'' Still, we will call it a higher-group symmetry.}
    Higher-group symmetries involving a $-1$-form symmetry recently also appeared in \cite{Brennan:2026ira}.  Crucially, although it is related to it, this is different from the ``anomalies in the space of coupling constants'' \cite{Cordova:2019jnf, Cordova:2019uob}.\footnote{An interplay between a higher-group symmetry and the space of defect coupling constants has been studied recently in \cite{Harder:2026csj}. Some other related works are \cite{Choi:2025ebk,Wen:2025xka,Copetti:2025sym,Komargodski:2025jbu}.}
    \item When the configuration of $B$ is topologically nontrivial, i.e., it winds around the (compact) space with winding number $K$, the $U(1) \times U(1)$ target space momentum symmetry is explicitly broken to a discrete $\mathbb{Z}_{|K|}\times\mathbb{Z}_{|K|}$ subgroup.  Moreover, this discrete momentum symmetry is realized projectively \cite{Seiberg:2024wgj, Seiberg:2024yig}.  This explicit $U(1) \times U(1)\to \mathbb{Z}_{|K|}\times\mathbb{Z}_{|K|}$ symmetry breaking is reminiscent of the translation symmetry of a particle on a torus in a constant magnetic field and the related problem of explicit chiral symmetry breaking due to the Adler-Bell-Jackiw (ABJ) anomaly \cite{Adler:1969gk,Bell:1969ts}.\footnote{See Appendix A in \cite{Seiberg:2024yig} and Appendix \ref{particleontorus} here for pedagogical discussions of this model from an anomaly perspective and for the relation between the path integral and the Hilbert space of this model.}

    \item When $B$ depends linearly on the spatial coordinate and has winding number $K$, the spatial translation symmetry is explicitly broken to a discrete $\mathbb{Z}_{|K|}$ subgroup. The algebra of this $\mathbb{Z}_{|K|}$ translation symmetry and the $\mathbb{Z}_{|K|}\times\mathbb{Z}_{|K|}$ target space momentum symmetry is extended by the $U(1) \times U(1)$ winding symmetry of the theory to form a non-Abelian symmetry \cite{Seiberg:2024wgj, Seiberg:2024yig}.   More general translations are resurrected as non-invertible symmetries \cite{Seiberg:2024wgj,Seiberg:2024yig}.\footnote{See \cite{McGreevy:2022oyu,Freed:2022qnc,Brennan:2023mmt,Shao:2023gho,Schafer-Nameki:2023jdn,Costa:2024wks,Kaidi:2026urc} for reviews of non-invertible symmetries, and \cite{Nagoya:2023zky, Heckman:2024obe, Bharadwaj:2024gpj,Furuta:2024vtc,Ando:2025hwb} for discussions on internal non-invertible global symmetries of the theory \eqref{eq:lagint} and related models.} For translations that shift the spatial coordinate by a rational multiple of the total length of the space, the corresponding non-invertible translation symmetry operators are reminiscent of the 3+1d non-invertible chiral symmetry operators that were studied in \cite{Choi:2022jqy, Cordova:2022ieu}.
\end{itemize}

\subsection{The lattice model}

One of our main results is the lattice realization of the continuum theory \eqref{eq:lagint} with all its intricate symmetry structure mentioned above on a finite lattice. The Euclidean lattice version of this modified Villain lattice model was presented in \cite{Katayama:2025pmz}. 
Here, we discuss a lattice quantum Hamiltonian formulation of the model.
The lattice Hamiltonian is
\ie\label{Hint}
H=&\frac{1}{2}\sum_{j=0}^{L-1} G^{IJ}
\left[ p_{j,I} -\frac{B}{2}\epsilon_{II'}
\left(n^{I'}_{j,j+1}+n^{I'}_{j-1,j}+\frac{\phi^{I'}_{j+1}-\phi^{I'}_{j-1}}{2\pi}\right)\right]\\
&\qquad\qquad\cdot\left[p_{j,J}-\frac{B}{2}\epsilon_{JJ'}
\left(n^{J'}_{j,j+1}+n^{J'}_{j-1,j}+\frac{\phi^{J'}_{j+1}-\phi^{J'}_{j-1}}{2\pi}\right)\right]\\
&+\frac{1}{2}\sum_{j=0}^{L-1} G_{IJ}\left(n^I_{j,j+1}+\frac{\phi^I_{j+1}-\phi^I_j}{2\pi}\right)\left(n^J_{j,j+1}+\frac{\phi^J_{j+1}-\phi^J_j}{2\pi}\right)\,, \\
&j\sim  j+L\,,
\fe
where $L$ is the number of lattice sites, $\epsilon_{IJ}$ is the epsilon tensor in the target space,\footnote{We use $\epsilon_{12}=+1$.} $\phi_j^I$ are real-valued variables at each site, $p_{j,I}$ are their conjugate momenta, and $n^I_{j,j+1}$ are $\mathbb{Z} \times \mathbb{Z}$ gauge fields on links. As we will explain later, this system should be constrained by appropriate $\mathbb{Z} \times \mathbb{Z}$ Gauss's laws, which effectively make the bosons $\phi_j^I$ compact. See also \cite{Cheng:2022sgb,Fazza:2022fss}. 

Importantly, not only does the lattice model \eqref{Hint} flow to the continuum theory \eqref{eq:lagint}, it also exhibits the four $U(1)$ global symmetries of the continuum theory as well as their 't Hooft anomalies \cite{Gorantla:2021svj, Cheng:2022sgb, Fazza:2022fss, Seifnashri:2023dpa, Katayama:2025pmz, Thorngren:2026ydw, Seifnashri:2026ema}.

Here, we are particularly interested in the dependence on $B$.  We will find the following properties of this lattice model.
\begin{itemize}
    \item The T-duality transformation $B \to B+1$ is realized exactly on the lattice and is implemented by a conjugation by a unitary operator.\footnote{See \cite{Gross:1990ub, Gorantla:2021svj, Cheng:2022sgb, Fazza:2022fss, Pace:2024oys, Katayama:2025pmz, Seifnashri:2026ema, Argurio:2026txf, Bacq:2026ega} for discussions on exact T-duality in lattice models.} In particular, $B$ is a periodic coupling constant of the Hamiltonian \eqref{Hint}.
    \item A broader view of the model \eqref{Hint} allows $B$ to be a position-dependent classical background field.  In particular, we let $B$ on the link $B_{j,j+1}$ depend on $j$. Importantly, we will make precise what it means for a periodic coupling constant on the lattice to have a nontrivial winding configuration.
    \item When $B$ has winding number $K$, the $U(1) \times U(1)$ momentum symmetry is broken to a $\bZ_{|K|}\times \bZ_{|K|}$ subgroup. When $B_{j,j+1}$ varies linearly as a function of $j$, the (unitary) $\bZ_L$ lattice translation symmetry is broken to a $\bZ_g$ subgroup where $g=\gcd(L,K)$.\footnote{We use the standard convention that $\mathrm{gcd}(a,b)$ is a positive integer for any $a,b\in \bZ$.} More general lattice translations, including the translation by one lattice site, are realized as non-invertible symmetries.
    \item The algebra obeyed by the invertible and non-invertible symmetry operators on the lattice matches precisely with the algebra of the corresponding symmetry operators of the continuum theory \eqref{eq:lagint}.
\end{itemize}

\subsection{Outline}

The rest of the paper is organized as follows.

In Section \ref{sec:continuum}, we will analyze the continuum $c=2$ theory \eqref{eq:lagint}, its $-1$-form symmetry, and its higher-group symmetry.  We will also discuss the various phenomena mentioned above that take place when $B$ is topologically nontrivial.

In Section \ref{sec:trasnlation_continuum}, we will specialize to $B$ varying linearly in the spatial coordinate $x$.  In this case, the system has a discrete unitary translation symmetry, which is combined with the internal symmetry to a non-Abelian symmetry group.  Also, there is a larger non-invertible translation symmetry.

Section \ref{oursystem} will discuss the Hamiltonian formulation of the continuum system.  In addition to being interesting in its own right, it will make it easier to match with the lattice discussion, which uses a Hamiltonian lattice description.
 
In Section \ref{sec:modifiedVillain}, we will study the lattice Hamiltonian \eqref{Hint}, and discuss the T-duality $B \to B+1$.  In Sections \ref{sec:windingBLat} and \ref{latticetrans}, we will construct the winding $B$ lattice model and will analyze its symmetries. 

Section \ref{comparingcontla} will be devoted to a detailed comparison between the lattice and the continuum theories.

We will summarize our results and suggest some future directions in Section \ref{sec:conclusion}.

In several appendices, we will present more background and details of our analysis.  

Appendix \ref{particleontorus} will be devoted to a pedagogical discussion of the problem of a particle on a torus in a constant magnetic field in different gauges.  This well-known system will demonstrate some of the issues we will face in the body of the paper.

Appendix \ref{app:bulk} will give more details on the bulk action of the continuum theory and its independence of the extension of the dynamical fields.

In Appendix \ref{app:spectrum}, we will analyze the spectrum of both the continuum and the lattice model in the presence of linearly winding $B$. 

Appendix \ref{app:noninv} will provide more details about the non-invertible translations in the continuum and on the lattice.  In particular, it will present the symmetry operator/defect as supporting a quantum-mechanical system of a fuzzy torus, thus making its locality manifest.

Appendix \ref{app:more_bosons} will extend our discussion to an arbitrary number of compact bosons in two dimensions.  And Appendix \ref{app:higherD} will generalize our discussion to higher-dimensional spacetime.  

\section{The 1+1d $c=2$ system with varying $B$} \label{sec:continuum}

\subsection{Review of the $c=2$ conformal field theory (constant $B$)}\label{cistwo}

We begin by studying the continuum theory \eqref{eq:lagint}, with the Lagrangian density
\begin{align} \label{eq:lag}
\begin{split}
        \frac{1}{4\pi} G_{IJ} d\varphi^I \wedge \star d\varphi^J + 2\pi B \mathsf{q} \,,
\end{split}
\end{align}
where\footnote{In Euclidean signature, $q={1\over (2\pi)^2}(\partial _\tau\varphi^1\partial_x \varphi^2-\partial _x \varphi^1\partial_\tau \varphi^2)$ and $\int d\tau dx\, q \in \bZ$ is the instanton number.}
\ie\label{qdef}
\mathsf{q} = q dt \wedge dx \,, \quad q={1\over (2\pi)^2}(\partial _t\varphi^1\partial_x \varphi^2-\partial _x\varphi^1\partial_t \varphi^2)\, .
\fe
At this stage, we take $G_{IJ}$ and $B$ to be constants.  Later, we will consider a position-dependent $B$.  This Lagrangian describes a family of CFTs with central charge $c=2$.

The theory has four $U(1)$ global symmetries with conserved currents
\begin{align} \label{eq:currents_forms}
\begin{split}
    \mathsf{J}_{I} = \frac{1}{2\pi} \left( G_{IJ} \star d \varphi^J + B \epsilon_{IJ} d \varphi^J \right) \,, \qquad \widetilde{\mathsf{J}}^I = \frac{1}{2\pi} d \varphi^I \,,
\end{split}
\end{align}
satisfying $d\mathsf{J}_{I} = 0$ and $d\widetilde{\mathsf{J}}^I = 0$.\footnote{Our convention for a conserved current is such that, as a differential form, the conservation equation is $d\mathsf{J} = 0$, and in components, it is $\partial_\mu J^\mu =0$, where $\mathsf{J} = \epsilon_{\mu\nu} J^\mu dx^\nu$.  Also, we take $\epsilon_{tx}=+1$. \label{formcong}}
The currents $\mathsf{J}_{I}$ generate the two momentum symmetries in the target space, and the currents $\widetilde{\mathsf{J}}^I$ generate the two winding symmetries, with quantized charges
\ie\label{eq:cont_charges}
Q_{I}=\oint \mathsf{J}_I\,, \qquad \widetilde{Q}^I=\oint \widetilde{\mathsf{J}}^I\,,
\fe
where the integrals are along a spatial slice.

Importantly, we will view $\mathsf{q}$ in \eqref{qdef} as the ``current'' of a $-1$-form symmetry \cite{Cordova:2019jnf,Cordova:2019uob},\footnote{The current $\mathsf{q}$ is a product of two winding symmetry currents, $\mathsf{q} = \widetilde{\mathsf{J}}^1 \wedge \widetilde{\mathsf{J}}^2$. Such ``derived'' currents were previously discussed in the literature, for instance, in \cite{Heidenreich:2020pkc, Brauner:2020rtz, Damia:2022seq, Seiberg:2024wgj, Seiberg:2024yig}. See also Appendix \ref{app:higherD}.} in which case the coupling constant $B$ is to be regarded as the background gauge field for the $-1$-form symmetry. Later, when we make $B$ position-dependent, such a viewpoint will become particularly natural.

Let us briefly recall the theory's Hamiltonian and the spatial momentum operator in the constant-$B$ case.  (In Section \ref{sec:modifiedVillain}, this will be compared with the lattice realization of the model in its modified Villain version \cite{Cheng:2022sgb, Fazza:2022fss}.)  We quantize the theory on a cylinder with a spatial coordinate $x \sim x + 2\pi$ and Lorentzian time $t \in \mathbb{R}_t$.
The momenta conjugate to $\varphi^I$ are
\begin{equation}
    \pi_I = \frac{1}{2\pi} G_{IJ} \partial_t \varphi^J + \frac{1}{2\pi} B \epsilon_{IJ} \partial_x \varphi^J \,, \qquad [\varphi^I (x) , \pi_J (y)] = i \delta^I_J \delta (x-y ) \,.
\end{equation}
Note that $\pi_I$ corresponds to the time component of the current $\mathsf{J}_I$ in \eqref{eq:currents_forms} (as in footnote \ref{formcong}).
We expand
\begin{align} \label{eq:modeexp_ham}
\begin{split}
    \varphi^I (x) = \widetilde{Q}^I x + \sum_{n\in \bZ} e^{inx} \varphi^I_n \,, \qquad \pi_I (x) = \frac{1}{2\pi} \sum_{n\in\mathbb{Z}} e^{inx} \pi_{n,I} \,,
\end{split}
\end{align}
where $\varphi_{-n}^I = (\varphi_n^I)^\dagger$, $\pi_{-n,I} = (\pi_{n,I})^\dagger$, and $[\varphi_m^I , \pi_{n,J}] = i\delta_{m,-n} \delta^I_J$.
In particular,
\begin{equation}
    \pi_{0,I} = \int_0^{2\pi} dx \, \pi_I = Q_{I} \,.
\end{equation}

Ignoring the normal ordering constant, the Hamiltonian is 
\ie\label{eq:Hamiltonian_continuum}
    H &= \int_0^{2\pi} dx \, \left[
        \pi\, G^{IJ} \left( \pi_I - \frac{1}{2\pi} B \epsilon_{II'} \partial_x \varphi^{I'} \right)\left( \pi_J - \frac{1}{2\pi} B \epsilon_{JJ'} \partial_x \varphi^{J'} \right) + \frac{1}{4\pi} G_{IJ} \partial_x \varphi^I \partial_x \varphi^J
    \right] \\
    &=\frac{1}{2} G^{IJ} \left( Q_{I} - B \epsilon_{I I'} \widetilde{Q}^{I'} \right)\left( Q_{J} - B \epsilon_{JJ'} \widetilde{Q}^{J'} \right) + \frac{1}{2} G_{IJ} \widetilde{Q}^{I} \widetilde{Q}^{J}  + \sum_{n \neq 0} \sum_{r= 1,2} |n|  a_{n,r}^\dagger a_{n,r}  \,.
\fe
Here, $a_{n,r}^\dagger$ and $a_{n,r}$ are standard creation and annihilation operators, i.e., $[a_{n,r}, a^\dagger_{m,s}]= \delta_{nm}\delta_{rs}$.
They are defined by
\begin{align} \label{eq:cont_osc}
\begin{split}
    a_{n,r}^\dagger &= \frac{1}{\sqrt{2|n|}} \left[
        E_r^I \left( \pi_{-n,I} + in B \epsilon_{I I'} \varphi_{-n}^{I'} \right) + i |n| \delta_{r s} E^s_I \varphi_{-n}^I
    \right] \,,\\
    a_{n,r} &= \frac{1}{\sqrt{2|n|}} \left[
        E_r^I \left( \pi_{n,I} - in B \epsilon_{I I'} \varphi_n^{I'} \right) - i |n| \delta_{r s} E^s_I \varphi_n^I
    \right] \,,
\end{split}
\end{align}
where $E_I^r$ is the target space vielbein, and $E^I_r$ is its inverse, and they satisfy 
\ie \label{eq:vielbein}
    \delta_{rs} E_I^r E_J^s = G_{IJ} \,, \quad \delta^{r s} E_r^I E_s^J=G^{I J} \,, \quad E_r^I E_I^s=\delta_r^s\,, \quad E_I^r E_r^J=\delta_I^J\,.
\fe

The energy-momentum tensor of the theory is\footnote{We use the Lorentzian metric $\eta_{tt} = +1$, $\eta_{xx} = -1$. And the reason for the hat in $\hat \Theta$ will be clear soon.}
\ie \label{eq:emtensor}
    \hat {\Theta}_{\mu\nu} = \frac{1}{2\pi} G_{IJ} \left(\partial_\mu \varphi^I \partial_\nu \varphi^J - \frac{1}{2} \eta_{\mu\nu} \partial_\rho \varphi^I \partial^\rho \varphi^J \right) \,.
\fe
The spatial momentum operator is then given by
\ie\label{eq:momentum_continuum}
    P = \int_0^{2\pi} dx \, \hat {\Theta}_{tx} = Q_{I} \widetilde{Q}^I - \sum_{n \neq 0}\sum_{r= 1,2} n  a_{n,r}^\dagger a_{n,r}
    \,.
\fe

We take a basis of the Hilbert space with basis states $\ket{m_I , w^I , N_{n,r} }$.
They are labeled by the eigenvalues $m_I$ and $w^I$ of the momentum and winding charges \eqref{eq:cont_charges}, respectively, and by the occupation numbers $N_{n,r}$ of the oscillators \eqref{eq:cont_osc}.  The corresponding energy \eqref{eq:Hamiltonian_continuum}  and momentum \eqref{eq:momentum_continuum} eigenvalues are
\begin{align}\label{eq:contspectrum}
\begin{split}
    E(m_I , w^I , N_{n,r} ) &= \frac{1}{2} G^{IJ} \left( m_I - B \epsilon_{I I'} w^{I'} \right)\left( m_J - B \epsilon_{JJ'} w^{J'} \right) + \frac{1}{2} G_{IJ} w^{I} w^{J} \\
    &~~~~~~~~+ \sum_{n \neq 0} \sum_{r= 1,2} |n| N_{n,r}  \\
 P(m_I , w^I , N_{n,r} ) &=
        m_I w^I - \sum_{n \neq 0} \sum_{r= 1,2}  n N_{n,r}   \,.
\end{split}
\end{align}

\subsection{Higher group and $-1$-form symmetry} \label{sec:higher_group}

Here, we will show that the four $U(1)$ global symmetries mix with the $-1$-form symmetry to form a higher group symmetry  \cite{Kapustin:2013uxa, Cordova:2018cvg, Benini:2018reh}.\footnote{We thank Brandon C. Rayhaun for several discussions on this point.}
Related discussions appear in \cite{Brennan:2026ira},\footnote{In \cite{Brennan:2026ira}, such a phenomenon was also referred to as a ``higher family'' of theories.} and higher-dimensional analogs, which do not involve a $-1$-form symmetry, appear in \cite{Brauner:2020rtz, Damia:2022seq}.
See also \cite{Heidenreich:2020pkc, Jacobson:2024muj, Manjunath:2026ezp}.\footnote{Discussions on higher-group symmetries in Hamiltonian lattice models can be found, for instance, in \cite{Chen:2020msl, Barkeshli:2022wuz, Barkeshli:2022edm, Delcamp:2023kew, Barkeshli:2023bta, Oishi:2026cvg, Lu:2026jnq, Brito:2026eqi}.}

One way to see the higher group structure is as follows.
Let us couple the theory \eqref{eq:lag} to background gauge fields $A^I$, $\widetilde{A}_I$ for the four $U(1)$ symmetries.
For the moment, we keep $B$ a constant.
The Lagrangian density becomes
\begin{align} \label{eq:lag2}
\begin{split}
    &\mathcal{L} = \mathcal{L}_G + \mathcal{L}_{B} + \mathcal{L}_{\widetilde{A}}  \,,\\
    &\mathcal{L}_G = \frac{1}{4\pi} G_{IJ} (d\varphi^I -A^I ) \wedge \star(d\varphi^J - A^J ) \,,\\
    &\mathcal{L}_{B} = \frac{1}{2\pi} B (d\varphi^1 -A^1 ) \wedge (d\varphi^2 - A^2 ) \,,\\
    &\mathcal{L}_{\widetilde{A}} = \frac{1}{2\pi} \widetilde{A}_I \wedge (d\varphi^I - A^I )\, .
\end{split}
\end{align}

This Lagrangian density is imprecise for two reasons.  First, if $A^I$ is topologically non-trivial, $\varphi^I$ should have singularities, which it winds around.  These are insertions of winding operators.  We will not write them explicitly, but assume that they are present.  Second, $\mathcal{L}_{\widetilde{A}}$ is not invariant under gauge transformations of $\widetilde A_I$.  And depending on the other fields, even the exponential of the action is not invariant under them.  This signals an anomaly, which we will discuss below.

In the presence of $A^I$ and $\widetilde A_I$ background gauge fields, the periodic identification $B \sim B+1$ is violated. However, the combined transformation
\begin{align} \label{eq:modified_periodicity}
\begin{split}
    \widetilde{A}_I &\rightarrow \widetilde{A}_I - k \epsilon_{IJ} A^J \,,\\
    B &\rightarrow B + k \,,
\end{split}
\end{align}
with $k \in \mathbb{Z}$, preserves the theory, modulo a term that depends only on the background gauge fields, which originates from the anomaly mentioned above.

The transformation \eqref{eq:modified_periodicity} is a manifestation of the Witten effect \cite{Witten:1979ey}, which states that under shifting $B\to B+k$ (or equivalently $\theta \to \theta +2\pi k$ with $\theta =2\pi B $ as in footnote \ref{fn:theta}), the momentum charges are shifted as 
\ie\label{Witteneff}
Q_{ I}\to Q_{ I}+ k\epsilon_{IJ} \widetilde{Q}^J \,.
\fe
This shift of the momentum charges is equivalent to \eqref{eq:modified_periodicity}.

We should also consider a nontrivial background $B$. Then, the transformation law \eqref{eq:modified_periodicity} indicates that the field strengths $d\widetilde{A}_I$ are not single-valued.
Instead, the combinations
\ie \label{eq:tildeF}
    \widetilde{F}_I = d\widetilde{A}_I + B \epsilon_{IJ} dA^J
\fe
are well-defined.
The modified periodicity \eqref{eq:modified_periodicity} and the modified field strengths \eqref{eq:tildeF} describe a higher-group symmetry, where the momentum, winding, and the $-1$-form symmetries mix.  

Now, when $B$ is a background field, which can wind, not only  $\mathcal{L}_{\widetilde{A}}$, but also $\mathcal{L}_B$ in \eqref{eq:lag2} should be defined carefully. One approach to studying such problems involves extending all the fields to a 2+1d bulk.
In our case, we write the bulk term\footnote{When the boundary fields are topologically nontrivial, there might not be an extension to the bulk with a single boundary.  One way to address this question is the following.  We add a bulk direction, $z$, and set the boundary at $z=0$.  Then, we can have another boundary at some $z_0<0$.  Assuming that $|z_0|$ is large, we can focus on the fields at finite $z$ and neglect the contribution from $z_0$. This approach is common in the study of symmetry-protected topological (SPT) phases with different boundaries, e.g., in the AKLT chain.  In this case, the anomaly of the boundary theory at $z=0$ is canceled by the anomaly of the degrees of freedom at $z_0$.\label{nobulk}}
\ie\label{bulktermg}
&\mathcal{L}_{bulk} = \frac{1}{2\pi} dB \wedge (d\varphi^1 - A^1 ) \wedge (d\varphi^2 - A^2 ) + \frac{1}{2\pi} \widetilde{F}_I \wedge (d\varphi^I - A^I ) \,,
\fe
with the field strengths $\widetilde{F}_I$ in \eqref{eq:tildeF}.
The bulk Lagrangian density ${\cal L}_{bulk}$ is well-defined (except for the needed singularities in $\varphi^I$ mentioned above), and it is invariant under \eqref{eq:modified_periodicity}.
It can be written as
\ie\label{Btotald}
&\mathcal{L}_{bulk} =  d \left[ \frac{1}{2\pi}
            B (d\varphi^1 - A^1 ) \wedge (d\varphi^2 - A^2 ) + \frac{1}{2\pi}  \widetilde A_I \wedge (d\varphi^I - A^I )
      \right] +\mathcal{L}_{anomaly}\,,\\
&\mathcal{L}_{anomaly} = -\frac{1}{2\pi} \widetilde{A}_I \wedge dA^I  \, .
\fe
The first term in ${\cal L}_{bulk}$ leads to $\mathcal{L}_B$ and $\mathcal{L}_{\widetilde{A}}$ in the 1+1d Lagrangian \eqref{eq:lag2}.
Importantly, the 2+1d action $\int {\cal L}_{bulk}$ does not depend on the extension of $\varphi^I$ to the bulk, and therefore the theory is still 1+1-dimensional.
We will discuss this fact in more detail in Appendix \ref{app:bulk}.
The remaining bulk term ${\cal L}_{anomaly}$ represents the mixed anomaly between the momentum and winding symmetries.
Moreover, under the transformation \eqref{eq:modified_periodicity}, ${\cal L}_{anomaly}$ changes by a boundary term, which cancels the variation of the Lagrangian \eqref{eq:lag2}.

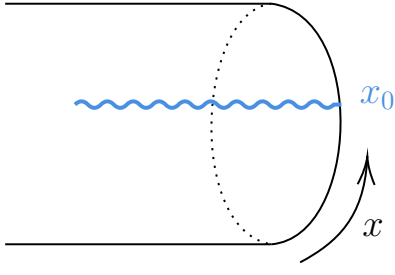
\begin{figure}
    \centering
\tikzset{every picture/.style={line width=0.75pt}}
\begin{tikzpicture}[x=0.75pt,y=0.75pt,yscale=-1.3,xscale=1.3]

\draw    (137,14) .. controls (174,19) and (171,104) .. (137,107) ;

\draw  [dash pattern={on 0.84pt off 2.51pt}]  (137,14) .. controls (108,17) and (104,102) .. (137,107) ;

\draw    (34,14) -- (137,14) ;

\draw    (34,107) -- (137,107) ;

\draw [color={rgb, 255:red, 74; green, 144; blue, 226 }  ,draw opacity=1 ][line width=1.5]    (61,53) .. controls (62.67,51.33) and (64.33,51.33) .. (66,53) .. controls (67.67,54.67) and (69.33,54.67) .. (71,53) .. controls (72.67,51.33) and (74.33,51.33) .. (76,53) .. controls (77.67,54.67) and (79.33,54.67) .. (81,53) .. controls (82.67,51.33) and (84.33,51.33) .. (86,53) .. controls (87.67,54.67) and (89.33,54.67) .. (91,53) .. controls (92.67,51.33) and (94.33,51.33) .. (96,53) .. controls (97.67,54.67) and (99.33,54.67) .. (101,53) .. controls (102.67,51.33) and (104.33,51.33) .. (106,53) .. controls (107.67,54.67) and (109.33,54.67) .. (111,53) .. controls (112.67,51.33) and (114.33,51.33) .. (116,53) .. controls (117.67,54.67) and (119.33,54.67) .. (121,53) .. controls (122.67,51.33) and (124.33,51.33) .. (126,53) .. controls (127.67,54.67) and (129.33,54.67) .. (131,53) .. controls (132.67,51.33) and (134.33,51.33) .. (136,53) .. controls (137.67,54.67) and (139.33,54.67) .. (141,53) .. controls (142.67,51.33) and (144.33,51.33) .. (146,53) .. controls (147.67,54.67) and (149.33,54.67) .. (151,53) .. controls (152.67,51.33) and (154.33,51.33) .. (156,53) .. controls (157.67,54.67) and (159.33,54.67) .. (161,53) -- (164,53) -- (164,53) ;

\draw    (148,114) .. controls (162.63,106.2) and (172.5,96.5) .. (173.91,74.7) ;
\draw [shift={(174,73)}, rotate = 92.49] [color={rgb, 255:red, 0; green, 0; blue, 0 }  ][line width=0.75]    (10.93,-3.29) .. controls (6.95,-1.4) and (3.31,-0.3) .. (0,0) .. controls (3.31,0.3) and (6.95,1.4) .. (10.93,3.29) ;

\draw (170,45) node [anchor=north west][inner sep=0.75pt]  [color={rgb, 255:red, 74; green, 144; blue, 226 }  ,opacity=1 ]  {{\large $x_{0}$}};

\draw (171,96.4) node [anchor=north west][inner sep=0.75pt]    {\large $x$};

\end{tikzpicture}
    \caption{A fixed time slice of the bulk manifold $M_3$ is a half-infinite cylinder with coordinates $z \in (-\infty , 0]$ and $x \sim x +2\pi$. The boundary $M_2 = \partial M_3$ is at $z=0$. We trivialize various bundles with transition functions at $x = x_0$ \eqref{BtildeAt}. Integrating by parts the bulk Lagrangian density $\mathcal{L}_{bulk}$ \eqref{Btotald} on $M_3$ leads to surface terms at $x=x_0$ (second and third rows of \eqref{finalaction}).}
    \label{fig:bulk}
\end{figure}

Let us explore \eqref{Btotald} in more detail.  While the total Lagrangian density ${\cal L}_{bulk}$ is well-defined, this is not true globally for the individual terms in \eqref{Btotald}.
In particular, we cannot simply integrate by parts the first term to find a boundary term because the object in the square brackets in \eqref{Btotald} is not single-valued.  One way to address this issue is to trivialize the bundles.  Concretely, let us take our spacetime $M_2 = \partial M_3$ to be an infinite cylinder parameterized by a non-compact time $t\in {\mathbb R}_t$ and a compact spatial coordinate $x\sim x+2\pi$, i.e., $M_2={\mathbb R}_t \times S_x^1$.
We parameterize the bulk by the coordinate $z\in {\mathbb R}_z^- =(-\infty , 0]$ such that $M_3={\mathbb R}_z^- \times {\mathbb R}_t \times S_x^1$.\footnote{We choose the orientation on $M_3$ such that the volume form is $dz \wedge dt \wedge dx$. The induced orientation on $M_2$ is such that its volume form is $dt \wedge dx$.\label{orientation}}
See Figure \ref{fig:bulk} for an illustration.

Next, we choose a point $x_0\in S^1_x$, and let $B$, $A^I$, and $\widetilde{A}_I$ be single-valued and real for $x\ne x_0$. We then postulate transition functions there,
\ie\label{BtildeAt}
&B(t,x_0^-,z)=B(t,x_0^+,z)+K\,,\\
&\widetilde A_I(t,x_0^-,z)=\widetilde A_I(t,x_0^+,z)+d\widetilde \Lambda_I(t,x_0,z)-K \epsilon_{IJ} A^J(t,x_0,z)\,,\\
&A^I(t,x_0^-,z)= A^I(t,x_0^+,z)\, .
\fe
For simplicity, we assumed that the transition functions of $A^I$ vanish at $x_0$, but we did include transition functions $\widetilde \Lambda_I$ for $\widetilde A_I$.
More general cases will be discussed in Appendix \ref{app:bulk}.
Here, $K$ is an integer, and it is the winding number of $B$ around the spatial circle,
\begin{equation} \label{eq:K_winding}
    K = \int_{x_0}^{x_0 + 2\pi} dx\, \partial_x B \in \bZ \,.
\end{equation}
Similarly, the bosons $\varphi^I$ are taken to be single-valued real functions for $x \neq x_0$, and their transition functions are given by the winding charges in \eqref{eq:cont_charges},
\begin{equation} \label{eq:winding_bosons}
    \varphi^I (t, x_0^- , z ) = \varphi^I (t, x_0^+ , z ) + 2\pi \widetilde{Q}^I \,.
\end{equation}

Under gauge transformations $\lambda^I$ of $A^I$, we have
\ie\label{Agauget}
&A^I\to A^I +d\lambda^I \,,~ \varphi^I \rightarrow \varphi^I + \lambda^I \,,\\
&\widetilde A_I\to \widetilde A_I \,,\\
&B\to B\,,\\
&\widetilde \Lambda_I(t,x_0,z)\to \widetilde \Lambda_I(t,x_0,z) +K \epsilon_{IJ} \lambda^J(t,x_0,z)\, .
\fe
The field strengths $\widetilde F_I=d\widetilde A_I +B \epsilon_{IJ}dA^J$ are invariant under the gauge transformations \eqref{Agauget}, and as was mentioned, they are also invariant under the shift $B \rightarrow B + k$, $\widetilde{A}_I \rightarrow \widetilde{A}_I - k \epsilon_{IJ} A^J$ for any $k\in \mathbb{Z}$.
Importantly, using \eqref{BtildeAt}, they are continuous at $x_0$.

With this understanding, we can analyze  ${\cal L}_{bulk}$.
Locally, it is the same as \eqref{Btotald}, but the integration by parts of the first term should be done carefully.  With the simplified assumptions above, we find contributions from the boundary at $z=0$, and from $x_0^+$ and $x_0^-+2\pi \sim x_0^-$.  (We drop the contribution from $z=-\infty$. See footnote \ref{nobulk}.)  From the boundary at $z=0$, we find
\ie
\int_{{\mathbb R}_t \times (x_0, x_0+2\pi)}\left(
    \frac{1}{2\pi}B (d\varphi^1 - A^1 ) \wedge (d\varphi^2 - A^2 ) + \frac{1}{2\pi} \widetilde{A}_I \wedge (d\varphi^I - A^I )
\right)\, .
\fe
This integral is within a single patch and is therefore well-defined.  The  contributions from $x_0^+$ and $x_0^-+2\pi \sim x_0^-$ combine to
\begin{align} \label{eq:x0}
\begin{split}
    \int_{\mathbb{R}_z^{-}\times \mathbb{R}_t} \left[ d\left(
        \frac{1}{2\pi} K \varphi^1 d\varphi^2 + \frac{1}{2\pi} \widetilde{\Lambda}_I d\varphi^I \right) - \frac{1}{2\pi} K A^1 \wedge A^2 - \frac{1}{2\pi} d\widetilde{\Lambda}_I \wedge A^I
     \right]\Big|_{x=x_0^+}\, .
\end{split}
\end{align}
Ignoring the contribution from $z=-\infty$, and adding  $\mathcal{L}_G$ in \eqref{eq:lag2}, we end up with the total action
\ie\label{finalaction}
S=&\int_{\partial M_3} \mathcal{L} +\int_{M_3}{\cal L}_{anomaly}\\
&+\int_{{\mathbb R}_t}\left(\frac{1}{2\pi}K \varphi^1  d\varphi^2 +\frac{1}{2\pi} \widetilde{\Lambda}_I d\varphi^I \right)\Big|_{x=x_0^+}\\
&+\int_{{\mathbb R}_z^- \times {\mathbb R}_t}\left( -\frac{1}{2\pi} K A^1 \wedge A^2 - \frac{1}{2\pi} d\widetilde{\Lambda}_I \wedge A^I  \right)\Big|_{x=x_0^+}\,,
\fe
where $\mathcal{L}$ is the 1+1d Lagrangian density \eqref{eq:lag2}.
The first term in the first row is integrated along $x \in (x_0,x_0+2\pi)$ and $t \in {\mathbb R}_t$, and the second term in the first row is integrated along $x \in (x_0,x_0+2\pi)$, $t \in {\mathbb R}_t$ and $z \in{\mathbb R}_z^-$.  Using our trivialization, they are well-defined.  The terms in the second and third rows are needed because of the transition functions at $x_0$.

Alternatively, we can view the action \eqref{finalaction} as follows.  The first and the third term depend on the dynamical fields $\varphi^I$, and are integrated over the two-dimensional spacetime $\partial M_3$ or its subspace ${\mathbb R}_t$ at fixed $x=x_0$.  The second and the fourth term depend only on the background fields $A^I$ and the transition functions $K$, $\widetilde{\Lambda}_I$, and are integrated over the bulk $M_3$ or its subspace ${\mathbb R}_z^- \times {\mathbb R}_t$.

Since we derived the action \eqref{finalaction}
using the well-defined terms, $\mathcal{L}_G$ and ${\cal L}_{bulk}$, it should be well-defined and gauge-invariant.
It is straightforward to see that the action is invariant under the periodicity of $B$ \eqref{eq:modified_periodicity} and the gauge transformations of $\widetilde{A}_I$.
Moreover, it is independent of our choice of trivialization, i.e., the value of $x_0$, as well as invariant under gauge transformations of $A^I$ \eqref{Agauget}.
Let us demonstrate these facts.

As we change $x_0\to x_0+\delta x_0$,
\ie \label{eq:x0_independence}
&\delta\int_{\partial M_3}{\cal L} = \int_{{\mathbb R}_t \times (x_0,x_0+\delta x_0)}\left[
     d\left(\frac{1}{2\pi} K \varphi^1 d\varphi^2 + \frac{1}{2\pi} \widetilde{\Lambda}_I d\varphi^I \right) -\frac{1}{2\pi}K A^1 \wedge A^2 - \frac{1}{2\pi} d\widetilde{\Lambda}_I \wedge A^I
\right] \,,\\
&\delta\int_{M_3}{\cal L}_{anomaly}=\int_{ {\mathbb R}_z^- \times {\mathbb R}_t \times (x_0,x_0+\delta x_0) } d\left(
     \frac{1}{2\pi} K A^1 \wedge A^2 +\frac{1}{2\pi} d\widetilde{\Lambda}_I \wedge A^I
\right) \, .
\fe
Integrating by parts, these terms cancel the change in $x_0$ in the terms in the second and third rows of \eqref{finalaction}. In fact, we could have reversed the logic as follows.  Naively, the action is the terms in the first row.  But since it is not well-defined, we choose a trivialization at $x_0$ and add the terms in the second and third rows as ``correction-terms,'' thereby ensuring that $S$ is independent of $x_0$.
See also \cite{Bauer:2004nh,Cordova:2019jnf,Seiberg:2024wgj,Seiberg:2024yig,Shao:2025qvf}.

Under the gauge transformations of $A^I$ \eqref{Agauget}, we get
\ie
&\delta\int_{\partial M_3} {\cal L}=0 \,, \\ &\delta\int_{M_3}{\cal L}_{anomaly}= 0 \,,\\
&\delta\int_{{\mathbb R}_t}\left(
    \frac{1}{2\pi} K \varphi^1 d\varphi^2 + \frac{1}{2\pi} \widetilde{\Lambda}_I d\varphi^I
\right)\Big|_{x=x_0^+} \\
&~~~~~~~~~~~~~~~~~~~~~~~~~~~~~~= \int_{\mathbb{R}_t} \left(
    \frac{1}{2\pi}K d(\varphi^1 \lambda^2 ) + \frac{1}{2\pi} K \lambda^2 d\lambda^1 + \frac{1}{2\pi} \widetilde{\Lambda}_I d\lambda^I
\right)\Big|_{x=x_0^+} \,,\\
&\delta\int_{{\mathbb R}_z^- \times {\mathbb R}_t}\left(
    -\frac{1}{2\pi} K A^1 \wedge A^2 - \frac{1}{2\pi} d\widetilde{\Lambda}_I \wedge A^I
\right)\Big|_{x=x_0^+} \\
&~~~~~~~~~~~~~~~~~~~~~~~~~~~~~~= \int_{{\mathbb R}_z^- \times {\mathbb R}_t} d\left( \frac{1}{2\pi} K \lambda^1 d\lambda^2 - \frac{1}{2\pi} \widetilde{\Lambda}_I d\lambda^I \right)\Big|_{x=x_0^+} \, .
\fe
Performing integration by parts, as above, dropping the contributions at $z=-\infty$ and $t=\pm \infty$, we find that the total action \eqref{finalaction} is gauge-invariant.

\subsection{The target space momentum symmetry} \label{sec:momentum}

To explore consequences of the mixing between the symmetries, let $B = B(x)$ depend on the spatial coordinate $x$, and keep it independent of time $t$.
As before, we pick $x_0 \in S^1_x$, and let $B(x)$ be a real function for $x \neq x_0$, with a transition function as in \eqref{BtildeAt}.
Alternatively, we can work in the covering space, and write
\ie \label{eq:winding_B}
 B(x+2\pi)=B(x)+K \,.
\fe
We now analyze the symmetries of the theory in the presence of such a winding configuration of $B(x)$, i.e., a configuration with nonzero $K$.

Specializing the action \eqref{finalaction} to this case and setting the background gauge fields for the momentum and winding symmetries to zero, we have
\begin{align} \label{eq:action1}
\begin{split}
    S &= \int dt \int_{x_0}^{x_0 + 2\pi} dx  \left[
        \frac{1}{4\pi} G_{IJ} (\partial_t \varphi^I \partial_t \varphi^J - \partial_x \varphi^I \partial_x \varphi^J) + 2\pi B(x)  q
    \right] \\
    &~~~~~~+\frac{K}{4\pi} \epsilon_{IJ} \int dt \, \varphi^I \partial_t \varphi^J \Big|_{x=x_0^+} \, .
\end{split}
\end{align}
Compared to \eqref{finalaction}, we wrote the terms localized at $x_0$ such that the two bosons are treated symmetrically, and we dropped a total time derivative.
Equivalently, after integrating by parts, the action \eqref{eq:action1} becomes
\begin{align} \label{eq:action2}
\begin{split}
    S = \int dt \int_{x_0}^{x_0 + 2\pi} dx
         \Bigg[&\frac{1}{4\pi} G_{IJ} (\partial_t \varphi^I \partial_t \varphi^J - \partial_x \varphi^I \partial_x \varphi^J) + \frac{1}{4\pi} \partial_x B(x) \epsilon_{IJ} \varphi^I \partial_t \varphi^J\\
        &+ 2\pi B( x_0^+ ) q
    \Bigg]  \, .
\end{split}
\end{align}
In Section \ref{oursystem}, we will give a more careful comparison between various actions, in particular \eqref{eq:action1} and \eqref{eq:action2}, and discuss the corresponding Hamiltonian formulations.

In the presentation \eqref{eq:action1}, the last term, which is localized at $x_0$, makes the action independent of the trivialization, i.e., the choice of $x_0$.
More physically, it represents a topological defect that implements the jump in $B(x)$ at $x=x_0$.
In the presentation \eqref{eq:action2}, there is no such localized term.  Instead, the independence of the trivialization is guaranteed by the explicit $x_0$ dependence of $B(x_0^+)$. (See \cite{Seiberg:2024wgj, Seiberg:2024yig} for more details.)
In both presentations, we see that the target space momentum symmetries are explicitly broken.
That is, the actions \eqref{eq:action1} and \eqref{eq:action2} are not invariant under the shifts $\varphi^I\to \varphi^I+\alpha^I$ by arbitrary constants $\alpha^I$.

The classical equations of motion of the two (equivalent) actions \eqref{eq:action1} and \eqref{eq:action2} are
\ie\label{classicaleom}
    G_{IJ} (\partial_t^2 \varphi^J - \partial_x^2 \varphi^J ) - \epsilon_{IJ}\partial_xB(x) \partial_t \varphi^J = 0 \, .
\fe
Note that the correction term at $x_0$ in \eqref{eq:action1} guarantees that the equations of motion are independent of $x_0$, and in \eqref{eq:action2}, the term that depends explicitly on $x_0$ does not contribute to the equations of motion.

One lesson from \eqref{classicaleom} is that the classical theory, which depends only on the equations of motion, is invariant under the $U(1) \times U(1)$ momentum symmetry.
Its violation in the quantum theory, which we now discuss in detail, can be interpreted as an ABJ anomaly.

To better understand the anomaly and the symmetry breaking, we examine the currents in \eqref{eq:currents_forms}.
From the equations of motion \eqref{classicaleom}, we see that the currents of the two momentum symmetries,
\ie\label{firstcur}
J_{\mu,I} = \frac{1}{2\pi} \left( G_{IJ} \partial_\mu \varphi^J - \epsilon_{IJ} \epsilon_{\mu \nu} B(x) \partial^\nu \varphi^J \right)\,,
\fe
are still conserved, $\partial^\mu J_{\mu, I} = 0$.
Importantly, when $B(x)$ has a nonzero winding number $K\ne0$ as in \eqref{eq:winding_B}, these currents are not single-valued around the spatial circle.

Alternatively, we could take the currents
\ie\label{secondcur}
J_{\mu,I}' = \frac{1}{2\pi} \left( G_{IJ} \partial_\mu \varphi^J +\epsilon_{IJ} \epsilon_{\mu \nu} \partial^\nu B(x) \varphi^J \right)\,,
\fe
which are also conserved.  Here, the explicit dependence on $x$ through $\partial^\nu B(x)$ is single-valued.  However, the dependence on $\varphi^J$ without a derivative means that these currents are not single-valued in the target space.\footnote{When a current $J_\mu$ is well-defined, it is subject to an ambiguity known as an improvement transformation $J_\mu \to J'_\mu=J_\mu +\epsilon_{\mu\nu}\partial^\nu X$ with a well-defined $X$.  The relation between our $J_{\mu,I}$ \eqref{firstcur} and $J'_{\mu, I}$ \eqref{secondcur} is almost such an improvement transformation with $X_I={1\over 2\pi}\epsilon_{IJ} B(x)\varphi^J$, except that in our case, all of $J_{\mu,I}$, $X_I$, and $J'_{\mu, I}$ are ill-defined.  Similarly, there are other options, using other $X$, e.g., $J''_{\mu,I}= \frac{1}{2\pi} \left( G_{IJ} \partial_\mu \varphi^J - \epsilon_{IJ} \epsilon_{\mu \nu} (B(x)-B_0) \partial^\nu \varphi^J \right)$ with constant $B_0$.}

Finally, we can also consider the current
\ie\label{thirdcur}
\hat J_{\mu,I} = \frac{1}{2\pi}  G_{IJ} \partial_\mu \varphi^J \, .
\fe
It is well-defined, but it is not conserved,
\ie
\partial^\mu \hat J_{\mu,I}={1\over 2\pi}\partial_xB(x) \epsilon_{IJ} \partial_t \varphi^J \, .
\fe

This situation with ill-defined conserved currents and well-defined non-conserved currents is familiar from the study of ABJ anomalies.  As in that case, the classical theory has a symmetry, which is absent in the quantum theory.

The continuous $U(1) \times U(1)$ momentum symmetry of the classical theory is broken to a discrete subgroup. 
The well-defined momentum symmetry operators are
\ie \label{eq:mosymm_ops}
    U_I &= \exp \left[
        \frac{2\pi i}{K} \left(
            \int_{x_0}^{x_0 + 2\pi}  J_{t,I} \, dx - \frac{K}{2\pi} \epsilon_{IJ} \varphi^J \Big|_{x=x_0^+}
        \right)
    \right] \\
    & =\exp \left[
        \frac{2\pi i}{K}
            \int_{x_0}^{x_0 + 2\pi}  \left(J'_{t,I} +  {B(x_0^+)\over 2\pi}\epsilon_{IJ}\partial_x\varphi^J\right)\, dx
    \right] \, .
    \fe
The second term in the first expression, which is localized at $x_0$, ensures that $U_I$ is independent of the trivialization and that the operator is conserved.  The second expression does not have a term localized at $x_0$.  Instead, it has a term that depends explicitly on $x_0$.  This explicit $x_0$ dependence cancels the dependence on the trivialization.  These two expressions for $U_I$ are related to the two expressions for the action \eqref{eq:action1} and \eqref{eq:action2}, respectively. (See \cite{Seiberg:2024wgj, Seiberg:2024yig} for more details.)

The operators \eqref{eq:mosymm_ops} generate a discrete $\bZ_{|K|}\times \bZ_{|K|}$ momentum symmetry. Furthermore, this symmetry is realized projectively \cite{Seiberg:2024wgj},\footnote{The fact that $U_I^K = 1$ can be derived as follows. 
We redefine $\varphi^I \rightarrow \varphi^I + 2\pi N^I \varTheta (t)$ in \eqref{eq:action1}.
Here, $N^I \in \mathbb{Z}$ are arbitrary integers and $\varTheta (t)$ is the Heaviside step function.  Since the bosons are $2\pi$-periodic, this does not affect the path integral, but it shifts the action \eqref{eq:action1} by
\ie \nonumber
    \delta S =  2\pi N^I \left(\int_{x_0}^{x_0 + 2\pi} dx \,
       J_{t,I} \Big|_{t=0} - \frac{K}{2\pi} \epsilon_{IJ} \varphi^J \Big|_{x= x_0^+, t=0}  \right) \,,
\fe
where we dropped a contribution from $t= + \infty$.
Since $N^I$'s are arbitrary, we derive the operator statement $U_I^K = 1$. In Section \ref{oursystem}, this will also be explained from a different perspective.}
\ie\label{momentumsymmmetryb}
    U_1^{K} = U_2^{K} = 1 \,, \qquad U_1 U_2 = e^{{2\pi i \over K }} U_2 U_1  \, .
\fe
This result originates from the need for the correction term in \eqref{eq:action1}.

At this point, one might ask the following question.  In the previous subsection, we coupled the action to background gauge fields for the various symmetries, $A^I$, $\widetilde A_I$, and $B$.  Then, we verified that the action is gauge invariant under the momentum symmetry gauge transformations, $A^I \to A^I +d\lambda^I$ in \eqref{Agauget}.
So, how come the theory lacks the global continuous momentum symmetries?

To answer this question, we recall that the currents $J_{\mu,I}$ are the derivatives of the action with respect to $A^I$, holding the other dynamical and background fields fixed.  Importantly, this definition of the currents cannot be used when $B(x)$ is topologically nontrivial. In such a case, $\widetilde A_I$ is shifted by $K\epsilon_{IJ}A^J$ across $x= x_0$ as in \eqref{BtildeAt}. Therefore, when $K \ne 0$, we cannot differentiate with respect to $A^I$ with fixed $\widetilde A_I$.  (Equivalently, when $K\ne 0$, we cannot have a nonzero $A^I$ with $\widetilde A_J=0$.)  In more detail, we can perform such a derivative away from $x= x_0$ and find the current \eqref{firstcur}.  Then, the fact that we cannot do it at $x_0$ is related to the jump of $J_{\mu,I}$ at that point.  Consequently, $J_{\mu,I}$ is not globally well-defined.

One might further ask why a constant gauge transformation $\lambda^I$ does not lead to a global symmetry.  The answer is that when $K\ne 0$, the constant $\lambda^I$ does not change $A^I$, but it does change the transition functions $\widetilde \Lambda_I$ of $\widetilde A_I$ \eqref{Agauget}.  Therefore, it does not lead to an ordinary global symmetry.

Finally, this explicit breaking of the continuous momentum symmetry has the following intuitive interpretation.  Consider an operator with winding charges $w^I$, and move it around our compact space.  Due to the Witten effect \cite{Witten:1979ey}, this shifts its momentum charges by $K\epsilon_{IJ} w^J$.  (See the discussion around \eqref{Witteneff}.)  Consequently, the momentum charges are well-defined only modulo $K$, and the continuous  $U(1)\times U(1)$ momentum symmetry is broken to $\bZ_{|K|}\times \bZ_{|K|}$.

\section{Linear $B$ and translation symmetries}\label{sec:trasnlation_continuum}

\subsection{The spatial translation symmetry}\label{spatialtran}

Clearly, a generic $B(x)$ violates the spatial translation symmetry.  Ignoring the coupling with $B(x)$, the energy-momentum tensor is given by \eqref{eq:emtensor}. It is well-defined.  However, it is not conserved,\footnote{$\hat {\Theta}_{\mu x}$ is analogous to $\hat J_{\mu,I}$ in \eqref{thirdcur}.  It is well-defined but not conserved.\label{thetahatcomment}}    
\ie \label{thetaisnotcon}
    \partial^\mu \hat {\Theta}_{\mu x} = - 2\pi \partial_xB(x) \, q \,,
\fe
where $q$ is the instanton density \eqref{qdef}.

Importantly, when $B(x)$ winds, even the complete translation around space, implemented by $ \exp \left( 2\pi i \int_{x_0}^{x_0 + 2\pi} \hat {\Theta}_{t x} \, dx \right)$, does not commute with the Hamiltonian. Instead, it maps the Hamiltonian with $B(x)$ to the Hamiltonian with $B(x)-K$.\footnote{In our convention throughout, a unitary operator $U$ acts on an operator $\cal O$ as $U: {\cal O} \to {\cal O}' = U{\cal O}U^{-1}$.\label{actionconvention}} This can be canceled by accompanying it with $\mathcal {B}^K$ where 
    \ie \label{eq:calB}
    \mathcal{B} = \exp \left(\left.\frac{i}{2 \pi} \int_{x_0}^{x_0+2 \pi} d x \,\varphi^1 \partial_x \varphi^2-i \widetilde{Q}^1 \varphi^2\right|_{x=x_0^+}\right)
    \fe
is the operator that shifts $B\to B+1$. In other words,
\ie\label{completetrn}
\exp \left(
       2\pi i
            \int_{x_0}^{x_0 + 2\pi} \hat {\Theta}_{t x} \, dx 
    \right)\mathcal{B}^K=1\,.
\fe
For later purposes, we also write \eqref{completetrn} as
\ie\label{twopishif}
\exp \left(
       2\pi i
            \int_{x_0}^{x_0 + 2\pi}  {\Theta}_{t x}\, dx -i K\widetilde{Q}^1 \varphi^2\Big|_{x= x_0^+}
    \right)=1\,,
\fe
where
\ie \label{eq:T}
    {\Theta}_{\mu x} = \hat {\Theta}_{\mu x} - \frac{1}{(2\pi)^2} K \epsilon_{\mu \nu} \varphi^1 \partial^\nu \varphi^2 \, .
\fe
These equations are intuitively obvious. For an explicit proof in the case of a linear $B$ profile below, see the discussion around \eqref{momentum.with.B}.

Let us specialize to the case where $B(x)$ is linear in $x$,
\begin{equation}
    B(x) = B(x_0^+) + \frac{K}{2\pi} (x- x_0 )\,, \qquad x_0 < x < x_0 +2\pi\,.
\end{equation}
Alternatively, as in \eqref{eq:winding_B}, we can write it in the covering space as 
\ie\label{eq:linear_B}
B(x)=B(0)+K{x\over 2\pi}\,.
\fe

Then, the equations of motion \eqref{classicaleom} do not have explicit $x$ dependence, and they preserve the spatial translation symmetry.  Therefore, the classical theory is translation invariant.  This is not the case in the quantum theory.

The lack of conservation of the energy-momentum tensor \eqref{thetaisnotcon} is now
\ie \label{eq:ABJ}
    \partial^\mu \hat {\Theta}_{\mu x} = - K q \,.
\fe
This equation is reminiscent of the ABJ anomaly.  In the classical theory, the instantons can be neglected, and the symmetry is present.  In the quantum theory, instantons break that symmetry.

Instead, we may consider the modified energy-momentum tensor $\Theta_{\mu x}$ in \eqref{eq:T}. It is conserved, but ill-defined.\footnote{$\Theta_{\mu x}$ is analogous to $J_{\mu,I}$ of \eqref{firstcur} and $J'_{\mu,I}$ of \eqref{secondcur}, which are conserved, but are not well-defined.  (Compare with footnote \ref{thetahatcomment}.)}  Again, we encounter here the standard situation with the ABJ anomaly in continuous symmetries.
There is an ill-defined conserved current and a well-defined non-conserved current.
See the related discussion about the translation symmetry in \cite{Seiberg:2024wgj, Seiberg:2024yig}.

Given that we lost the continuous translation symmetry, we can attempt to find a discrete subgroup of it.  In \eqref{completetrn} and \eqref{twopishif}, we combined a complete translation around space with a shift of $B$ by $K$.  This motivates us to take the ``$K$'th root'' of this operation to find a $\bZ_{|K|}$ translation symmetry.

This symmetry is generated by \cite{Seiberg:2024wgj}
\ie \label{eq:symm_ops}
    \mathcal{P} 
    &=\exp \left[
        \frac{2\pi i}{K}
            \int_{x_0}^{x_0 + 2\pi} {\hat{\Theta}}_{t x} \, dx
    \right]\cal B \\
    &=\exp \left[
        \frac{2\pi i}{K} \left(
            \int_{x_0}^{x_0 + 2\pi} {\Theta}_{t x} \, dx - \frac{K}{2\pi}  \widetilde{Q}^1 \varphi^2 \Big|_{x=x_0^+}
        \right)
    \right] \, .
\fe
As in the first expression for $U_I$ in \eqref{eq:mosymm_ops}, the discrete translation operator $\cal P$  includes a correction term localized at $x = x_0$.  It guarantees the conservation of $\cal P$ and its independence of the trivialization, i.e., the choice of $x_0$.

Using the fact that the momentum symmetry operator $U_I$ shifts $\varphi^I \rightarrow \varphi^I + 2\pi / K$, we find that the algebra \eqref{momentumsymmmetryb} is embedded in
\begin{align} \label{eq:cont-translation}
\begin{split}
    U_1^{K} &= U_2^{K} = \mathcal{P}^K = 1 \,, \qquad U_1 U_2 = e^{ \frac{2\pi i}{K}} U_2 U_1  \,, \qquad U_I \mathcal{P} = e^{ \frac{2\pi i}{K} \epsilon_{IJ} \widetilde{Q}^J } \mathcal{P} U_I   \,.
\end{split}
\end{align}
In particular, the two momentum $\bZ_{|K|}$ and the translation $\bZ_{|K|}$ symmetries are extended by the two $U(1)$ winding symmetries and form a non-Abelian symmetry group \cite{Seiberg:2024wgj}.

To summarize, the breaking of the spatial translation symmetry arises from the explicit $ x$-dependence of $B(x)$.  For linear $B(x)$ \eqref{eq:linear_B}, the $\bZ_{|K|}$ symmetry is seen from the action \eqref{eq:action1} by combining $x\to x+{2\pi\over K}$ and $B\to B+1$ (see footnote \ref{footnote:shiftingB}). This can also be made more transparent by replacing the defect at $x_0$ with $|K|$ equally spaced topological defects, with $B$ changing by one at each. 

The same conclusion follows from the presentation \eqref{eq:action2}. Here, the Lagrangian does not have explicit $x$ dependence.  However, as explained in \cite{Seiberg:2024wgj, Seiberg:2024yig}, the theory is not translation-invariant, as evidenced by its dependence on $x_0$.  

\subsection{Mode expansion} \label{sec:modeexp}

Here, we will show that even though the quantum theory lacks the continuous spatial translation symmetry, a standard Fourier decomposition still decouples modes with different spatial momenta. This decoupling stems from the underlying non-invertible translation symmetry, discussed in Section \ref{sec:nonInvertibleCont}, together with the fact that the Hamiltonian is local and quadratic.

For simplicity, we place the trivialization point at $x_0=0$. We use the mode expansion
\ie\label{modeexpa}
    \varphi^I(x,t) =  \widetilde{Q}^{I}\,x + \sum_{n \in \bZ} e^{inx} \varphi_n^I (t) \,, \qquad (\varphi_n^I)^*=\varphi_{-n}^I\,,
\fe
in the Lagrangian that follows from the Lorentzian action \eqref{eq:action2}, 
\ie \label{Lagrangian}
    L = \int_{0}^{2\pi} dx
         \Bigg[&\frac{1}{4\pi} G_{IJ} (\partial_t \varphi^I \partial_t \varphi^J - \partial_x \varphi^I \partial_x \varphi^J) + \frac{K}{8\pi^2} \epsilon_{IJ} \varphi^I \partial_t \varphi^J + 2\pi B(0^+) q
    \Bigg]  \, ,
\fe
to find
\ie\label{LagrsumwithF}
    L = L_0 + \sum_{n > 0} L_n + \frac{dF}{dt}\,,
\fe
where
\ie \label{mode.decoupling}
    L_0 &= \frac{1}{2} G_{IJ}\left(\partial_t \varphi_0^I \partial_t \varphi_0^J-\widetilde{Q}^I\widetilde{Q}^J\right) + \frac{K}{4\pi} \epsilon_{IJ} \varphi_0^I \partial_t \varphi^J_0 + \left( B(0^+) -\frac{K}{4} \right) \epsilon_{IJ} \partial_t \varphi_0^I \widetilde{Q}^J \,,\\
    L_n & = G_{IJ} \left( \partial_t \varphi_n^I \partial_t \varphi_{-n}^J - n^2\varphi^I_n \varphi^J_{-n} \right) + \frac{K}{4\pi}\epsilon_{IJ} \left( \varphi_n^I \partial_t \varphi^J_{-n} + \varphi_{-n}^I \partial_t \varphi^J_{n} \right) \,,\\
    F &= \frac{K}{4\pi} \epsilon_{IJ} \widetilde{Q}^I \sum_{n \neq 0} \frac{\varphi_n^J}{in}- \frac{iB(0^+)}{2} \epsilon_{IJ} \sum_{n \neq 0} n \varphi_n^I \varphi^J_{-n}\,.
\fe
$F$ is well-defined and single-valued since it depends only on the winding charges and the non-compact non-zero modes. Therefore, we can drop the term $dF/dt$ in $L$. This is to be contrasted with the last term in $L_0$, which is a total derivative of a function that is not single-valued. This is a $\theta$-term for the zero modes. However, since $K\ne 0$, without loss of generality, we can shift $\varphi_0^I$ by a constant proportional to $\widetilde Q^I$ and drop that $\theta$-term.

As we see from \eqref{mode.decoupling}, even though the theory is not translation invariant, different Fourier modes decouple from one another. (Note that $L_n$ is independent of $\widetilde{Q}^I$, so this decoupling holds in every winding sector.)  The zero modes in $L_0$ describe a quantum mechanical particle moving on a 2-torus with metric $G_{IJ}$ in the presence of a constant magnetic field with $K$ units of flux.  (See Appendix \ref{particleontorus}.) And the non-zero modes in $L_n$ describe particles on ${\mathbb R}^2$  with a harmonic potential and a constant background magnetic field.

The classical Lagrangian has the translation symmetry $x \to x + 2\pi \xi$ acting on the modes as $\varphi_0^I \to \varphi_0^I + 2\pi \xi \widetilde{Q}^I$ and $\varphi_n^I \to e^{2\pi i n \xi}\varphi_n^I$. One way to see the breaking of the symmetry in the quantum theory is to consider the Hamiltonian formulation and to note that due to the non-commutative algebra of the zero modes, no operator implements this transformation for generic $\xi$. We can also see this directly in the Lagrangian formulation. The translation $x \to x + 2\pi \xi$ shifts the Euclidean action (to be discussed in Section \ref{oursystem}) by 
\ie
    \delta S = -2\pi i K \xi \epsilon_{IJ} \widetilde{Q}^I \widetilde{Q}_{\tau}^J = +2\pi i K \xi \int d\tau dx \, q\,,
\fe
where $\widetilde{Q}_{\tau}^I$ is the winding number of $\varphi^I$ along the Euclidean time direction $\tau$ \cite{Seiberg:2024wgj,Seiberg:2024yig}. Thus, there is an unbroken $\bZ_{|K|}$ translation symmetry with $K\xi \in \bZ$. 

In the next subsection, we will project onto sectors with constrained winding charges $\widetilde{Q}^I$, and identify non-invertible translation symmetries. The Hamiltonian formulation of \eqref{Lagrangian} will be derived in Section \ref{oursystem} and will be diagonalized in Appendix \ref{contspec}.

\subsection{Non-invertible translation symmetry}
\label{sec:nonInvertibleCont}

When $B(x)$ is linear \eqref{eq:linear_B}, a spatial translation $x \to x + 2\pi \xi$ changes $B(x) \to B(x) - K \xi$.\footnote{Note that translating the operators as ${\cal O}(x) \to {\cal O}(x+2\pi\xi)$ is equivalent to shifting the background fields in the opposite direction, i.e., $B(x) \to B(x-2\pi\xi)$. \label{footnote:shiftingB}}
As in Section \ref{spatialtran}, for $\xi \in \frac{1}{K}\mathbb{Z}$, we could turn this into a symmetry by accompanying it with an appropriate power of the operator \eqref{eq:calB} that shifts $B(x)$ by an integer.

For a generic real $\xi$, one can still turn this into a symmetry, albeit a non-invertible one.  Since the states with zero winding charges are not affected by a constant shift of $B(x)$, one can accompany the translation $x \to x + 2\pi \xi$ with a projection operator on $\widetilde{Q}^I = 0$ to find a non-invertible symmetry \cite{Seiberg:2024wgj, Seiberg:2024yig}.

Suppose we perform a translation $x \to x + 2\pi \xi$ by an arbitrary rational number
\ie \label{eq:fraction.t}
\xi={s\over rK}\,,\qquad  \gcd(s,r)=1\,.
\fe
This leads to a shift $B(x) \to B(x) - \frac{s}{r}$.
This shift of $B$ is invisible in the subspace of states whose winding charges vanish modulo $r$.
Therefore, for such rational translations, we can find a symmetry by accompanying the translation with a weaker projection operator that sets $\widetilde{Q}^I = 0\bmod r$.

Let us consider the rational translations $x \to x + 2\pi {s\over rK}$ in more detail, and without loss of generality, take  $r \geq 1$.

One way to describe the non-invertible translation symmetry operator is to introduce auxiliary topological degrees of freedom on the operator \cite{Kaidi:2021xfk, Choi:2022jqy, Cordova:2022ieu, Choi:2022fgx}.\footnote{As discussed in \cite{Kaidi:2021xfk, Choi:2022jqy}, the choice of auxiliary degrees of freedom on the operator is, in general, not unique. Below, we make a minimal choice so that the symmetry operator does not depend on decoupled quantum-mechanical degrees of freedom.}
We take
\ie \label{eq:noninvT}
    &\mathcal{D}_{\frac{s}{rK}} = \int [Du Dv] \\
    &~~\exp \left[
        i \int_{x_0}^{x_0 + 2\pi} dx \left(
            \frac{2\pi s}{rK} \hat{\Theta}_{tx} + \frac{r}{2\pi} u \partial_x v - \frac{s}{2\pi} u \partial_x \varphi^1 + \frac{1}{2\pi} v \partial_x \varphi^2
        \right) + \cdots
    \right] \, .
\fe
The ellipses represent localized terms at $x= x_0$, similar to those in \eqref{eq:mosymm_ops} and \eqref{eq:symm_ops}, which we will soon describe explicitly.

Let us unpack \eqref{eq:noninvT}.
Suppose the operator $\mathcal{D}_{s\over rK}$ is inserted at some fixed time slice.
Then, $u (x)$ and $v (x)$ are auxiliary $2\pi$-periodic scalar fields ($u \sim u + 2\pi$, $v \sim v + 2\pi$), which reside on the operator, i.e., on the fixed time slice.
The definition of the operator \eqref{eq:noninvT} involves a path integral over all possible configurations of these auxiliary fields, including topologically nontrivial configurations with nonzero winding numbers of $u$ and $v$ along the spatial circle.
The coupling between $u$ and $v$ is such that they define a fuzzy torus quantum mechanics.
These quantum mechanical degrees of freedom on the operator are then coupled to the 1+1-dimensional fields $\varphi^I$ in a particular way \eqref{eq:noninvT}.

Suppose the auxiliary fields $u$, $v$, as well as the bulk fields $\varphi^I$ were all non-compact, real-valued scalar fields.
Then, in \eqref{eq:noninvT}, it is straightforward to integrate out $u$ and $v$.
The equation of motion of $u$ sets $v =  {s\over r} \varphi^1 + const.$, and integrating out $u$ and $v$ leaves behind the integral of the conserved, modified energy-momentum tensor ${\Theta}_{tx} =\hat{\Theta}_{tx} + \frac{K}{(2\pi)^2} \varphi^1 \partial_x \varphi^2$ \eqref{eq:T} along the spatial circle.
The couplings in \eqref{eq:noninvT} are designed such that this is the case.

The role of the auxiliary fields $u$ and $v$ in \eqref{eq:noninvT} is intuitively as follows. 
The ``naive''  translation operator is $\exp \left[\frac{2\pi i s}{rK} \int_{x_0}^{x_0 + 2\pi} dx  \, \hat{\Theta}_{tx} \right]$.  As we discussed, it shifts $B(x) \to B(x-2\pi\frac{s}{rK})=B(x) - \frac{s}{r}$.
In the non-invertible translation symmetry operator \eqref{eq:noninvT}, this shift of $B(x)$ is compensated by an opposite shift induced by the couplings between the auxiliary fields $u$, $v$, and the bosons $\varphi^I$.
This is analogous to the 3+1d non-invertible chiral symmetry operators studied in \cite{Choi:2022jqy, Cordova:2022ieu} (see also \cite{Kaidi:2021xfk, Choi:2021kmx, Choi:2022fgx}), where the shift of the $\theta$-angle due to an anomalous chiral rotation of fermions is compensated by auxiliary topological degrees of freedom living on the non-invertible chiral symmetry operator.\footnote{Here, if we view $x$ along the operator as Euclidean time, the $(u,v)$ system describes a fuzzy torus with a projectively realized $\bZ_r\times \bZ_r$ global symmetry. Let $C_1$ and $C_2$ be discrete $\bZ_r$ background gauge fields, normalized such that $\oint C_i$ on closed 1-cycles are integers modulo $r$.  The projective realization of the symmetry corresponds to an 't Hooft anomaly described by the two-dimensional inflow action $\exp \left[ \frac{2\pi i}{r} \int C_1 \cup C_2 \right]$. (Our convention is such that we put the inflow action to the future in real time $t$ of the fuzzy torus system.) In the non-invertible translation symmetry operator \eqref{eq:noninvT}, the couplings between $u$, $v$ and the bulk fields $\varphi^I$ are such that we have effectively turned on the $\bZ_r\times \bZ_r$ background fields through $C_1 = - \frac{s}{2\pi} d\varphi^1$ and $C_2 = \frac{1}{2\pi} d\varphi^2$. The inflow action becomes $\exp \left[ \frac{2\pi i}{r} \int C_1 \cup C_2 \right] = \exp \left[ -\frac{is}{2\pi r} \int d\varphi^1 \wedge d\varphi^2 \right]$, and this induces the shift $B(x) \to B(x) + \frac{s}{r}$.}

Let us perform a more careful analysis, taking into account topologically nontrivial winding configurations of $\varphi^I$ as well as those of the auxiliary fields $u$ and $v$.
At the end, we will find a simpler description of the non-invertible symmetry operator \cite{Seiberg:2024wgj, Seiberg:2024yig}.

The non-invertible translation symmetry operator \eqref{eq:noninvT}, including the correction terms at $x = x_0$, is
\begin{align} \label{eq:noninvT_corr}
\begin{split}
    \mathcal{D}_{\frac{s}{rK}} = \int [Du Dv] \exp \Bigg[
        i \int_{x_0}^{x_0 + 2\pi} dx &\Bigg(
            \frac{2\pi s}{rK} \hat{\Theta}_{tx} + \frac{r}{2\pi} u \partial_x v - \frac{s}{2\pi} u \partial_x \varphi^1 + \frac{1}{2\pi} v \partial_x \varphi^2 \Bigg) \\
        & -  i\left( r \widetilde{Q}^{(u)} v  - s \widetilde{Q}^{(u)} \varphi^1  + \widetilde{Q}^{(v)} \varphi^2  \right)\Big|_{x=x_0^+}
    \Bigg] \,,
\end{split}
\end{align}
where
\ie \label{eq:winding_uv}
    \widetilde{Q}^{(u)} &= \frac{1}{2\pi} \int_{x_0}^{x_0 + 2\pi} \partial_x u \, dx = \frac{1}{2\pi} \left[ u (x_0^- ) - u (x_0^+ ) \right] \,,\\
    \widetilde{Q}^{(v)} &= \frac{1}{2\pi} \int_{x_0}^{x_0 + 2\pi} \partial_x v \, dx = \frac{1}{2\pi} \left[ v (x_0^- ) - v (x_0^+ ) \right] \,,
\fe
are the winding numbers of the auxiliary fields. 
Here, similar to the bosons $\varphi^I$ \eqref{eq:winding_bosons}, we take the auxiliary fields $u$ and $v$ to be single-valued and real for $x \neq x_0$, and their transition functions at $x=  x_0$ are given by the winding numbers.

The total expression \eqref{eq:noninvT_corr} including the correction terms is independent of the choice of the reference point $x_0$.
The path integral over $u$ and $v$ involves two infinite sums over the winding numbers \eqref{eq:winding_uv}.
Explicitly integrating out $u$ and $v$ (see Appendix \ref{app:noninv.cont}), we obtain\footnote{The overall normalization of the operator $\mathcal{D}_{s\over rK}$ is determined such that its quantum dimension is given by the partition function of the fuzzy torus quantum mechanics of $u$ and $v$. It is related to the fact that the Hilbert space of the fuzzy torus is $r$-dimensional.\label{factorofr}}
\ie \label{eq:noninv_proj}
    \mathcal{D}_{\frac{s}{rK}} = r \,\mathbb{P}_r^1 \mathbb{P}_r^2 \exp \left[
        \frac{2\pi i s}{rK} \left(
            \int_{x_0}^{x_0+2\pi} {\Theta}_{tx}\, dx  - \frac{K}{2\pi } \widetilde{Q}^1 \varphi^2 \Big|_{x=x_0^+}
        \right)
    \right] \,,
\fe
where
\begin{equation} \label{eq:projectors_cont}
    \mathbb{P}_r^I = \frac{1}{r}\sum_{l=0}^{r-1} e^{\frac{2 \pi i l}{r} \widetilde{Q}^I} \,, \quad I = 1,2 \,,
\end{equation}
are projection operators.

The expression \eqref{eq:noninv_proj} with projection operators is consistent with the intuitive picture we presented earlier in this section.
Note that without the projection operators, the exponential in \eqref{eq:noninv_proj} is not well-defined.
Therefore, the operator $\mathcal{D}_{s\over rK}$ does not factorize as a projection operator times another well-defined operator.

Finally, when $r = 1$, $\mathcal{D}_{s\over rK}$ coincides with the invertible translation symmetry operator $\mathcal{P}^s$ \eqref{eq:symm_ops}.

The presentation \eqref{eq:noninv_proj} of the operator $\mathcal{D}_{s\over rK}$ makes it manifest that it is preserved under time evolution.  On the other hand, the earlier presentation \eqref{eq:noninvT} with auxiliary fields makes the operator's locality manifest.

Let us discuss the fusion algebra obeyed by the non-invertible symmetry operators $\mathcal{D}_{s\over rK}$.
First, for any fixed $s$ and $r$, we have
\ie \label{eq:Dalgebra}
    \mathcal{D}_{\frac{s}{rK}}^\dagger &= \mathcal{D}_{-\frac{s}{rK}} \,,\\
    \mathcal{D}_{\frac{s}{rK}}^\dagger \times \mathcal{D}_{\frac{s}{rK}} &= r^2 \mathbb{P}_r^1 \mathbb{P}_r^2 = \left(
        \sum_{l_1=0}^{r-1} e^{\frac{2\pi i l_1}{r} \widetilde{Q}^1}
    \right) \left(
        \sum_{l_2=0}^{r-1} e^{\frac{2\pi i l_2}{r} \widetilde{Q}^2}
    \right) \,,\\
    e^{\frac{2\pi i }{r} \widetilde{Q}^I} \mathcal{D}_{\frac{s}{rK}} &= \mathcal{D}_{\frac{s}{rK}} e^{\frac{2\pi i }{r} \widetilde{Q}^I} = \mathcal{D}_{\frac{s}{rK}} \,,\\
    U_I \mathcal{D}_{\frac{s}{rK}} &= e^{ \frac{2\pi i s }{rK} \epsilon_{IJ} \widetilde{Q}^J} \mathcal{D}_{\frac{s}{rK}} U_I \,.
\fe
We find that the algebra involving the non-invertible translation symmetry and the target space momentum symmetry is extended by the winding symmetry. This is similar to the extension of the invertible translations \eqref{eq:cont-translation}.

Next, let us compute the general fusion rule of two non-invertible translation symmetry operators $\mathcal{D}_{s_1 \over r_1 K} \times \mathcal{D}_{s_2 \over r_2 K}$ where $\mathrm{gcd}(s_1 , r_1) = \mathrm{gcd}(s_2 , r_2) = 1$ and $r_1, r_2 \geq 1$.
Write
\begin{equation}
    \frac{s_1}{r_1} + \frac{s_2}{r_2} = \frac{s_{12}}{r_{12}}
\end{equation}
with $\mathrm{gcd}(s_{12} , r_{12}) = 1$ and $r_{12} \geq 1$.
Then,
\ie \label{eq:genFusion}
    \mathcal{D}_{\frac{s_1}{r_1 K}} \times \mathcal{D}_{\frac{s_2}{r_2 K}} &= \frac{r_1 r_2}{r_{12}} \, \mathbb{P}^1_{\mathrm{lcm}(r_1 , r_2)} \, \mathbb{P}^2_{\mathrm{lcm}(r_1 , r_2)} \, \mathcal{D}_{\frac{s_{12}}{r_{12}K}} \\ &=\frac{r_{12}\,\mathrm{gcd}(r_1 , r_2)}{\mathrm{lcm}(r_1 , r_2)}  \left(
        \sum_{l_1,l_2=0}^{\frac{\mathrm{lcm}(r_1 , r_2)}{r_{12}}-1} e^{ \frac{2\pi i }{\mathrm{lcm}(r_1 , r_2)} (l_1 \widetilde{Q}^1 + l_2 \widetilde{Q}^2 ) } \, \mathcal{D}_{\frac{s_{12}}{r_{12}K}}
    \right) \, .
\fe
This is a linear combination of simple symmetry operators, as can be seen by expanding the projection operators in the first line.
In the second line, the fusion coefficients ${r_{12}\,\mathrm{gcd}(r_1, r_2)\over \mathrm{lcm}(r_1, r_2)}$ are always positive integers.\footnote{See \cite{Copetti:2023mcq} for an analogous fusion algebra obeyed by non-invertible chiral symmetry operators in 3+1d.}

As we mentioned earlier, a continuous non-invertible translation symmetry shifting $x \to x + 2\pi \xi$ with an arbitrary real $\xi$ can also be defined by using projection operators that enforce  $\widetilde{Q}^I=0$. A presentation of such an operator similar to that in \eqref{eq:noninvT} requires using auxiliary quantum mechanical degrees of freedom with an infinite-dimensional Hilbert space.
See \cite{Arbalestrier:2024oqg, Karasik:2022kkq, GarciaEtxebarria:2022jky, Brennan:2024fgj, Antinucci:2024zjp} for related discussions on 3+1d chiral symmetries.

In Section \ref{sec:modeexp}, we emphasized the decoupling of the Fourier modes.  Let us discuss it in more detail. The non-invertible translation operator $\mathcal{D}_{\frac{s}{rK}}$ annihilates the states with $\widetilde{Q}^I\neq 0 \bmod r$, so as an operator equation, the fact that it commutes with the Hamiltonian constrains only the states with $\widetilde{Q}^I=0\bmod r$. However, as we stressed, $\mathcal{D}_{\frac{s}{rK}}$ has the local expression \eqref{eq:noninvT}. Therefore, the fact that it commutes with the Hamiltonian cannot depend on the global winding sector and thus constrains the Hamiltonian and the spectrum also in sectors with $\widetilde{Q}^I\ne 0\bmod r$. For a quadratic Hamiltonian, like ours, this implies that the Fourier modes decouple in every winding sector, as we found in \eqref{mode.decoupling}.\footnote{To see that this conclusion depends on the Hamiltonian being quadratic, consider deforming it by $\int dx\,\cos(K\varphi^1)$.  The Hamiltonian still commutes with $U_I$, $\mathcal{P}$, and all the $\mathcal{D}_{\frac{s}{rK}}$, but it makes the theory interacting. In the sector with $\widetilde{Q}^I=0$, it still conserves the spatial momentum of the oscillators, as required by the non-invertible translations. However, in the sector with $\widetilde{Q}^1=1$, $\widetilde{Q}^2=0$, which these non-invertible operators annihilate, it transfers $K$ units of spatial momentum between the oscillators and the zero modes, which is consistent with the invertible translation $\mathcal{P}$.}

Finally, one might try to repeat this reasoning to find a non-invertible target space momentum symmetry larger than $\bZ_{|K|}\times \bZ_{|K|}$ by projecting on certain values of $\widetilde{Q}^I$.  However, even if we constrain the winding number of a state, we can act on it with a pair of winding operators with winding charges $\pm 1$.  Then, as in the discussion at the end of Section \ref{sec:momentum}, we can move one of these operators around the space and change the momentum charge of the state by an arbitrary multiple of $K$.  This shows that even with fixed $\widetilde{Q}^I$, there are transitions that change the momentum charges by $K$ units, thus showing that there is no non-invertible momentum symmetry larger than $\bZ_{|K|}\times \bZ_{|K|}$.

\section{Hamiltonian formulation of the continuum system with varying $B$}\label{oursystem}

So far, we have followed mostly a path-integral approach based on the Lagrangian formulation of the theory.\footnote{Notable exceptions have been the Hamiltonian of the continuum theory for constant $B$ in
Section \ref{cistwo}, and some aspects of the symmetries with varying $B$ in Section \ref{sec:trasnlation_continuum}.} 
To prepare for the comparison with the lattice models in the later sections, we provide additional details on the Hamiltonian formulation of the continuum theory.

We begin with the Euclidean action (see Appendix \ref{app:bulk}),
\begin{align} \label{eq:action1Euc}
\begin{split}
    S &= \int_{\tau_0}^{\tau_0 + 2\pi} d\tau \int_{x_0}^{x_0 + 2\pi} dx  \left[
         \frac{1}{4\pi} G_{IJ} ( \partial_\tau \varphi^I \partial_\tau \varphi^J + \partial_x \varphi^I \partial_x \varphi^J ) - 2\pi iB(x)  q
    \right] \\
    &~~-\frac{iK}{2\pi} \int_{\tau_0}^{\tau_0 + 2\pi} d\tau \, \varphi^1 \partial_\tau \varphi^2 \Big|_{x=x_0^+} + iK \widetilde{Q}_{\tau}^1 \varphi^2 (x_0^+ , \tau_0^+ ) \, .
\end{split}
\end{align}
As in \eqref{eq:winding_B}, we have a space-dependent $B(x)$ with the winding number $K\in\mathbb{Z}$, and
\begin{equation} \label{eq:winding_time}
    \widetilde{Q}_{\tau}^I = \frac{1}{2\pi} \left[ 
        \varphi^I (\tau_0^- ) - \varphi^I (\tau_0^+ ) 
    \right] = \frac{1}{2\pi} \int_{\tau_0}^{\tau_0 + 2\pi} d\tau \,\partial_\tau \varphi^I \in \mathbb{Z} 
\end{equation}
are the winding numbers of the bosons along the Euclidean time direction.

Integrating by parts, we can rewrite the action \eqref{eq:action1Euc} as
\begin{align} \label{eq:action1Euc2}
\begin{split}
    S &= \int_{\tau_0}^{\tau_0 + 2\pi} d\tau \int_{x_0}^{x_0 + 2\pi} dx  \left[
         \frac{1}{4\pi} G_{IJ} ( \partial_\tau \varphi^I \partial_\tau \varphi^J + \partial_x \varphi^I \partial_x \varphi^J ) - 2\pi iB(x)  q
    \right] \\
    &~~-\frac{iK}{4\pi} \int_{\tau_0}^{\tau_0 + 2\pi} d\tau \, \epsilon_{IJ} \varphi^I \partial_\tau \varphi^J \Big|_{x=x_0^+} + \frac{iK}{2} \epsilon_{IJ} \widetilde{Q}_{\tau}^I \varphi^J (x_0^+ , \tau_0^+ ) - i\pi K \widetilde{Q}_{\tau}^1 \widetilde{Q}_{\tau}^2 \,,
\end{split}
\end{align}
or as
\begin{align} \label{eq:action2Euc}
\begin{split}
    S &= \int_{\tau_0}^{\tau_0 + 2\pi} d\tau \int_{x_0}^{x_0 + 2\pi} dx  \left[
         \frac{1}{4\pi}  G_{IJ} ( \partial_\tau \varphi^I \partial_\tau \varphi^J + \partial_x \varphi^I \partial_x \varphi^J ) \right.\\
        &\qquad \qquad\qquad\qquad\qquad\qquad\left.-\frac{i}{4\pi} \partial_x B(x) \epsilon_{IJ} \varphi^I \partial_\tau \varphi^J \, -  2\pi i B(x_0^+ ) q
    \right] \\
    &~~ + \frac{i}{2} \int_{x_0}^{x_0 + 2\pi} dx  \,\partial_x B(x) \epsilon_{IJ} \widetilde{Q}_{\tau}^I  \varphi^J \Big|_{\tau = \tau_0^+} - i\pi K \widetilde{Q}_{\tau}^1 \widetilde{Q}_{\tau}^2 \, .
\end{split}
\end{align}
(In the Euclidean actions, we omit the terms that are valued in $2\pi i \mathbb{Z}$.)

The three Euclidean actions \eqref{eq:action1Euc}, \eqref{eq:action1Euc2}, and \eqref{eq:action2Euc} are all identical, configuration by configuration, in a functional integral.
In Section \ref{sec:continuum}, we discussed two Lorentzian actions, \eqref{eq:action1} and \eqref{eq:action2}.
They are obtained from the Euclidean actions \eqref{eq:action1Euc2} and \eqref{eq:action2Euc}, respectively, by
making Euclidean time non-compact, dropping the terms that depend on $\widetilde{Q}_{\tau}^I$, and continuing to Lorentzian signature.

In the Hamiltonian formulation of the theory, the terms in the Euclidean actions that depend on $\widetilde{Q}_{\tau}^I$, or equivalently, the correction terms that are localized near $\tau = \tau_0$, play an important role. In particular, we will see that we can embed the Hilbert space of the theory in a bigger Hilbert space $\mathscr H$, where the bosons are taken to be non-compact, and coupled to a $\mathbb{Z} \times \mathbb{Z}$ gauge field. 
The correction terms near $\tau = \tau_0$ are related to the $\bZ \times \bZ$ Gauss's law, which effectively makes the bosons compact.  As we will explain in Section \ref{comparingcontla}, this point of view facilitates a detailed comparison between the continuum theory and its lattice realization using the modified Villain formulation \cite{Cheng:2022sgb, Fazza:2022fss}.

We will discuss the Hamiltonian formulation of the theory based on the actions \eqref{eq:action1Euc2} and \eqref{eq:action2Euc}.
In the two cases, we are led to two different Hilbert spaces, $\cal H$ and $\mathcal{H}'$, both of which are subspaces of the bigger Hilbert space $\mathscr H$.
We will find that the two Hilbert spaces are mapped into each other by a unitary operator acting on $\mathscr H$.

Some of the subtleties below are familiar from the quantum mechanics of a particle on a torus with a constant magnetic field, which we review in Appendix \ref{particleontorus} as a warm-up to this section.

\subsection{Hamiltonian formulation of the action \eqref{eq:action1Euc2}} \label{sec:H1}

We begin by studying the Hamiltonian formulation of the system based on the Euclidean action \eqref{eq:action1Euc2}.
To do so, we first drop the terms near $\tau = \tau_0$, and continue to Lorentzian time.
The corresponding Lorentzian action is \eqref{eq:action1}.
The momenta conjugate to $\varphi^I$ that follow from \eqref{eq:action1} are
\ie \label{eq:pi_app}
\pi_I(x)=\frac{1}{2 \pi} G_{I J} \partial_t \varphi^J+\frac{B(x)}{2 \pi} \epsilon_{I J} \partial_x \varphi^J-\frac{K}{4 \pi} \epsilon_{I J} \varphi^J\left(x_0^+\right) \delta\left(x-x_0^+\right) \,.
\fe
The Hamiltonian takes the form 
\ie \label{eq:Happ1}
H_{x_0} &= \int_{x_0}^{x_0+2 \pi} d x\left[\pi\, G^{I J}\left(\pi_I-\frac{B\left(x\right)}{2 \pi} \epsilon_{I I'} \partial_x \varphi^{I'}+\frac{K}{4 \pi} \epsilon_{I I'} \varphi^{I'} (x_0^+ ) \delta (x- x_0^+ ) \right)\right. \\
&~ \left.\times\left(\pi_J-\frac{B\left(x\right)}{2 \pi} \epsilon_{J J'} \partial_x \varphi^{J'}+\frac{K}{4 \pi} \epsilon_{J J'} \varphi^{J'} (x_0^+ ) \delta (x-x_0^+ ) \right)+\frac{1}{4 \pi} G_{I J} \partial_x \varphi^I \partial_x \varphi^J\right] \,.
\fe
It contains singular terms with the delta function $\delta (x- x_0^+ )$.
Such singularities are avoided and made precise in the lattice realization of the model that will be discussed in Section \ref{sec:windingBLat}.
Also, in an alternative description of the theory in Section \ref{sec:H2}, these singular terms will not be present.

Let us denote the wavefunctional of the system by $\Psi \left[\varphi^1 (x), \varphi^2 (x)\right]$, which is a functional of the compact boson fields $\varphi^I (x)$.
We claim that the correction terms in the Euclidean action \eqref{eq:action1Euc2} that are localized at $\tau = \tau_0$ imply that the wavefunctional must satisfy
\begin{align} \label{eq:wfn1}
\begin{split}
    \Psi \left[\varphi^1 (x) + 2\pi , \varphi^2 (x)\right] &= \exp \left(
        \frac{iK}{2} \varphi^2 (x_0^+ )
    \right) \Psi \left[\varphi^1 (x), \varphi^2 (x) \right] \,,\\
    \Psi \left[\varphi^1 (x) , \varphi^2 (x) + 2\pi \right] &= \exp \left(
        -\frac{iK}{2} \varphi^1 (x_0^+ )
    \right) \Psi \left[\varphi^1 (x) , \varphi^2 (x) \right] \,.
\end{split}
\end{align}
We denote the Hilbert space of wavefunctionals satisfying \eqref{eq:wfn1} by $\cal H$.

As we review in Appendix \ref{particleontorus}, when restricted to the zero modes of $\varphi^I$, \eqref{eq:wfn1} is the familiar fact that the wavefunction of a particle on a torus with a constant magnetic field is not a function, but rather a section of a line bundle.

Another way to understand \eqref{eq:wfn1} is as follows.
In the Euclidean action \eqref{eq:action1Euc2}, the bosons jump discontinuously by $2\pi \widetilde{Q}_{\tau}^I$ at $\tau = \tau_0$ as in \eqref{eq:winding_time}, and the terms that depend on $\widetilde{Q}_{\tau}^I$ are related to the phase that the wavefunctional acquires when the bosons are shifted by $2\pi\widetilde{Q}_{\tau}^I$.
In fact, from \eqref{eq:wfn1}, it follows that
\begin{equation} \label{eq:arbitrary_shifts}
    \Psi \left[\varphi^I (x) + 2\pi \widetilde{Q}_{\tau}^I  \right] = \exp \left( \frac{iK}{2} \epsilon_{IJ} \widetilde{Q}_{\tau}^I \varphi^J (x_0^+  ) - i\pi K \widetilde{Q}_{\tau}^1 \widetilde{Q}_{\tau}^2 \right) \Psi \left[ \varphi^I (x) \right] \,,
\end{equation}
for arbitrary integers $\widetilde{Q}_{\tau}^I \in \mathbb{Z}$.
The phase factor on the right-hand side is precisely the correction terms near $\tau = \tau_0$ in the Euclidean action \eqref{eq:action1Euc2}.

The Hilbert space $\mathcal{H}$ can equivalently be characterized as follows.
Let $\mathscr H$ be the Hilbert space of arbitrary wavefunctionals of the bosons $\varphi^I (x)$, where we do not impose \eqref{eq:wfn1}.
On this Hilbert space, the conjugate momenta \eqref{eq:pi_app} act as functional derivatives,
\begin{equation}
    \pi_I (x) = -i \frac{\delta}{\delta \varphi^I (x)}  \qquad ~~~~\text{on $\mathscr H$} \,.
\end{equation}
Then, the Hilbert space $\mathcal{H}$ of wavefunctionals satisfying the conditions \eqref{eq:wfn1} is a subspace of $\mathscr H$ on which the following constraints are satisfied:
\begin{align} \label{eq:C1}
\begin{split}
    \mathcal{C}_I &= \exp \left[ 
        2\pi i \left( \int_{x_0}^{x_0 + 2\pi} \pi_I (x) \, dx - \frac{K}{4\pi} \epsilon_{IJ} \varphi^J \Big|_{x=x_0^+} \right)
    \right] \\
    &= \exp \left[ 
        2\pi i \left( \int_{x_0}^{x_0 + 2\pi} \left( -i \frac{\delta}{\delta \varphi^I (x)}\right) \, dx - \frac{K}{4\pi} \epsilon_{IJ} \varphi^J \Big|_{x=x_0^+} \right)
    \right] \\
    &= 1  \qquad~~~~\text{on $\mathcal{H} \subset {\mathscr H}$}\,.
\end{split}
\end{align}
Importantly, in the exponents of the constraints \eqref{eq:C1}, the additional term after the spatial integral of $\pi_I$ is dictated by the correction terms near $\tau = \tau_0$ in \eqref{eq:action1Euc2}.

The constraints \eqref{eq:C1} are related to the target space momentum symmetry operators $U_I$, in particular the first line of \eqref{eq:mosymm_ops}, as
\begin{equation} \label{eq:CU}
    \mathcal{C}_I=U_I^K = \exp \left[ 
        2\pi i \left( \int_{x_0}^{x_0 + 2\pi} J_{t,I} \, dx - \frac{K}{2\pi}\epsilon_{IJ} \varphi^J \Big|_{x=x_0^+} \right)
    \right] \,,
\end{equation}
with the currents $J_{\mu,I}$ given in \eqref{firstcur}.
Note that $U_I^K$ is the operator that shifts the boson field $\varphi^I$ by $2\pi$, and the constraints \eqref{eq:C1} implement the $2\pi$-periodicities of the bosons in the bigger Hilbert space $\mathscr H$.

A natural physical interpretation of the larger Hilbert space $\mathscr{H}$ and the constraints \eqref{eq:C1} is as follows.
We may equivalently realize our theory of two compact bosons $\varphi^I$ by first letting them be non-compact, and then gauging a $\mathbb{Z} \times \mathbb{Z}$ momentum symmetry of the non-compact bosons.
Then, the winding charges $\widetilde{Q}^I$ correspond to the holonomy of the dynamical $\mathbb{Z} \times \mathbb{Z}$ gauge field around the space, and the Hilbert space $\mathscr{H}$ is that of the two non-compact bosons coupled to the $\mathbb{Z} \times \mathbb{Z}$ gauge field.
The constraints \eqref{eq:C1} are the $\mathbb{Z} \times \mathbb{Z}$ Gauss's law, which restricts the unconstrained Hilbert space $\mathscr{H}$ to the physical subspace $\mathcal{H} \subset \mathscr{H}$.

Finally, we note that the Hamiltonians $H_{x_0}$ \eqref{eq:Happ1} for different values of $x_0$ are related to each other by unitary transformations acting on the Hilbert space $\mathscr{H}$.
This will be demonstrated later with the lattice realization of the system.

\subsection{Hamiltonian formulation of the action \eqref{eq:action2Euc}} \label{sec:H2}

Let us now discuss the Hamiltonian formulation of the theory, based on the other, equivalent expression \eqref{eq:action2Euc} of the action.
Again, we first drop the terms near $\tau = \tau_0$, and continue to the Lorentzian action in \eqref{eq:action2}. 
The momenta conjugate to $\varphi^I$ that follow from \eqref{eq:action2} are
\ie \label{app:pi.tilde}
\pi_I(x)=\frac{1}{2 \pi} G_{I J} \partial_t \varphi^J+\frac{B\left(x_0^+\right)}{2 \pi} \epsilon_{I J} \partial_x \varphi^J-\frac{\partial_x B(x)}{4 \pi} \epsilon_{I J} \varphi^J\,.
\fe
For later convenience, we introduce
\ie \label{eq:BigPicont}
    \Pi_I (x) & = \hat{J}_{t,I} = \frac{1}{2\pi}G_{I J} \partial_t \varphi^J \\
    &= \pi_I (x) - \frac{B\left(x_0^+\right)}{2 \pi} \epsilon_{I J} \partial_x \varphi^J + \frac{\partial_x B(x)}{4 \pi} \epsilon_{I J} \varphi^J \,.
\fe
Here, $\hat{J}_{\mu,I}$ are the well-defined, not-conserved currents for the target space momentum symmetry that were given in \eqref{thirdcur}.
The Hamiltonian is 
\ie \label{eq:Happ2}
H'_{x_0}& =\int_{x_0}^{x_0+2 \pi} d x\left[\pi\, G^{I J}\left(\pi_I (x) -\frac{B\left(x_0^+\right)}{2 \pi} \epsilon_{I I'} \partial_x \varphi^{I'}+\frac{\partial_x B(x)}{4 \pi} \epsilon_{I I'} \varphi^{I'}\right)\right. \\
&\qquad \quad \left.\times\left(\pi_J (x) -\frac{B\left(x_0^+\right)}{2 \pi} \epsilon_{J J'} \partial_x \varphi^{J'}+\frac{\partial_x B(x)}{4 \pi} \epsilon_{J J'} \varphi^{J'}\right)+\frac{1}{4 \pi} G_{I J} \partial_x \varphi^I \partial_x \varphi^J\right] \\
&= \int_{x_0}^{x_0+2 \pi} d x \left[ \pi\, G^{IJ} \Pi_I \Pi_J + \frac{1}{4 \pi} G_{I J} \partial_x \varphi^I \partial_x \varphi^J\right] \,.
\fe

The expressions for the conjugate momenta $\pi_I (x)$ in \eqref{eq:pi_app} and those in \eqref{app:pi.tilde} are different as functions of $\partial_t \varphi^I$ and $\varphi^I$. 
The reason is that $\varphi^I(x)$ and $\pi_I(x)=-i {\delta \over \delta\varphi^I(x)}$ are the same in the two cases, while in one presentation,  $\partial_t \varphi^I(x)=i[H_{x_0},\varphi^I(x)]$, and in the other, $\partial_t \varphi^I(x)=i[ H'_{x_0},\varphi^I(x)]$.

Let us denote the wavefunctionals of the system by $\Psi'\left[\varphi^1 (x) , \varphi^2 (x) \right]$.
From the correction terms near $\tau = \tau_0$ in the Euclidean action \eqref{eq:action2Euc}, we deduce that
\ie \label{eq:wfn2}
    \Psi' \left[\varphi^1 (x) + 2\pi , \varphi^2 (x) \right] &= \exp \left( 
        \frac{i}{2} \int_{x_0}^{x_0 + 2\pi} dx \, \partial_x B(x) \varphi^2
    \right) \Psi' \left[\varphi^1 (x) , \varphi^2 (x) \right] \,,\\
    \Psi' \left[\varphi^1 (x) , \varphi^2 (x) + 2\pi  \right] &= \exp \left( 
        -\frac{i}{2} \int_{x_0}^{x_0 + 2\pi} dx \, \partial_x B(x) \varphi^1
    \right) \Psi' \left[\varphi^1 (x) , \varphi^2 (x) \right] \,.
\fe
We denote the Hilbert space of wavefunctionals satisfying \eqref{eq:wfn2} by $\mathcal{H}'$.
Similar to \eqref{eq:arbitrary_shifts}, it is straightforward to check that the phase factor that the wavefunctional acquires when the bosons $\varphi^I$ are shifted by $2\pi \widetilde{Q}_{\tau}^I$ agrees with the correction terms near $\tau = \tau_0$ in the Euclidean action \eqref{eq:action2Euc}.

The Hilbert space $\mathcal{H}'$ can be characterized as a subspace of the bigger Hilbert space $\mathscr H$, on which the following constraints are satisfied:
\ie \label{eq:C2}
    \mathcal{C}'_I &= \exp \left[
        2\pi i \int_{x_0}^{x_0 + 2\pi} \,dx \left( \pi_I (x) 
            - \frac{\partial_x B(x)}{4\pi} \epsilon_{IJ} \varphi^J (x)
        \right)
    \right] \\
    &= \exp \left[
        2\pi i \int_{x_0}^{x_0 + 2\pi} \,dx \left( -i \frac{\delta}{\delta \varphi^I (x)}
            - \frac{\partial_x B(x)}{4\pi} \epsilon_{IJ} \varphi^J (x)
        \right)
    \right] \\
    &= 1 ~~~~\text{on $\mathcal{H}' \subset {\mathscr H}$} \,.
\fe
As before, the form of the constraints \eqref{eq:C2} is determined by the correction terms near $\tau = \tau_0$ in the Euclidean action \eqref{eq:action2Euc}.

The constraints \eqref{eq:C2} are again related to $U_I$; using the second line of \eqref{eq:mosymm_ops}, we have
\begin{equation} \label{eq:tCU}
    \mathcal{C}'_I = U_I^K = \exp \left[ 
        2\pi i  \int_{x_0}^{x_0 + 2\pi}\,dx \left( J'_{t,I} + \frac{B(x_0^+ )}{2\pi} \epsilon_{IJ} \partial_x \varphi^J \right) 
    \right] \,,
\end{equation}
with the currents $J'_{\mu ,I}$ defined in \eqref{secondcur}.\footnote{Note that, the two expressions for $U_I$ in \eqref{eq:mosymm_ops} are the same when written in terms of $\varphi^I$, $\partial_t \varphi^I$ and $\partial_x \varphi^I$. However, as we said after \eqref{eq:Happ2}, in the Hamiltonian formulation, $\partial_t \varphi^I$ depends on the Hamiltonian, either $H_{x_0}$ of \eqref{eq:Happ1} or $H'_{x_0}$ of \eqref{eq:Happ2}.
Therefore, to be precise, $U_I$ in \eqref{eq:tCU} is a different operator from $U_I$ in \eqref{eq:CU}, but for conciseness, we denote both of them by $U_I$. (The $U_I$'s in the two cases are related by the conjugation by the unitary operator $\cal U$ that we will introduce below in \eqref{eq:U_cont}.)}

Again, the constraints \eqref{eq:C2} can be interpreted as the $\mathbb{Z} \times \mathbb{Z}$ Gauss's law, imposed on the unconstrained Hilbert space $\mathscr{H}$ of non-compact bosons.
Then, $\mathcal{H}' \subset \mathscr{H}$ is the physical subspace of states satisfying Gauss's law.

\subsection{Relation between the two Hamiltonian formulations} \label{sec:rel}

The two Hamiltonian formulations given in Sections \ref{sec:H1} and \ref{sec:H2} are related by a unitary operator ${\cal U}:{\mathscr H} \to {\mathscr H}$,
\ie \label{eq:U_cont}
    {\cal U} &=   \exp \Bigg[
        -\frac{i}{2\pi} \int_{x_0}^{x_0 + 2\pi} dx \left( (B(x) - B(x_0^+ )) \varphi^1 \partial_x \varphi^2 + \frac{1}{2} (\partial_x B) \varphi^1 \varphi^2 \right) \\
        &\qquad\qquad + \frac{iK}{4\pi} \varphi^1 (x_0^+ ) \varphi^2 (x_0^+ ) + iK \widetilde{Q}^1 \varphi^2 (x_0^+ )
    \Bigg]\,.
\fe
Given a wavefunctional $\Psi \in \mathcal{H} \subset \mathscr H$ satisfying \eqref{eq:wfn1}, $\Psi' = {\cal U}\Psi$ satisfies \eqref{eq:wfn2} and is hence an element of $\mathcal{H}' \subset \mathscr H$.
That is, ${\cal U}$ maps the subspace $\mathcal{H} \subset \mathscr H$ to the subspace $\mathcal{H}' \subset \mathscr H$.
Moreover, the two Hamiltonians \eqref{eq:Happ1} and \eqref{eq:Happ2}, as well as the $\bZ \times \bZ$ Gauss's laws \eqref{eq:C1} and \eqref{eq:C2}, are related by
\begin{equation} \label{eq:unitary_transformation_cont}
    H'_{x_0} = {\cal U} H_{x_0} {\cal U}^{-1} \,, \quad {\cal C}'_I = {\cal U} \mathcal{C}_I {\cal U}^{-1} \,.
\end{equation}

This picture with the unitary operator ${\cal U}$ can also be seen from the relation between the functional integrals based on the two equivalent Euclidean actions \eqref{eq:action1Euc2} and \eqref{eq:action2Euc}.
These functional integrals represent a trace over the Hilbert space.
The functional integral based on the action \eqref{eq:action1Euc2} is a trace over $\cal H$ of evolution using the Hamiltonian $H_{x_0}$ of \eqref{eq:Happ1}. 
On the other hand, the functional integral based on the action \eqref{eq:action2Euc} is a trace over ${\cal H}'$ of evolution using the Hamiltonian $H'_{x_0}$ of \eqref{eq:Happ2}.

Hence, the same Euclidean partition function can be interpreted in several ways, 
\ie\label{insertintrace}
    \Tr_{\cal H} \left( e^{-2\pi H_{x_0}}\right) = \Tr_{{\cal H}'} \left(e^{-2\pi  H'_{x_0} }\right) = \Tr_{\cal H} \left({\cal U}^{-1} e^{-2\pi  H'_{x_0}}\ {\cal U} \right) \,.
\fe
In the last expression, we take the trace over the Hilbert space $\cal H$ that follows from the action \eqref{eq:action1Euc2}, but we evolve using the Hamiltonian $H'_{x_0}$ that is obtained from the action \eqref{eq:action2Euc}, and we have an additional conjugation by ${\cal U}$.

\section{A modified Villain realization of the system} \label{sec:modifiedVillain}

\subsection{The model and its symmetries}

Here, we discuss a lattice realization of our system using the Hamiltonian formulation of the modified Villain model \cite{Cheng:2022sgb, Fazza:2022fss}. We take space to be a periodic chain with $L$ sites. At each site, there are two real-valued fields $\phi_j^{I}$ and their conjugate momenta $p_{j,I}$, such that
\ie \label{eq:lat_canonical}
\left[\phi_j^{I},p_{j',J}\right]=i \delta_{jj'}\delta^I_J\, .
\fe

Following Villain \cite{Villain:1974ir}, we make ${\phi}_j^I$ effectively compact, by gauging a $\bZ \times \bZ$ symmetry that shifts ${\phi}_j^I$ by integer multiples of $2\pi$.  This is achieved by adding on the links $\bZ\times\bZ$ gauge fields ${n}_{j,j+1}^{I}\in \bZ$ and their ``electric fields'' ${E}_{j,j+1;I} \sim {E}_{j,j+1;I}+2\pi$ with
\ie
\left[E_{j,j+1;I},n_{j',j'+1}^{J}\right]=i \delta_{jj'}\delta_I^J\, .
\fe
This $\bZ\times \bZ$ gauge theory implies the
Gauss's law constraints
\ie\label{ZZGauss}
&e^{2\pi i q_{j,I}}=1 \,,\\
&q_{j,I}=p_{j,I}-\frac{E_{j,j+1;I}-E_{j-1,j;I}}{2\pi}\, .
\fe
We find it convenient to replace the discrete $n^I_{j,j+1}$ and the circle-valued ${E}_{j,j+1;I}$ by real fields and impose additional constraints
\ie\label{anotherconst}
e^{2\pi i n_{j,j+1}^{I}} = 1\, .
\fe

Then, we take the Hamiltonian
\ie\label{Hintersofphi}
H=&\frac{1}{2}\sum_j G^{IJ}
\left[ p_{j,I} -\frac{B}{2}\epsilon_{II'}
\left(n^{I'}_{j,j+1}+n^{I'}_{j-1,j}+\frac{\phi^{I'}_{j+1}-\phi^{I'}_{j-1}}{2\pi}\right)\right]\\
&\qquad\qquad\cdot\left[p_{j,J}-\frac{B}{2}\epsilon_{JJ'}
\left(n^{J'}_{j,j+1}+n^{J'}_{j-1,j}+\frac{\phi^{J'}_{j+1}-\phi^{J'}_{j-1}}{2\pi}\right)\right]\\
&+\frac{1}{2}\sum_j G_{IJ}\left(n^I_{j,j+1}+\frac{\phi^I_{j+1}-\phi^I_j}{2\pi}\right)\left(n^J_{j,j+1}+\frac{\phi^J_{j+1}-\phi^J_j}{2\pi}\right).
\fe
Importantly, this $H$ commutes with the $\bZ \times \bZ$ Gauss's law \eqref{ZZGauss} and the constraints \eqref{anotherconst}.  We will identify $G_{IJ}$ and $B$ with these parameters in the continuum model \eqref{eq:lag}.

The key point of the modified Villain model \eqref{Hintersofphi} (as opposed to the standard Villain model) is that we take the $\bZ\times\bZ$ gauge coupling $g$ to zero \cite{Gross:1990ub, Sulejmanpasic:2019ytl, Gorantla:2021svj, Cheng:2022sgb, Fazza:2022fss}.  This means that we do not include in the Hamiltonian the usual $-g^2 \sum_{j, I}\cos\left({E}_{j,j+1;I}\right)$ kinetic term.

This model has two $U(1)$ (target space) momentum and two $U(1)$ winding symmetries associated with the conserved charges\footnote{The standard Villain model has only the two momentum symmetries.  By setting the $\bZ\times\bZ$ gauge coupling $g$ to zero, we gained the two winding symmetries.}
\ie
Q_{I} = \sum_j q_{j,I}\,, \qquad \widetilde{Q}^{I} = \sum_j n_{j,j+1}^{I}\,.
\fe
The local charge densities ${q}_{j,I}$ and ${n}_{j,j+1}^I$ have integer eigenvalues.  However, they are not $\bZ\times \bZ$ gauge invariant.
Alternatively, we can write the charges as
\ie\label{gaugeinvachar}
Q_{I} = \sum_j {p}_{j,I} \,, \qquad \widetilde{Q}^I = \sum_j {\Delta}_{j,j+1}^I\,,
\fe
where
\ie\label{DeltaIdef}
\Delta_{j,j+1}^{I} = \frac{\phi_{j+1}^{I}-\phi_{j}^{I}}{2\pi} + n_{j,j+1}^{I}\, .
\fe
Then, the four local charges\footnote{The local charges are the time components of the lattice conserved currents.} ${p}_{j,I}$ and ${\Delta}_{j,j+1}^I$ are gauge-invariant, but their eigenvalues are not integer-valued.  (The eigenvalues of the total charges $Q_{I} $ and $ \widetilde{Q}^I$ are still quantized.)

These lattice symmetries have mixed 't Hooft anomalies between them \cite{Cheng:2022sgb, Seifnashri:2023dpa, Seifnashri:2026ema}.  These anomalies can be seen from the commutation relations
\ie\label{pDeltac}
&[p_{j,I},p_{j',J}]=[\Delta_{j,j+1}^I,\Delta_{j',j'+1}^J]=0\,,\\
&[p_{j,I},\Delta_{j',j'+1}^J] = \frac{i}{2 \pi}\left(\delta_{j, j'}-\delta_{j, j'+1}\right) \delta_I^J \,,
\fe
with the lattice Schwinger term \cite{Cheng:2022sgb}.

The Hamiltonian  \eqref{Hintersofphi} can be expressed using the gauge invariant variables $p_{j,I}$ and $\Delta_{j,j+1}^{I}$ as
\ie \label{eq:latH}
H = & \frac{1}{2} \sum_j G^{IJ} \left(p_{j,I} - \frac{B}{2}\epsilon_{II'}\left(\Delta_{j,j+1}^{I'} + \Delta_{j-1,j}^{I'}\right)\right)
\left(p_{j,J} - \frac{B}{2}\epsilon_{JJ'}\left(\Delta_{j,j+1}^{J'} + \Delta_{j-1,j}^{J'}\right)\right)\\
& + \frac{1}{2} \sum_j G_{IJ}\, \Delta_{j,j+1}^{I} \Delta_{j,j+1}^{J}\, .
\fe

In view of that, it is natural to ask whether the whole model can be described using $p_{j,I}$ and $\Delta_{j,j+1}^I$.  To answer this question, we should check whether there exist additional gauge-invariant operators that cannot be expressed in terms of $p_{j,I}$ and $\Delta_{j,j+1}^I$, and whether these operators are constrained.

First, note that there are two physical  operators (i.e., operators that commute with
\eqref{ZZGauss} and \eqref{anotherconst}) that cannot be expressed in terms of $p_{j,I}$ and $\Delta_{j,j+1}^I$. We can take them to be $e^{i \phi_j^I}$ and $e^{i  E_{j,j+1;I}}$ at a chosen point, say $j=0$.  (Changing away from $j=0$ is achieved by shifting the exponents by appropriate linear combinations with integer coefficients of $2\pi i\Delta_{j,j+1}^I$ or $2\pi i p_{j,I}$.)  Alternatively, using a linear combination with fractional coefficients, we can write
\ie\label{Phid}
    e^{i\Phi^I} = \exp{\left[ i \phi_0^I - \frac{2\pi i}{L}\sum_{j=0}^{L-1} j \Delta^I_{j,j+1} \right]} = \exp{\left[ \frac{i}{L}\sum_{j=0}^{L-1} \left( \phi_j^I - 2\pi j \, n^I_{j,j+1} \right) \right]}
\fe
and
\ie\label{Phitd}
    &e^{i \widetilde{\Phi}_{I}} = \exp{\left[ i E_{0,1;I} -  \frac{2\pi i}{L} \sum_{j=1}^{L} j p_{j,I} \right]} = \exp{\left[ \frac{i}{L}\sum_{j=1}^{L} \left( E_{j,j+1;I} - 2\pi j \, q_{j,I} \right) \right]}\,.
\fe
These $\Phi^I$ and $\widetilde \Phi_I$ are close to the continuum zero modes of $\varphi^I$ (denoted $\varphi^I_0$ in \eqref{eq:modeexp_ham} and \eqref{modeexpa}) and its T-dual $\widetilde\varphi_I$.  They are circle-valued variables that commute with all the nonzero modes of $p_{j,I}$ and $\Delta_{j,j+1}^I$, and are conjugate to their zero modes $Q_{I}$ and $\widetilde{Q}^J$.

The lattice translation symmetry $T$, which acts on general operators as
\ie\label{Tactionge}
{\cal O}_j\to T{\cal O}_jT^{-1} = {\cal O}_{j+1}\, , \quad {\cal O}_{L}={\cal O}_0\,,
\fe
does not leave the ``zero modes'' \eqref{Phid} and \eqref{Phitd} invariant.  It acts on them as:
\ie\label{TonPhi}
    T e^{i\Phi^I} T^{-1} = e^{i\Phi^I + \frac{2\pi i}{L} \widetilde{Q}^I}\,, \qquad  T e^{i\widetilde{\Phi}_{J} } T^{-1} = e^{i\widetilde{\Phi}_{J} + \frac{2\pi i}{L} Q_{J}} \,.
\fe
Equivalently, $T$ acts on these operators the same way as $e^{2\pi i Q_{I}\widetilde{Q}^I/L}$.  This behavior is the same as in the continuum where shifting $x \to x + \xi$ shifts the zero mode in \eqref{modeexpa} as $\varphi^I_0 \to \varphi^I_0 + \widetilde{Q}^I \xi$.

Next, we will verify that the set of physical operators generated by $p_{j,I}$, $\Delta_{j,j+1}^I$, $e^{i\Phi^I}$, and $e^{i\widetilde{\Phi}_{I} }$ is complete and unconstrained, besides the quantization conditions on the eigenvalues of $Q_{I} = \sum_j p_{j, I} $ and $\widetilde{Q}^J = \sum_j \Delta_{j,j+1}^J $, or equivalently, the fact that $\Phi^I$ and $\widetilde \Phi_{I}$ are circle-valued.

We will do this by finding a unitary change of variables that simplifies the constraints \eqref{ZZGauss} and \eqref{anotherconst} and decouples the constraints on sites from those on links. Consider the unitary transformation
\ie
    \mathsf{U} = \exp\left[ 2\pi i \sum_{j=1}^{L-1} \sum_{i=0}^{j-1} n^I_{i,i+1} p_{j,I} \right]\,,
\fe
which acts non-trivially only on $\phi_j^I$ and $E_{j,j+1;I}$ as
\ie
    \mathsf{U} \phi_j^I \mathsf{U}^{-1} & = \hat{\phi}_j^I = \phi_j^I +2\pi \sum_{i=0}^{j-1} n^I_{i,i+1} \,, \\
    \mathsf{U} E_{j,j+1; I} \mathsf{U}^{-1} &= \hat{E}_{j,j+1; I} = E_{j,j+1; I} + 2\pi \sum_{i=j+1}^{L-1} p_{i,I} \,.
\fe
Note that $\hat{\phi}_{j+1}^I-\hat{\phi}_j^I$ is gauge-invariant. 

In terms of the new variables $(\hat{\phi},p)$ and $(\hat{E},n)$, the constraints \eqref{ZZGauss} and \eqref{anotherconst} are simplified to
\ie
    &e^{2\pi i n_{j,j+1}^I} = 1 \,,  &&\text{for } j = 0,1,\cdots, L-1 \,,\\
    &e^{i (\hat{E}_{j,j+1;I} - \hat{E}_{j-1,j;I})} = 1\,, \quad &&\text{for } j =1,2, \cdots, L-1 \, , \\
    &e^{2\pi i Q_{I}}=1 \, .
\fe
For each flavor $I=1,2$, the $L$ rotors associated with $(\hat{E}_{j,j+1;I},n_{j,j+1}^I)$ are all identified with each other, which can be taken to be the rotor $(\widetilde{\Phi}_{I}, \widetilde{Q}^I)$. The remaining operators $(\hat{\phi}_j^I,p_{j,I})$ are subject to a single constraint $e^{2\pi iQ_{I}}=1$ for each flavor.  It shifts $\hat{\phi}_j^I\to \hat{\phi}_j^I +z^I$ with $z^I\in 2\pi\bZ $ simultaneously for all $j$. Thus, for each flavor, we are left with a single rotor associated with $(\Phi^I,Q_{I})$ and $L-1$ unconstrained and real gauge invariant operators 
\ie
    & \hat{\phi}_{j+1}^I-\hat{\phi}_j^I = 2\pi \Delta_{j,j+1}^I \qquad \text{for } j=0,1, \cdots, L-2 \,,\\
    & \hat{\phi}_0^I-\hat{\phi}_{L-1}^I = 2\pi \Delta_{L-1,0}^I-2\pi \widetilde{Q}^I \,,
\fe
whose conjugate variables are spanned by $p_{j,I}$. 

The upshot is that instead of using the modified Villain variables $\phi_j^I$, $p_{j,I}$, $E_{j,j+1;I}$, and  $n_{j,j+1}^I$, and imposing the constraints \eqref{ZZGauss} and \eqref{anotherconst}, we can have a purely gauge invariant description using $p_{j,I}$, $\Delta_{j,j+1}^I$, $e^{i\Phi^I}$, and $e^{i\widetilde{\Phi}_{I} }$.  In terms of these, the Hamiltonian is \eqref{eq:latH}.  The only unusual aspect of these variables is that their commutation relations \eqref{pDeltac} and those with the zero modes (which quantize the charges) are nonstandard.\footnote{Note that the gauge-invariant operators found here are not all local. In particular, $e^{i\Phi^I}$, and $e^{i\widetilde{\Phi}_{I}}$ are nonlocal.  Therefore, even though we can use the local ones to write the Hamiltonian as \eqref{eq:latH}, they cannot be taken as ``the fundamental degrees of freedom.''  Still, as we will see, they can be useful in finding the spectrum.}

Clearly, this description of the theory is very similar to the continuum description in \eqref{eq:Hamiltonian_continuum}.

\subsection{The spectrum}

To find the spectrum, we Fourier transform the gauge-invariant variables
\ie \label{eq:non-zero.modes}
&\frak{p}_{n,I} = \frac{1}{\sqrt L}\sum_{j=0}^{L-1} e^{-2\pi i n j/L}p_{j,I} ~ ,
\qquad && \frak{p}_{n,I}=\frak{p}_{n+L,I}= \frak{p}_{-n,I}^\dagger ~ , \\
&\Delta_{n}^{I} = \frac{1}{\sqrt L}\sum_{j=0}^{L-1} e^{-2 \pi i n(j+\frac{1}{2}) / L}\Delta_{j,j+1}^{I}~ , \qquad && \Delta^I_{n}=-\Delta_{n+L}^I=(\Delta_{-n}^I)^\dagger ~ .
\fe
Their commutators are
\ie
\relax[\frak{p}_{n,I},\Delta_m^J] = -\frac{\sin (\pi n / L)}{\pi}\delta_{n,-m}\delta_I^J\,,
\qquad
\relax[\frak{p}_{n,I},\frak{p}_{m,J}] = \relax[\Delta_n^I,\Delta_m^J]=0\, .
\fe
Then \eqref{eq:latH} becomes
\ie
H & =
\frac{1}{2}\sum_{n=0}^{L-1} G^{IJ}\left(\frak{p}_{-n,I} -B \cos (\pi n / L)\epsilon_{II'}\Delta_{-n}^{I'}\right)
\left( \frak{p}_{n,J} -B \cos (\pi n / L)\epsilon_{JJ'}\Delta_{n}^{J'} \right)\\
& \qquad +\frac{1}{2}\sum_{n=0}^{L-1}G_{IJ}\,\Delta_{-n}^{I}\Delta_{n}^{J} \, .
\fe
For the zero modes,
\ie
\Delta_0^{I} = \frac{1}{\sqrt L}\widetilde{Q}^{I}\,,
\qquad
\frak{p}_{0,I} = \frac{1}{\sqrt L}Q_{I}\, .
\fe
For the nonzero modes, we let $\omega_n=\frac{\sin(\pi n/L)}{\pi}$, and define creation and annihilation operators for $n=1, \cdots , L -1$,
\ie \label{eq:lat_osc}
a_{n,r}^{\dagger} &=
\frac{1}{\sqrt{2\omega_n}} \biggl[E_{r}^{I} \left( \frak{p}_{-n,I} -B \cos (\pi n / L)\epsilon_{IJ}\Delta_{-n}^{J} \right) -\delta_{r s} E_I^s \Delta_{-n}^I\biggr]\,, \\
a_{n,r} &=
\frac{1}{\sqrt{2\omega_n}}
\biggl[ E_{r}^{I}\left(\frak{p}_{n,I}-B \cos (\pi n / L)\epsilon_{IJ}\Delta_{n}^{J} \right) -\delta_{r s} E_I^s \Delta_{n}^I \biggr]\,,
\fe
where $E_r^I$, $E_I^r$ are the target space vielbein and its inverse \eqref{eq:vielbein}.
We have
\ie
[a_{m,r},a_{n,s}^\dagger]=\delta_{mn}\delta_{rs}\, .
\fe
In terms of these operators,  up to a normal-ordering constant, the Hamiltonian \eqref{eq:latH} is
\ie \label{eq:latH2}
H & = \frac{1}{2L} \left[
G^{IJ}\left(Q_{I}-B\epsilon_{II'} \widetilde{Q}^{I'}\right) \left(Q_{J}-B\epsilon_{JJ'} \widetilde{Q}^{J'}\right)
+G_{IJ}\,\widetilde{Q}^{I}\widetilde{Q}^{J}
\right]\\
& \qquad + \sum_{n=1}^{L-1}\sum_{r=1}^2\frac{\sin(\pi n/L)}{\pi}a_{n,r}^\dagger a_{n,r}\,.
\fe
Let us compare this to the Hamiltonian of the continuum theory in \eqref{eq:Hamiltonian_continuum}.
First, the zero mode part of the lattice Hamiltonian \eqref{eq:latH2} matches precisely with that of the continuum Hamiltonian \eqref{eq:Hamiltonian_continuum}. Next, in \eqref{eq:latH2}, the oscillators $a_{n,r}$ with $n$ near either $0$ or $L$ give us an energy spectrum approximately linear in $n$ when $L$ is large.
In the continuum limit, these two sets of oscillators from $n$ near 0 or $L$ give rise to the oscillators $a_{n,r}$ in \eqref{eq:Hamiltonian_continuum} with $n$ positive and negative, respectively. In summary, $\lim_{L\to\infty} (L H_{\rm lattice}) = H_{\rm continuum}$.

The eigenstates of \eqref{eq:latH2}, which we denote by $\ket{m_I,w^I,N_{n,r}}$, are labeled by the eigenvalues $m_I$ and $w^I$ of the momentum and winding charges \eqref{gaugeinvachar}, and by the occupation numbers $N_{n,r}$ of the oscillators \eqref{eq:lat_osc} with $n=1,\cdots,L-1$ and $r=1,2$.  Their energies are
\ie \label{eq:latspectrum}
E(m_I,w^I,N_{n,r}) =& \, \frac{1}{2L} \left[
G^{IJ}\left(m_I-B\epsilon_{II'} w^{I'}\right) \left(m_J-B\epsilon_{JJ'} w^{J'}\right)
+G_{IJ}\,w^{I}w^{J}
\right] \\
&+ \sum_{n=1}^{L-1}\sum_{r=1,2}\frac{\sin(\pi n/L)}{\pi} N_{n, r}\,,
\fe
and the translation symmetry acts as
\ie \label{eq:latmomentum}
 T\ket{m_I,w^I,N_{n,r}} = \exp\left[
\frac{2\pi i}{L}\left( m_I w^I - \sum_{n=1}^{L-1}\sum_{r=1,2} n N_{n, r} \right)
\right] \ket{m_I,w^I,N_{n,r}}\, .
\fe
This matches with \eqref{eq:contspectrum} in the continuum.

\subsection{A special T-duality: $B\to B+1$}

The continuum models with different $G_{IJ}$ and $B$ can be related by T-duality. Some of these T-duality transformations are present on the lattice \cite{Cheng:2022sgb, Seifnashri:2026ema, Choi:2026}.

Let us focus on the unitary operator that relates $B$ and $B+k$ for $k \in \bZ$.  It is given by the product of local unitary operators $\B^k$ with 
\ie \label{eq:lat-B-shift}
&\B=\prod_j e^{i h_{j,j+1}}\,,\\
& h_{j,j+1} =\pi\,{n}_{j,j+1}^{1}{n}_{j,j+1}^{2}
+\frac{1}{2}\left(\phi_j^{I}+\phi_{j+1}^{I}\right)
\epsilon_{IJ}\,n_{j,j+1}^{J}
+\frac{1}{4\pi}\epsilon_{IJ}\,\phi_j^{I}\phi_{j+1}^{J}\, .
\fe
It acts on the gauge-invariant fields as
\ie\label{Bongau}
\B^k\,\Delta_{j,j+1}^{I}\,\B^{-k} &=\Delta_{j,j+1}^{I}\,,
\qquad&
\B^k\,p_{j,I}\,\B^{-k} &=p_{j,I}-\frac{k}{2}\epsilon_{IJ}
\left(\Delta_{j,j+1}^{J}+\Delta_{j-1,j}^{J}\right)\,,\\
\B^k \widetilde{Q}^I \B^{-k}&=\widetilde{Q}^I\,, \qquad &
\B^k Q_{ I} \B^{-k}&=Q_{ I}-k \epsilon_{I J} \widetilde{Q}^J \,,\\
\B^k e^{i \Phi^I} \B^{-k}&=e^{i \Phi^I} \,, \qquad & \B^k e^{i \widetilde{\Phi}_{ I}} \B^{-k}&=e^{i\left(\widetilde{\Phi}_{I}-k \epsilon_{I J} \Phi^J+\frac{\pi k}{L} \epsilon_{I J} \widetilde{Q}^J\right)}
\,.
\fe
Importantly, this operator preserves the constraints \eqref{ZZGauss} and \eqref{anotherconst}, because
\ie\label{Bconjq}
& \B^k q_{j,1}\B^{-k}= q_{j,1} -k n_{j,j+1}^{2}\,,\qquad
\B^k q_{j,2}\B^{-k}= q_{j,2} +k n_{j-1,j}^{1}\,,\\
& \B^k n_{j,j+1}^{I}\B^{-k}= n_{j,j+1}^{I}\,,
\fe
and the shift in ${q}_{j,I}$ does not affect $e^{2\pi i {q}_{j,I}}$.

We conclude that the lattice model has the desired $B\sim B+k$ duality. Namely, conjugating the Hamiltonian \eqref{eq:latH} by the unitary operator $\B^k$ has the effect of shifting $B \to  B+k$:
\ie
    {\cal B}^k \;H(B) \; {\cal B}^{-k} = H(B + k ) \,.
\fe 
In addition, the background field $B$ is odd under parity and time-reversal, and therefore all the models with $B \in \frac{\bZ}{2}$ are invariant under these symmetries. More precisely, these symmetry operators should be accompanied by an appropriate power of $\cal B$. See also \cite{Choi:2026}.

\section{Winding $B$ on the lattice}
\label{sec:windingBLat}

So far, $B$ was position-independent, and the Hamiltonian was \eqref{Hintersofphi}, or equivalently, \eqref{eq:latH}.  Here, we will study a position-dependent $B$, by generalizing \eqref{eq:latH} to
\ie\label{eq:latHj}
H=
&\,\frac{1}{2} \sum_j G^{IJ} \left[p_{j,I} - \frac{\epsilon_{II'}}{2}\left(B_{j,j+1}\Delta_{j,j+1}^{I'} + B_{j-1,j}\Delta_{j-1,j}^{I'}\right)\right]\\
& \qquad\qquad\cdot\left[p_{j,J} - \frac{\epsilon_{JJ'}}{2}\left(B_{j,j+1}\Delta_{j,j+1}^{J'} + B_{j-1,j}\Delta_{j-1,j}^{J'}\right)\right]\\
&\, + \frac{1}{2} \sum_j G_{IJ}\,\Delta_{j,j+1}^{I} \Delta_{j,j+1}^{J}\,,\\
\Delta_{j,j+1}^{I} =& \frac{\phi_{j+1}^{I}-\phi_{j}^{I}}{2\pi} + n_{j,j+1}^{I}\,.
\fe
On every link, we have an independent coupling constant $B_{j,j+1}$.

Moreover, we will study the lattice counterpart of the topologically nontrivial winding configuration of $B$ in the continuum \eqref{eq:winding_B}, and will see how the modified Villain model captures the phenomena we saw in the continuum.

\subsection{The $B$-defect}

In the continuum, $B(x)$ is a continuous, circle-valued function, and its winding number is $\oint dB=K$. Here, we focus on situations with nonzero $K$.  In Section \ref{sec:continuum}, we treated it as a continuous real-valued function for $x\neq x_0$, with the winding number captured by the nontrivial transition function \eqref{BtildeAt}. This approach cannot work on the lattice.  Here, we can have a position-dependent $B_{j,j+1}$ and treat it as a real coupling constant.  But since there is no notion of continuity, how can we determine when it ``jumps'' by an integer?

To answer this question, we view the change in $B$ due to a transition function of the continuum theory as a topological defect.  Then, we should find the lattice version of this topological defect.

Let us first place the theory on an infinite lattice with a possibly $j$-dependent $B_{j,j+1}$.  Then, we construct the defect at $j=j_0$ by acting on the system with a truncated version of ${\cal B}^K$ with $\cal B$ of \eqref{eq:lat-B-shift},
\ie
\prod_{j<j_0}\,e^{i K h_{j,j+1}}\,.
\fe
This has the effect of shifting $B_{j,j+1}\to B_{j,j+1}+K$ for all $j<j_0$. It also modifies Gauss's law \eqref{ZZGauss} to
\ie \label{eq:localqModified}
e^{2 \pi i\left(q_{j, I}-\delta_{j,j_0}\frac{K}{4 \pi} \epsilon_{I J} \phi_{j_0}^J\right)}=1\, .
\fe
These constraints are analogous to the constraints ${\cal C}_I$ in the continuum \eqref{eq:C1}. (See more details in Section \ref{comparingcontla}). Correspondingly, we replace the operator $p_{j,I}$ of \eqref{eq:lat_canonical} (which commutes with the constraint \eqref{ZZGauss}) with 
\ie\label{pofdefect}
p_{j, I}+\delta_{j,j_0}\frac{K}{4 \pi} \epsilon_{IJ} \phi_{j_0}^J\, .
\fe

Motivated by this description of the defect in an infinite chain, we study the finite periodic chain with $L$ sites, and place the defect at site $j_0$.
Then the Hamiltonian with the topological defect can be written as
\ie \label{eq:H1}
H_{j_0} (&B_{0,1},\cdots,B_{L-1,0})= \\
\,\,& \frac{1}{2} \sum_{{j=0\atop j\ne j_0}}^{L-1} G^{IJ} \left[p_{j,I} - \frac{\epsilon_{II'}}{2}\left(B_{j,j+1}\Delta_{j,j+1}^{I'} + B_{j-1,j}\Delta_{j-1,j}^{I'}\right)\right]\\
& \qquad\qquad\cdot\left[p_{j,J} - \frac{\epsilon_{JJ'}}{2}\left(B_{j,j+1}\Delta_{j,j+1}^{J'} + B_{j-1,j}\Delta_{j-1,j}^{J'}\right)\right]\\
& + \frac{1}{2} G^{IJ}\left[p_{j_0,I}+\frac{K}{4\pi}\epsilon_{II'}\phi_{j_0}^{I'} - \frac{\epsilon_{II'}}{2}\left(B_{j_0,j_0+1}\Delta_{j_0,j_0+1}^{I'} + B_{j_0-1,j_0}\Delta_{j_0-1,j_0}^{I'}\right)\right]\\
& \qquad\qquad\cdot\left[p_{j_0,J}+\frac{K}{4\pi}\epsilon_{JJ'}\phi_{j_0}^{J'} - \frac{\epsilon_{JJ'}}{2}\left(B_{j_0,j_0+1}\Delta_{j_0,j_0+1}^{J'} + B_{j_0-1,j_0}\Delta_{j_0-1,j_0}^{J'}\right)\right]\\
& + \frac{1}{2} \sum_{j=0}^{L-1} G_{IJ}\,\Delta_{j,j+1}^{I} \Delta_{j,j+1}^{J}\, .
\fe
The defect corresponds to a modification of the Hamiltonian density at $j_0$, without changing it elsewhere.  Comparing with the continuum, the modification on the defect plays the role of the correction term localized at $x_0$ in \eqref{eq:action1}. (Also, compare with the continuum Hamiltonian \eqref{eq:Happ1}.)

To demonstrate that this defect is topological, we conjugate it by the local unitary operator $e^{iKh_{j_0,j_0+1}}$.  This shifts $j_0$ in Gauss's law  \eqref{eq:localqModified} and changes the Hamiltonian as
\ie
e^{iKh_{j_0,j_0+1}}H_{j_0}(B_{0,1}, &\cdots, B_{j_0,j_0+1}, B_{j_0+1,j_0+2},\cdots) e^{-iKh_{j_0,j_0+1}}\\
&=H_{j_0+1}(B_{0,1}, \cdots, B_{j_0,j_0+1}+K, B_{j_0+1,j_0+2},\cdots) \,,
\fe
such that the defect moves from $j_0$ to  $j_0+1$, without changing the spectrum.
This is the lattice version of relating different trivialization points $x_0$ in the continuum action \eqref{eq:action1}.

Finally, let us summarize the answer to the question above, about the definition of the winding of $B$ on a discrete lattice.  We define
\ie \label{eq:beta_def}
\beta_j=B_{j,j+1}-B_{j-1,j}+K\delta_{j, j_0}\,,
\fe
such that $\beta_j$ and $B_{j_0,j_0+1}$ are the lattice counterparts of $\partial_xB(x)$ and $B(x_0)$ in the continuum, respectively.  Correspondingly, $\sum_j\beta_j=K$ is the lattice analog of $\oint dB=K$ in the continuum.

\subsection{The target space momentum symmetry}

In this section, we discuss the fate of the $U(1) \times U(1)$ target space momentum symmetry in the presence of the topological defect, with the modified constraint \eqref{eq:localqModified} and the Hamiltonian \eqref{eq:H1}.

In the presence of the defect, the conserved momentum charges include a correction term on the defect,  
\ie
Q_{j_0,I}
=\sum_j q_{j,I}-\frac{K}{4\pi}\epsilon_{IJ}\phi_{j_0}^{J}
=Q_{I}-\frac{K}{4\pi}\epsilon_{IJ}\phi_{j_0}^{J}\,.
\fe
This is the lattice analog of $\int_{x_0}^{x_0 + 2\pi} \pi_I (x) \, dx - \frac{K}{4\pi} \epsilon_{IJ} \varphi^J \Big|_{x=x_0^+} = \int_{x_0}^{x_0 + 2\pi}  J_{t,I} \, dx - \frac{K}{2\pi} \epsilon_{IJ} \varphi^J \Big|_{x=x_0^+}$ in \eqref{eq:mosymm_ops}, \eqref{eq:C1} or \eqref{eq:CU}. Because of the term with $\phi_{j_0}^{J}$, $Q_{j_0,I}$ does not commute with the modified Gauss's law \eqref{eq:localqModified}
\ie
\left[2\pi i \left(q_{j_0, I}-\frac{K}{4 \pi} \epsilon_{I I'} \phi_{j_0}^{I'}\right), Q_{j_0,J}\right]=K\epsilon_{IJ}\,.
\fe
Nonetheless, the operators
\ie\label{UsubI}
U_I=\exp\!\left(\frac{2\pi i}{K}\,Q_{j_0,I}\right)\,,\qquad I=1,2\,,
\fe
commute with the constraints \eqref{anotherconst} and \eqref{eq:localqModified}, and are therefore well-defined operators on the constrained Hilbert space. Using the modified constraints \eqref{eq:localqModified} and the commutator
\ie
[Q_{j_0,I},Q_{j_0,J}]=-\frac{iK}{2\pi}\epsilon_{IJ}\,,
\fe
one finds
\ie \label{eq:momlat}
U_1^K=U_2^K=1\,,
\qquad
U_1U_2=e^{2\pi i\over K}\,U_2U_1\, .
\fe
Interestingly, the modified constraint \eqref{eq:localqModified} at the defect explicitly breaks the continuous $U(1) \times U(1)$ momentum symmetry down to a discrete $\bZ_{|K|} \times \bZ_{|K|}$ subgroup, which is realized projectively. This is the lattice counterpart of \eqref{momentumsymmmetryb} in the continuum.

\subsection{An alternative presentation}\label{sec:alt}

To simplify the expressions, from this point on, we will place the defect at $j_0 = 0$.

Here, we present another version of our model \eqref{eq:H1}, which is related to it by the unitary transformation
\ie\label{summationbyparts}
{\cal U}=\exp\!\left[i \sum_{j=0}^{L-1}(B_{0,1}-B_{j,j+1}) h_{j,j+1}\right]\,,
\fe
i.e., we study
\ie \label{eq:H2}
H'_{j_0=0}(B_{0,1},&\cdots,B_{L-1,0})= {\cal U}H_{j_0=0}(B_{0,1},\cdots,B_{L-1,0})\,{\cal U}^{-1} \\
=& \frac{1}{2}\sum_{j=0}^{L-1}G^{IJ}\left(p_{j,I}+\frac{\beta_j}{4\pi}\,\epsilon_{II'}\phi_j^{I'}-\frac{B_{0,1}}{2}\epsilon_{II'}\left(\Delta_{j,j+1}^{I'}+\Delta_{j-1,j}^{I'}\right)\right)\\
& \qquad\qquad \cdot\left(p_{j,J}+\frac{\beta_j}{4\pi}\,\epsilon_{JJ'}\phi_j^{J'}-\frac{B_{0,1}}{2}\epsilon_{JJ'}\left(\Delta_{j,j+1}^{J'}+\Delta_{j-1,j}^{J'}\right)\right)\\
& +\frac{1}{2}\sum_{j=0}^{L-1}G_{IJ}\,\Delta_{j,j+1}^{I}\Delta_{j,j+1}^{J}\, .
\fe
(Compare this with the continuum Hamiltonian \eqref{eq:Happ2}.)
Here, $\beta_j$ is as in \eqref{eq:beta_def}. And the unitary transformation ${\cal U}$ maps Gauss's law \eqref{eq:localqModified} to
\ie \label{eq:qtilde}
& \exp\!\left(2\pi i \left(q_{j,1}-\frac{\beta_j}{4\pi}\,\phi_j^{2}- (B_{0,1}-B_{j,j+1})n_{j,j+1}^{2}\right)\right)=1\,,\\
& \exp\!\left(2\pi i \left(q_{j,2}+\frac{\beta_j}{4\pi}\,\phi_j^{1} +(B_{0,1}-B_{j-1, j})n_{j-1,j}^{1}\right)\right)=1\,.
\fe
Note that \eqref{anotherconst} is not modified.  These constraints are analogous to the constraints ${\cal C}'_I$ in the continuum \eqref{eq:C2}. (See more details in Section \ref{comparingcontla}). Related to that, we can replace $p_{j,I}$ by the local, physical operators
\ie\label{pshif}
p_{j, I}+\frac{\beta_j}{4 \pi} \epsilon_{I J} \phi_j^J\,,
\fe
without changing $\Delta_{j,j+1}^I$. (Compare with \eqref{pofdefect}.)

We will later find it convenient to consider
\ie\label{eq:BigPiLat}
\Pi_{j,I}=p_{j, I}+\frac{\beta_j}{4 \pi} \epsilon_{I J} \phi_j^{J}-\frac{B_{0,1}}{2} \epsilon_{I J}\left(\Delta_{j, j+1}^{J}+\Delta_{j-1, j}^{J}\right) \,.
\fe
Then,
the Hamiltonian \eqref{eq:H2} is
\ie \label{eq:H2short}
H'=\frac{1}{2} \sum_{j=0}^{L-1} G^{IJ} \Pi_{j,I}\Pi_{j,J}+ \frac{1}{2} \sum_{j=0}^{L-1} G_{IJ}\,\Delta_{j,j+1}^{I} \Delta_{j,j+1}^{J}\,.
\fe
(We omit the dependence on $j_0 =0$ and $B_{j,j+1}$ in $H'$ for notational brevity.)
It depends only on the combinations $\Delta_{j,j+1}^I$ and $\Pi_{j,I}$, which commute with the constraints and are therefore physical operators.  Their commutation relations are 
\ie \label{eq:BigPiLatCom}
\left[\Pi_{j, I}, \Delta_{j',j'+1}^J\right]=\frac{i}{2 \pi}\left(\delta_{j, {j'}}-\delta_{j, {j'}+1}\right) \delta_I^J\,, \quad\left[\Delta_{j,j+1}^I, \Delta_{j',j'+1}^J\right]=0\,, \quad\left[\Pi_{j, I}, \Pi_{{j'}, J}\right]=\frac{i \beta_j}{2 \pi} \epsilon_{I J} \delta_{j {j'}} \,.
\fe

Now, the momentum symmetry $\bZ_{|K|} \times \bZ_{|K|}$ is generated by 
\ie 
\mathcal{U} U_I \mathcal{U}^{-1} = \exp\left(\frac{2 \pi i}{K}\left[Q_{ I}-\epsilon_{I J} \sum_{j=0}^{L-1}\left(\frac{\beta_j}{4 \pi} \phi_j^J+\left(B_{0,1}-B_{j, j+1}\right) n_{j, j+1}^J\right)\right]\right)\,,
\fe 
and they obey the same algebra as in \eqref{eq:momlat}. To avoid cluttering the notation, in the rest of the paper we will denote them by $U_I$, even when we discuss ${\cal U} U_I{\cal U}^{-1}$.

\section{Lattice translation symmetries with linear $B$}\label{latticetrans}

For simplicity of the expressions, in this section, we work with the alternative presentation of the model, discussed in Section \ref{sec:alt}.

When $B$ is constant, our system is manifestly invariant under a $\bZ_L$ lattice translation symmetry.  It is generated by $T$, which acts on local operators as \eqref{Tactionge}.
When $B$ is position-dependent, this $T$ does not preserve the local constraints, i.e., it does not act within the physical Hilbert space. Instead, we study
\ie\label{Thatd}  
\hat T = \exp \left(-i \sum_{j=0}^{L-1} \beta_j h_{j, j+1}\right)T\,.
\fe 
It is well-defined and acts in the physical Hilbert space of \eqref{anotherconst} and \eqref{eq:qtilde}.  

However, $\hat T$ does not commute with the Hamiltonian $H'$ of \eqref{eq:H2}. It satisfies  
\ie\label{TtoLBtoK} 
\hat T^L \B^K = 1\,,
\fe 
which represents complete translation around the chain.  This is the lattice counterpart of the continuum relation \eqref{completetrn}.

Next, we examine the translation symmetry of the system when $B_{j,j+1}$ varies linearly in $j$,
\ie\label{constantbeta}
&B_{j,j+1}=B_{0,1}+\frac{jK}{L}\,,\qquad j=0,\dots,L-1\,,\\
&\beta_j={K\over L}\,.
\fe
Now,
\ie \label{eq:H2Linear}
H' =\,& \frac{1}{2}\sum_{j=0}^{L-1}G^{IJ}\left(p_{j,I}+\frac{K}{4\pi L}\epsilon_{II'}\phi_j^{I'}-\frac{B_{0,1}}{2}\epsilon_{II'}\left(\Delta_{j,j+1}^{I'}+\Delta_{j-1,j}^{I'}\right)\right)\\
& \qquad\qquad \cdot\left(p_{j,J}+\frac{K}{4\pi L}\epsilon_{JJ'}\phi_j^{J'}-\frac{B_{0,1}}{2}\epsilon_{JJ'}\left(\Delta_{j,j+1}^{J'}+\Delta_{j-1,j}^{J'}\right)\right)\\
& +\frac{1}{2}\sum_{j=0}^{L-1}G_{IJ}\,\Delta_{j,j+1}^{I}\Delta_{j,j+1}^{J}\, .
\fe
The corresponding Gauss's laws \eqref{eq:qtilde} simplify to
\ie \label{eq:G2Linear}
& \exp \!\left(2\pi i\left(q_{j,1}-\frac{K}{4\pi L}\phi_j^{2} +\frac{jK}{L}n_{j,j+1}^{2}\right)\right)=1\,,\\
& \exp \!\left(2\pi i\left(q_{j,2}+\frac{K}{4\pi L}\phi_j^{1} -\frac{(j-1)K}{L}n_{j-1,j}^{1}\right)\right)=1\, ,
\fe
and the gauge-invariant momentum operators \eqref{pshif} are
\ie
p_{j,I}+\frac{K}{4\pi L}\epsilon_{IJ}\phi_j^{J}\,.
\fe

\subsection{The unitary lattice translation symmetry}

We start with the special case where $L$ is a multiple of $K$.

The Hamiltonian \eqref{eq:H2Linear} is invariant under the lattice translation $T$ in \eqref{Tactionge}. However, $T$ does not preserve the constraint \eqref{eq:G2Linear}. In contrast, the constraints are preserved under the translation by $L/K$ sites,\footnote{One can verify the transformation of the constraints at the site $j$,
\ie\nonumber
T^{\ell}\exp\!&\left(2\pi i\left(q_{j,1}-\frac{K}{4\pi L}\phi_j^{2} +\frac{jK}{L}n_{j,j+1}^{2}\right)\right)T^{-\ell}\\
&=\exp\!\left(2\pi i\left(q_{j+\ell,1}-\frac{K}{4\pi L}\phi_{j+\ell}^{2} +\frac{(j+\ell) K}{L}n_{j+\ell,j+\ell+1}^{2}\right)\right)\exp\!\left(-2\pi i\,\frac{\ell K}{L}n_{j+\ell,j+\ell+1}^{2}\right)\,,\\
T^{\ell}\exp\!&\left(2\pi i\left(q_{j,2}+\frac{K}{4\pi L}\phi_j^{1} -\frac{(j-1) K}{L}n_{j-1,j}^{1}\right)\right)T^{-\ell}\\
&=\exp\!\left(2\pi i\left(q_{j+\ell,2}+\frac{K}{4\pi L}\phi_{j+\ell}^{1} -\frac{(j+\ell-1) K}{L}n_{j+\ell-1,j+\ell}^{1}\right)\right)\exp\!\left(2\pi i\,\frac{\ell K}{L}n_{j+\ell-1,j+\ell}^{1}\right)\,.
\fe
The last additional factor in each line equals 1 for arbitrary $n_{i,i+1}^I$ only when $\ell$ is a multiple of $L/K$.\label{ft:constraintpreserving}}
\ie 
{\sf P} = \hat{T}^{L \over K} \B= T^{\frac{L}{K}}\, .
\fe
As in \eqref{TtoLBtoK}, ${\sf P}^K = 1$, and the lattice translation symmetry is reduced to $\bZ_{|K|}$.
This is the lattice counterpart of the $\bZ_{|K|}$ translation symmetry of the continuum theory \eqref{eq:symm_ops}.

The translation and momentum symmetries combine to the algebra
\ie\label{algwhen.div}
 U_1^K = U_2^K={\sf P}^K=1\,,
\qquad
U_1U_2 =e^{\frac{2\pi i}{K}}U_2U_1\, , \qquad U_I \mathsf{P} = e^{ \frac{2\pi i}{K} \epsilon_{IJ} \widetilde{Q}^J } \mathsf{P} U_I \,.
\fe
This is the lattice version of the symmetry algebra  \eqref{eq:cont-translation} of the continuum theory.

For general $L$ which may not be divisible by $K$, the unitary lattice translation becomes $\bZ_g$, where
\ie
g = \mathrm{gcd}(L,K)\,.
\fe
It is generated by 
\ie \label{eq:latUniTranslation}
{\sf P} = \hat{T}^{L \over g}\B^{K \over g}= T^{\frac L g}\,. 
\fe
Then the algebra \eqref{algwhen.div} becomes
\ie \label{eq:latalg}
& {\sf P}^g=1\,, \quad U_1^K = U_2^K=1\,,
\quad
U_1U_2 =e^{\frac{2\pi i}{K}}U_2U_1 \,,  \quad U_I \mathsf{P} = e^{ \frac{2\pi i}{g} \epsilon_{IJ} \widetilde{Q}^J } \mathsf{P} U_I \,.
\fe

\subsection{Non-invertible translation symmetry}

More general lattice translations turn into non-invertible symmetries.
Here, we will show this explicitly. See Appendix \ref{app:noninv.lattice} for more details.

We start with the non-invertible lattice translation by one site\footnote{The normalization of the operator is chosen as in footnote \ref{factorofr}.} 
\ie \label{eq:noninv1site}
\mathsf{D} = \frac{L}{g} \, \mathbb{P}_{\frac{L}{g}}^1 \, \mathbb{P}_{\frac{L}{g}}^2 \,\hat T\, {\cal V}_{\frac{K}{L}}\,.
\fe
 Here, $\hat T$ is as in \eqref{Thatd}, $\mathbb{P}_{L\over g}^I $
are projection operators as in the continuum \eqref{eq:projectors_cont},
and
\ie\label{calVdef} 
{\cal V}_{\frac{s}{r}} & = \exp \left(i \frac{s}{2 \pi r} \sum_{j=1}^L\left(\phi_j^1+2 \pi \mathcal{N}_j^1\right)\left(\phi_{j+1}^2-\phi_j^2+2 \pi n_{j, j+1}^2\right)-\frac{i s}{r} \widetilde{Q}^1 \phi_1^2\right)\\
&\qquad \cdot\exp\left(\frac{i \pi s}{r} \sum_{j=0}^{L-1} \Delta_{j, j+1}^1 \Delta_{j, j+1}^2\right) \, , \qquad  s,r \in \mathbb{Z}\,,\\
\mathcal{N}_j^I&=\sum_{j^{\prime}=0}^{j-1} n_{j^{\prime}, j^{\prime}+1}^I\,, \quad I=1,2\,.
\fe
It satisfies
\ie\label{VtorBtos} 
({\cal V}_{\frac{s}{r}})^{r} = \B^s\,.
\fe 

Importantly, the operator ${\cal V}_{K \over L}$ (which appears in \eqref{eq:noninv1site}) does not commute with the constraints \eqref{eq:G2Linear}. However, adding the projection operators, $\mathsf{D}={L\over g}\mathbb{P}_{\frac{L}{g}}^1\mathbb{P}_{\frac{L}{g}}^2\hat{T}{\cal V}_{K \over L}$ does.  It satisfies
\ie
    \mathsf{D}^{\frac{L}{g}} = \left(\frac{L}{g}\right)^{\frac{L}{g}} \, \mathbb{P}_{\frac{L}{g}}^1\mathbb{P}_{\frac{L}{g}}^2T^{L \over g}=\left(\frac{L}{g}\right)^{\frac{L}{g}} \, \mathbb{P}_{\frac{L}{g}}^1\mathbb{P}_{\frac{L}{g}}^2 {\sf P}\,.
\fe
It is related to the invertible translation by $L/g$ sites \eqref{eq:latUniTranslation}.  

We note that $\mathsf{D}$ of \eqref{eq:noninv1site} can also be expressed in terms of $T$ as 
\ie \label{eq:noninv1site2}
{\sf D} = \frac{L}{g} \, \mathbb{P}_{\frac{L}{g}}^1 \, \mathbb{P}_{\frac{L}{g}}^2 \,T\, \exp \left(\frac{2 \pi i K}{L} \sum_{j_2=1}^{L-1} \sum_{j_1=0}^{j_2-1} n_{j_1, j_1+1}^1 n_{j_2, j_2+1}^2\right) \,.
\fe
In this form, it is manifest that $\mathsf{D}$ commutes with the Hamiltonian.

The lattice non-invertible translation operator \eqref{eq:noninv1site}, or equivalently \eqref{eq:noninv1site2}, is to be compared with the non-invertible translation operator \eqref{eq:noninv_proj} in the continuum.
Both on the lattice and in the continuum, these operators contain projection operators that restrict the two winding charges.
The lattice counterpart of the continuum operator \eqref{eq:noninvT}, which contains auxiliary degrees of freedom on the operator, will be discussed in Appendix \ref{app:noninv.lattice}.

The non-invertible symmetry operator $\mathsf{D}$ satisfies the fusion algebra
\ie \label{eq:Dfusion}
&\mathsf{D}^{\dagger} \times \mathsf{D} = \left(\frac{L}{g}\right)^2 \mathbb{P}^1_{\frac{L}{g}} \mathbb{P}^2_{\frac{L}{g}} = \sum_{l_1,l_2=0}^{L/g-1} e^{\frac{2 \pi i g}{L} (l_1\widetilde{Q}^1 +l_2\widetilde{Q}^2)}\, .
\fe
This is similar to a Kramers-Wannier type fusion rule arising from self-duality under gauging the $\bZ_{L/g} \times \bZ_{L/g}$ subgroup of the winding symmetry \cite{Frohlich:2004ef, Thorngren:2021yso, Choi:2021kmx, Kaidi:2021xfk, Seiberg:2024gek}.
Moreover, the fusion of $\mathsf{D}$ with itself generates all the other non-invertible translations. For a positive integer $\ell \in \bZ_{>0}$,
\ie
    \mathsf{D}^\ell &= \left(\frac{L}{g}\right)^{\ell-1} \mathrm{gcd}(L/g,\ell) \; \mathbb{P}^1_{\frac{L}{g}}  \mathbb{P}^2_{\frac{L}{g}} \, \mathsf{D}_\ell \\
    &= \left(\frac{L}{g}\right)^{\ell-1} \frac{1}{\mathrm{gcd}(L/g,\ell)} \sum_{l_1,l_2=0}^{\mathrm{gcd}(L/g,\ell)-1} e^{2 \pi i\frac{ g}{L} (l_1\widetilde{Q}^1 +l_2\widetilde{Q}^2)} \, \mathsf{D}_\ell \, ,
\fe
where $\mathsf{D}_\ell$ denotes the non-invertible translation symmetry by $\ell$ sites,
\ie \label{eq:noninvLatdef1}
    \mathsf{D}_{\ell} = r \, \mathbb{P}_r^1 \, \mathbb{P}_r^2 \, \hat{T}^{\ell} \mathcal{V}_{\frac{\ell K}{L}} \,, \quad \text{ with } \quad r = \frac{L}{\mathrm{gcd}(L,\ell K)} = \frac{L/g}{\mathrm{gcd}(L/g,\ell)}\, .
\fe

The general non-invertible translation operators $\mathsf{D}_{\ell}$ satisfy the fusion algebra
\ie \label{eq:LatNoniveAlg1}
& \mathsf{D}_{\ell}^{\dagger} = \mathsf{D}_{-\ell} \,,\\
&\mathsf{D}_{\ell}^{\dagger} \times \mathsf{D}_{\ell} = \sum_{l_1,l_2=0}^{L/\mathrm{gcd}(L,\ell K)-1} e^{2 \pi i\frac{ \mathrm{gcd}(L,\ell K)}{L} (l_1\widetilde{Q}^1 +l_2\widetilde{Q}^2)}  \,,\\
& e^{2 \pi i\frac{ \mathrm{gcd}(L,\ell K)}{L} \widetilde{Q}^{I}} \, \mathsf{D}_{\ell} = \mathsf{D}_{\ell} \, e^{2 \pi i\frac{ \mathrm{gcd}(L,\ell K)}{L} \widetilde{Q}^{I}}=\mathsf{D}_{\ell}\,,\\
& {U}_I \mathsf{D}_{\ell} =e^{\frac{2\pi i \ell }{L} \epsilon_{IJ}\widetilde{Q}^J} \mathsf{D}_{\ell}{U}_I\,.
\fe
Since the identity operator occurs once in $\mathsf{D}_{\ell}^{\dagger} \times \mathsf{D}_{\ell}$, the operator $\mathsf{D}_{\ell}$ is simple, i.e., it cannot be further decomposed. Moreover, 
\ie
&\mathsf{D}_{L/g}=T^{L/g}={\sf P} \,,\\
&{\sf P} \mathsf{D}_{\ell} = \mathsf{D}_{\ell+L/g}\,,
\fe
with ${\sf P}$ the ordinary invertible translation symmetry of \eqref{eq:latUniTranslation}.

This matches with the continuum algebra \eqref{eq:Dalgebra} by identifying the continuum and lattice operators as 
\ie
\mathsf{D}_\ell\to \mathcal{D}_{\ell \over L}  \,,
\fe
and using\footnote{This substitution ensures $\gcd(s,r)=1$, as in \eqref{eq:fraction.t}.} 
\ie \label{eq:substitution}
    r = \frac{L}{\mathrm{gcd}(L,\ell K)} \,, \qquad s = \frac{\ell K}{\mathrm{gcd}(L,\ell K)} \, .
\fe

Finally, general non-invertible translations fuse as
\ie \label{eq:LatNoniveAlg2}
\mathsf{D}_{\ell_1} \times \mathsf{D}_{\ell_2} &= \frac{r_1 r_2}{r_{12}} \, \mathbb{P}_{\mathrm{lcm}(r_1,r_2)}^1 \mathbb{P}_{\mathrm{lcm}(r_1,r_2)}^2 \, \mathsf{D}_{\ell_1+\ell_2} \\
&= \frac{r_{12}\gcd(r_1,r_2)}{\mathrm{lcm}(r_1,r_2)}
\sum_{l_1,l_2=0}^{\frac{\mathrm{lcm}(r_1,r_2)}{r_{12}}-1}
e^{\frac{2\pi i}{\mathrm{lcm}(r_1,r_2)}
(l_1\widetilde{Q}^1+l_2\widetilde{Q}^2)} \,
\mathsf{D}_{\ell_1+\ell_2}\,,
\fe
where $r_1$, $r_2$, and $r_{12}$ are given by \eqref{eq:substitution} with $\ell$ replaced by $\ell_1$, $\ell_2$, and $\ell_1+\ell_2$
\ie
    r_i = \frac{L}{\mathrm{gcd}(L,\ell_i K)}\,, \qquad r_{12} = \frac{L}{\mathrm{gcd}(L,(\ell_1+\ell_2) K)} \, .
\fe

To summarize, on the lattice, the translation by any number of sites $\ell$ remains a symmetry for any winding number $K$, and $K$ determines the projection that accompanies it.  The amount of the translation is set by $\ell$ alone, while $K$ enters \eqref{eq:noninvLatdef1} through \eqref{eq:substitution}, which fixes the projector $\mathbb{P}_r^1\mathbb{P}_r^2$ onto the states with $\widetilde{Q}^{I}=0 \bmod r$.

As a check, comparing with the continuum, every symmetry of the lattice model leads to a symmetry of the continuum theory.  Clearly, the converse is not true.  In the continuum, every translation $x\to x+2\pi\xi$ acts as an (invertible or non-invertible) symmetry, while on the lattice, only those with  $\xi=\ell/L$ are realized.

\section{Comparing the continuum and the lattice}\label{comparingcontla}

In Section \ref{sec:continuum}, we used an explicit trivialization to describe the continuum theory of two compact scalars $\varphi^I$ in terms of real $\varphi^I$ with transition functions in $\bZ\times\bZ$ between the patches.  We can view it as starting with real $\varphi^I$ and performing an orbifold by $\bZ\times \bZ$. Equivalently, we coupled the non-compact $\varphi^I$ to a dynamical $\bZ\times\bZ$ gauge theory.  

Specifically, placing the transition functions at $x=x_0$, the real scalars $\varphi^I$ change there by the winding charges $\widetilde{Q}^I$ as in \eqref{eq:winding_bosons}, and these winding charges can be viewed as the $\bZ \times \bZ$ holonomy of the gauge fields around space.

In Section \ref{oursystem}, we studied the Hamiltonian formulation of our theory.  Here, we used this gauge theory interpretation by embedding the Hilbert space of the theory in a larger Hilbert space $\mathscr H$.  Then, the physical Hilbert space ${\cal H}\subset {\mathscr H}$ was found by imposing $\bZ\times \bZ$ Gauss's law constraints ${\cal C}_I=1$. 

\begin{table}[t]
    \centering
     \begin{tabular}{|c c|c c|}
        \hline
        \multicolumn{2}{|c|}{Lattice expression} & \multicolumn{2}{c|}{Continuum expression} \\
        \hline\hline

        $\phi_j^I$ & \eqref{eq:lat_canonical} & $\varphi^I(x)$  & \eqref{eq:lag} \\

        $p_{j,I}$ & \eqref{eq:lat_canonical} & $\frac{2\pi}{L}\pi_I(x)$ & \eqref{eq:pi_app} \\

        $\Delta_{j,j+1}^I$ & \eqref{DeltaIdef} & $\frac{1}{L}\partial_x\varphi^I(x)$ & \eqref{eq:lag} \\

        $\beta_j$ & \eqref{eq:beta_def} & $\frac{2\pi}{L}\partial_x B(x)$ & \eqref{eq:action2} \\

        $\Pi_{j,I}$ & \eqref{eq:BigPiLat}  & $\frac{2\pi}{L}\Pi_I(x)$ & \eqref{eq:BigPicont} \\

        $\mathcal U$& \eqref{summationbyparts} & ${\cal U}$   & \eqref{eq:U_cont} \\

        $H_{j_0}\!\left(B_{0,1},\ldots,B_{L-1,0}\right)$ & \eqref{eq:H1} & $\frac{1}{L}H_{x_0}$ & \eqref{eq:Happ1} \\

        $H'_{j_0}\!\left(B_{0,1},\ldots,B_{L-1,0}\right)$ & \eqref{eq:H2} & $\frac{1}{L}H'_{x_0}$ & \eqref{eq:Happ2} \\

        Product of constraints  & \eqref{eq:localqModified}
            & $\mathcal C_I=1$ & \eqref{eq:C1} \\

        Product of constraints & \eqref{eq:qtilde}
            & ${\mathcal C}'_I=1$ & \eqref{eq:C2} \\
        \hline
        
        $Q_{I}$ & \eqref{gaugeinvachar}
            & $Q_{I}$ & \eqref{eq:cont_charges} \\

        $\widetilde{Q}^I$ & \eqref{gaugeinvachar}
            & $\widetilde{Q}^I$ & \eqref{eq:cont_charges} \\

        $\mathcal B$ & \eqref{eq:lat-B-shift}
            & $\mathcal B$ & \eqref{eq:calB} \\

        $U_I$ & \eqref{UsubI} & $U_I$ &  \eqref{eq:mosymm_ops} \\
        
        $\hat T$ & \eqref{Thatd}
            & $\exp\!\left[\frac{2\pi i}{L}
                \int dx\,\hat\Theta_{tx}\right]$
            & \eqref{eq:symm_ops} \\

        $\mathsf P=T^{L/g}$ & \eqref{eq:latUniTranslation}
            & $\mathcal P^{K/g}$ & \eqref{eq:symm_ops} \\

        $\mathsf D$ & \eqref{eq:noninv1site}
            & $\mathcal D_{1/L}$ & \eqref{eq:noninv_proj} \\

        $\mathsf D_\ell$ & \eqref{eq:noninvLatdef1}
            & $\mathcal D_{\ell/L}$
            & \eqref{eq:noninv_proj} \\
        \hline 
    \end{tabular}
    \caption{Comparison between the lattice and continuum expressions.  Here, $L$ is the number of sites in the lattice, $K$ is the winding number of $B$, and $g=\gcd(L,K)$.  For the other quantities, see the referenced equations.  The continuum limit corresponds to $L\to \infty$ with ${2\pi\over L}\to dx$. Equivalently, the lattice spacing is $2\pi\over L$. See Appendix \ref{app:spectrum} for a comparison between the continuum and the lattice spectra.}
    \label{tab:latcont}
\end{table}

Starting with a different Lagrangian density, we used unitarily equivalent constraints ${\cal C}'_I=1$ to find another physical Hilbert space ${\cal H}'\subset  {\mathscr H} $.  These descriptions are related by a unitary transformation ${\cal U}$ acting in the large Hilbert space $\mathscr H$.  It maps the Hamiltonian and the constraints as in \eqref{eq:unitary_transformation_cont}.
In particular, the canonical momenta $\pi_I(x)=-i{\delta\over \delta\varphi^I(x)}$ act in $\mathscr H$. Neither $\pi_I(x) $ nor ${\cal U}$ preserves the physical Hilbert spaces $\cal H$ or ${\cal H}'$.

The key point of the lattice construction is the use of the modified Villain formulation. It replaces the compact scalars $\phi^I_j$ of the standard XY model with non-compact $\phi^I_j$, and achieves the compactness by gauging $\bZ\times \bZ$.  As is standard in lattice gauge theory,  every lattice site defines a patch, and the link variables are the transition functions between them.  The latter are $\bZ \times \bZ$ gauge fields $n^I_{j,j+1}$ on the links.  Finally, to make it similar to the continuum $\bZ\times \bZ$ gauge theory, we set the gauge coupling of the lattice model to zero.

In view of that, to compare with the continuum, we fix a gauge such that all $n^I_{j,j+1}$ vanish except on one link, $n^I_{j_0,j_0+1}$. Then, we use the local constraints to eliminate the independent link variables $E_{j,j+1; I}$, except for one compact degree of freedom conjugate to $\widetilde{Q}^I$. The residual constraints are the products of the local constraints. Then the match between the lattice and the continuum is implemented by taking the continuum limit ${2\pi\over L} \to dx$. 

We summarize the relation between lattice expressions and continuum expressions in Table \ref{tab:latcont}.

\section{Conclusion}
\label{sec:conclusion}

In this paper, we have studied various topological aspects of the continuum $c=2$ CFT of two compact bosons and its lattice realization based on the modified Villain formulation.
In the continuum, we found that the model's $-1$-form symmetry mixes with other global symmetries to form a higher group.
When the periodic coupling constant $B$ is spatially varying with a nonzero winding number, the target space momentum symmetry is explicitly broken to a discrete subgroup. When $B$ varies linearly, the generic spatial translation symmetry becomes non-invertible, and only a discrete subgroup remains invertible.  This discrete subgroup is combined with the internal symmetry to form a non-Abelian group.

Our lattice model realizes the $B$-periodicity exactly on a finite lattice.
The T-duality transformation $B \to B+1$ is implemented by conjugation by a locality-preserving unitary operator.
This unitary operator allows us to realize a nontrivial winding configuration of $B$ on the lattice by inserting the corresponding topological defect.  Then, we showed that the subtle effects of the continuum model's symmetry are already present on the lattice.

Both in the continuum and on the lattice, the theories are exactly solvable even when $B$ is position-dependent.  This follows from the fact that the Lagrangian (and therefore also the Hamiltonian) is quadratic.  What is more interesting is that when $B$ varies linearly, the exact solution uses a standard Fourier decomposition.   Even though the spatial translation symmetry is explicitly broken to a discrete subgroup, the different Fourier modes do not mix. (See Section \ref{sec:modeexp} and Appendix \ref{app:spectrum}.)
The underlying reason for that is the non-invertible translation symmetry of the model. Since the non-invertible symmetry has a local expression, it constrains the local, quadratic Hamiltonian in every winding sector, including those that the symmetry operator annihilates.

Perhaps the most surprising aspect of this investigation is that the intricate topological aspects of the continuum theory can be realized on a discrete lattice.  This is achieved by using the Villain formulation, which relies on $\bZ\times\bZ$ gauge fields.  The other element is setting its gauge coupling constant to zero.  With these two elements, we find a detailed comparison between the lattice and the continuum theories, which we summarized in Section \ref{comparingcontla}, and in particular, in Table \ref{tab:latcont}.

There are a few natural directions for future investigations.
The continuum $c=2$ theory is invariant under a rich set of T-duality transformations. 
In this work, we focused only on those generated by $B \to B+1$.
It will be interesting to find lattice realizations of other duality transformations.
Some of them will be discussed in \cite{Choi:2026}.
Also, many of the phenomena we have found in this paper are not specific to the $c=2$ theory of compact bosons.
There exists a large class of continuum theories in various dimensions that exhibit similar peculiarities.  Some of these are discussed in \cite{Heidenreich:2020pkc, Brauner:2020rtz, Damia:2022seq, Seiberg:2024wgj, Seiberg:2024yig, Katayama:2025pmz, Brennan:2026ira} and in Appendix \ref{app:more_bosons} and Appendix \ref{app:higherD}. It would be nice to explore additional examples, both on the lattice and in the continuum, and perhaps uncover deeper underlying principles leading to these phenomena.

\section*{Acknowledgments}

We are grateful to Meng Cheng, Juan Maldacena, Greg Moore, Zohar Komargodski, Salvatore D. Pace, and Brandon C. Rayhaun for interesting discussions. YC was supported by NSF Grant PHY-2514611 and the Fund for Memberships in Natural Sciences at the Institute for Advanced Study. The work of NS was supported in part by DOE grant DE-SC0009988. WZ was supported in part by the Princeton University Department of Physics. This work was also supported by the Simons Collaboration on Ultra-Quantum Matter, which is a grant from the Simons Foundation (651444, NS). SS also gratefully acknowledges support from the Ambrose Monell Foundation and an anonymous donor at the Institute for Advanced Study.

\appendix

\section{Warm-up: a particle on a torus with a magnetic field}
\label{particleontorus}

In this appendix, we review the well-known problem of a particle on a torus with a constant magnetic field. There are two reasons to do it here. First, this simple problem demonstrates, in a familiar setting, some of the issues we discussed above, especially about the role of the correction terms in Euclidean time.  Second, the continuum zero modes $\varphi^I_0$ in \eqref{modeexpa} and the lattice zero modes $\Phi^I$ in \eqref{Phid} behave like this simple quantum mechanical example.

\subsection{The Euclidean action}

We parameterize the torus by $\varphi^I \sim \varphi^I+2\pi$,\footnote{Comparing with our discussion above, these are  $\varphi_0^I$ of \eqref{modeexpa}.} and denote the target-space gauge potential by $\cal A$. We start by picking a trivialization, where
\ie
    \mathcal{A}(\varphi^1 + 2\pi , \varphi^2) &= \mathcal{A}(\varphi^1 , \varphi^2) + \mathrm{d}\Lambda_1 \,, \\
    \mathcal{A}(\varphi^1 , \varphi^2 + 2\pi) &= \mathcal{A}(\varphi^1 , \varphi^2) + \mathrm{d}\Lambda_2 \,.
\fe
We use `$\mathrm{d}$' to denote the exterior derivative in the target space. The transition functions $\Lambda_I$ satisfy the cocycle condition
\ie
\Lambda_1(\varphi^1, \varphi^2+2\pi) -\Lambda_1(\varphi^1, \varphi^2) -\Lambda_2(\varphi^1+2\pi, \varphi^2) +\Lambda_2(\varphi^1, \varphi^2)=2\pi K
\fe
with $K=c_1\in \bZ$.

We will be interested in a configuration with a constant magnetic field with total flux $K$; i.e., $\mathrm{d}{\cal A}  = \frac{K}{2\pi} \mathrm{d}\varphi^1 \wedge \mathrm{d}\varphi^2$, and we will compare the gauge field and transition functions in the Landau gauge denoted by $(\mathrm{L})$ and the symmetric gauge denoted by $(\mathrm{S})$ 
\ie\label{A1mag}
&\mathcal{A}^{(\mathrm{L})}={K\over 2\pi }\varphi^1 \mathrm{d}\varphi^2\,, \qquad &&\Lambda_1^{(\mathrm{L})}=K\varphi^2\,, \qquad &&\Lambda_2^{(\mathrm{L})}=0 \,, \\
&\mathcal{A}^{(\mathrm{S})}={K\over 4\pi} \epsilon_{IJ}\varphi^I \mathrm{d}\varphi^J\,, \qquad &&\Lambda_1^{(\mathrm{S})}={K\over 2}\varphi^2\,, \qquad &&\Lambda_2^{(\mathrm{S})}=-{K\over 2}\varphi^1\,.
\fe
Here, we used the freedom to pick the origin of $\varphi^I$.

These two configurations are related by
\ie\label{Uchanget}
&\mathcal{A}^{(\mathrm{S})}=\mathcal{A}^{(\mathrm{L})} + i U\mathrm{d}U^{-1} \,, \\
&U=\exp\left(-i{K\over 4\pi }\varphi^1\varphi^2\right)\,.
\fe
Importantly, this is not a gauge transformation because it is not single-valued on our torus.  Instead, it transforms the transition functions $\Lambda_I^{(\mathrm{L})}$ to $\Lambda^{(\mathrm{S})}_I$.

We consider a charge-one particle in compact Euclidean time $\tau \sim \tau +2\pi$.  Its world-line is characterized by two instanton numbers $\widetilde{Q}_{\tau}^I$. Again, we choose a trivialization and consider it for $\tau\in (\tau_0, \tau_0+2\pi )$, where $\varphi^I(\tau_0^-) = \varphi^I(\tau_0^+) + 2\pi \widetilde{Q}_{\tau}^I$.  The holonomy/Wilson line of the gauge field along the world-line of the particle is
\ie\label{holonomypar}
{\cal W}=&\exp \left( \frac{iK}{2\pi}\int_{\tau_0}^{\tau_0+2\pi} d\tau \, \varphi^1 \partial_\tau \varphi^2
     - iK \widetilde{Q}_{\tau}^1 \varphi^2 (\tau_0^+ ) \right) \\
    =& \exp\left(\frac{iK}{4\pi}\int_{\tau_0}^{\tau_0 + 2\pi} d\tau \, \epsilon_{IJ} \varphi^I \partial_\tau \varphi^J
      -\frac{iK}{2}\epsilon_{IJ} \widetilde{Q}_{\tau}^I \varphi^J (\tau_0^+ ) + i \pi K \widetilde{Q}_{\tau}^1 \widetilde{Q}_{\tau}^2\right)\,.
\fe
In the first expression, we used ${\cal A}^{(\mathrm{L})}$ of \eqref{A1mag} and in the second expression, we used  ${\cal A}^{(\mathrm{S})}$.  In both cases, we included localized terms associated with the transition functions. It is easy to check, using integration by parts (while paying attention to the surface term), that these two expressions are equivalent.

Next, we make the particle dynamical and consider two actions corresponding to the two expressions for the holonomy \eqref{holonomypar}
\begin{align} \label{eq:S1}
\begin{split}
    S &= \int_{\tau_0}^{\tau_0 + 2\pi} d\tau \,\left( \frac{1}{2}G_{IJ} \partial_\tau \varphi^I \partial_\tau \varphi^J - \frac{iK}{2\pi}\varphi^1 \partial_\tau \varphi^2
    \right) + iK \widetilde{Q}_{\tau}^1 \varphi^2 (\tau_0^+ ) \,,
\end{split}
\end{align}
and its equal expression
\begin{align} \label{eq:S2}
\begin{split}
    S &= \int_{\tau_0}^{\tau_0 + 2\pi} d\tau \,\left( \frac{1}{2}G_{IJ} \partial_\tau \varphi^I \partial_\tau \varphi^J - \frac{iK}{4\pi}\epsilon_{IJ} \varphi^I \partial_\tau \varphi^J
    \right) 
    \\ &\qquad\qquad 
    + \frac{iK}{2}\epsilon_{IJ} \widetilde{Q}_{\tau}^I \varphi^J (\tau_0^+ ) - i \pi K \widetilde{Q}_{\tau}^1 \widetilde{Q}_{\tau}^2
    \,.
\end{split}
\end{align}
These actions are related to the Euclidean actions in Section \ref{oursystem}.

\subsection{Hamiltonian formulations \label{app:HS}}

Next, we make Euclidean time $\tau$ non-compact, such that we can drop the terms localized near $\tau=\tau_0$.  Then, we continue the theory to Lorentzian time $t\in \mathbb{R}$ to find the Lagrangians
\ie \label{eq:Lz1}
    &L^{(\mathrm{L})} = \frac{1}{2}G_{IJ} \partial_t \varphi^I \partial_t \varphi^J + \mathcal{A}_I^{(\mathrm{L})} \partial_t\varphi^I \,, \qquad &&\mathcal{A}_I^{(\mathrm{L})} = \frac{K}{2\pi} \delta_{I2} \varphi^1 \,,\\
    &L^{(\mathrm{S})} = \frac{1}{2}G_{IJ} \partial_t \varphi^I \partial_t \varphi^J + \mathcal{A}_I^{(\mathrm{S})} \partial_t\varphi^I \,, \qquad &&\mathcal{A}_I^{(\mathrm{S})} = -\frac{K}{4\pi} \epsilon_{IJ} \varphi^J \,,
\fe
where $\mathcal{A}^{(\mathrm{L})}=\mathcal{A}_I^{(\mathrm{L})} \mathrm{d} \varphi^I$ and $\mathcal{A}^{(\mathrm{S})}=\mathcal{A}_I^{(\mathrm{S})} \mathrm{d} \varphi^I$ as given in \eqref{A1mag}. The two Lagrangians differ by a total time derivative of $\frac{K}{4\pi}\varphi^1\varphi^2$, which is not globally well-defined. 

The conjugate momenta and Hamiltonians are
\ie \label{eq:Hz}
    & \pi_I = \frac{\partial L^{(\mathrm{L})}}{\partial (\partial_t \varphi^I )} = G_{IJ} \partial_t \varphi^J + \mathcal{A}_I^{(\mathrm{L})}\, , \qquad && 
    H^{(\mathrm{L})} = \frac{1}{2} G^{IJ} \left( \pi_I - \mathcal{A}_I^{(\mathrm{L})} \right) \left( \pi_J - \mathcal{A}_J^{(\mathrm{L})} \right)\,, \\
    & \pi_I = \frac{\partial L^{(\mathrm{S})}}{\partial (\partial_t \varphi^I )} = G_{IJ} \partial_t \varphi^J + \mathcal{A}_I^{(\mathrm{S})} \,, \qquad &&
    H^{(\mathrm{S})} = \frac{1}{2} G^{IJ} \left( \pi_I - \mathcal{A}_I^{(\mathrm{S})} \right) \left( \pi_J - \mathcal{A}_J^{(\mathrm{S})}\right) \,.
\fe
In both cases, we use the same $\varphi^I$ and $\pi_I$. Importantly, $\varphi^I$ are not well-defined operators, but $\partial_t\varphi^J$ are well-defined. Therefore, in the problem with ${\cal A}_I^{(\mathrm{L})}$, $\pi_2$ is ill-defined, and in the problem with ${\cal A}^{(\mathrm{S})}_I$, both $\pi_I$ are ill-defined.  Therefore, it is convenient to define the well-defined operators
\ie\label{PidefQM}
&\Pi_I^{(\mathrm{L})}=\pi_I - \mathcal{A}_I^{(\mathrm{L})}\,,\\
&\Pi^{(\mathrm{S})}_I=\pi_I - \mathcal{A}_I^{(\mathrm{S})}\,,
\fe
which are like covariant derivatives in the target space.  These ``momenta'' satisfy 
\ie
  &  [\varphi^I, \Pi_{J}^{(\mathrm{L})}] = i \delta^I_J \,, \qquad [\Pi_{I}^{(\mathrm{L})}, \Pi_{J}^{(\mathrm{L})}] = i\frac{K}{2\pi} \epsilon_{IJ}\,,\\
   &  [\varphi^I, \Pi_{J}^{(\mathrm{S})}] = i \delta^I_J \,, \qquad [\Pi_{I}^{(\mathrm{S})}, \Pi_{J}^{(\mathrm{S})}] = i\frac{K}{2\pi} \epsilon_{IJ}\, .
\fe
In terms of them, the Hamiltonians are
\ie
&H^{(\mathrm{L})} = \frac{1}{2} G^{IJ} \Pi_I^{(\mathrm{L})}\Pi_J^{(\mathrm{L})}\,, \\
    &H^{(\mathrm{S})} = \frac{1}{2} G^{IJ} \Pi^{(\mathrm{S})}_I\Pi_J^{(\mathrm{S})} \,.
\fe

The two Hamiltonians, $H^{(\mathrm{L})}$ and $H^{(\mathrm{S})}$ of \eqref{eq:Hz} are different functions of the same variables $\varphi^I$ and $\pi_I$.  Even though $\varphi^I$ are the same, the velocities,  $\partial_t\varphi^I$, which are determined dynamically by the Hamiltonians, are different in the two cases.  Relatedly, $\Pi_I^{(\mathrm{L})}=G_{IJ}\partial_t\varphi^J$ and $\Pi_I^{(\mathrm{S})}=G_{IJ}\partial_t\varphi^J$ are different operators.

Let us denote the wavefunction of the particle in the two descriptions by $\Psi^{(\mathrm{L})}(\varphi^1, \varphi^2 )$ and $\Psi^{(\mathrm{S})}(\varphi^1, \varphi^2 )$. As is well-known, the wavefunctions $\Psi^{(\mathrm{L})}(\varphi^1, \varphi^2 ),\Psi^{(\mathrm{S})}(\varphi^1, \varphi^2 )$ are not $2\pi$-periodic functions of $\varphi^I$'s, but rather, they are sections of line bundles.  

This is related to the nontrivial terms localized near $\tau = \tau_0$.  In particular, these terms in the Euclidean actions \eqref{eq:S1} and \eqref{eq:S2} tell us that the wavefunctions of the particle satisfy\footnote{These boundary conditions are consistent with the statement above that in the problem with ${\cal A}_I^{(\mathrm{L})}$, $\pi_2=-i{\partial \over \partial \varphi^2}$ is ill-defined, and in the problem with ${\cal A}^{(\mathrm{S})}_I$, both $\pi_I=-i{\partial \over \partial \varphi^I}$ are ill-defined.}
\ie \label{eq:wf}
    &\Psi^{(\mathrm{L})} (\varphi^1 + 2\pi , \varphi^2 ) = e^{i\Lambda_1^{(\mathrm{L})}} \Psi^{(\mathrm{L})} (\varphi^1 , \varphi^2 ) \,, \quad && \Psi^{(\mathrm{L})} (\varphi^1 , \varphi^2 + 2\pi) = e^{i\Lambda_2^{(\mathrm{L})}} \Psi^{(\mathrm{L})}(\varphi^1 , \varphi^2 ) \,, \\
    &\Psi^{(\mathrm{S})} (\varphi^1 + 2\pi , \varphi^2 ) = e^{i\Lambda_1^{(\mathrm{S})}} \Psi^{(\mathrm{S})}(\varphi^1 , \varphi^2 ) \,, \quad && \Psi^{(\mathrm{S})} (\varphi^1 , \varphi^2 + 2\pi) = e^{i\Lambda_2^{(\mathrm{S})}} \Psi^{(\mathrm{S})}(\varphi^1 , \varphi^2 ) \,.
\fe
We denote the Hilbert spaces of wavefunctions satisfying these boundary conditions by $\mathcal{H}^{(\mathrm{L})}$ and $\mathcal{H}^{(\mathrm{S})}$.
 
In more detail, the Euclidean functional integral evaluates a trace of the Euclidean evolution operator over the Hilbert space.  And a configuration in the functional integral that winds $\widetilde{Q}_{\tau}^I$ times around Euclidean time means that before evaluating the trace we change a state with wavefunction $\Psi(\varphi^1, \varphi^2) $ to $\Psi(\varphi^1 + 2\pi \widetilde{Q}_{\tau}^1, \varphi^2 + 2\pi \widetilde{Q}_{\tau}^2) $. The correction terms in \eqref{eq:S1} and \eqref{eq:S2} implement the identification \eqref{eq:wf}.

One way to describe this system with Hilbert spaces $\mathcal{H}^{(\mathrm{L})}, \mathcal{H}^{(\mathrm{S})}$ is to compare it with the system of a particle on a plane with the same magnetic field.  Here, the Lagrangians are still \eqref{eq:Lz1}, but the coordinates $\varphi^I$ are non-compact, and correspondingly, the wavefunctions are arbitrary functions of $\varphi^I$. We denote the Hilbert space of these wavefunctions by $\mathscr H$. 
The canonical momenta $\pi_I$ are represented as
\begin{equation} \label{eq:piz1rep}
    \pi_I = - i \frac{\partial}{\partial \varphi^I} ~~~~\text{on $\mathscr H$} \,.
\end{equation}

Then, the problem above with a torus target space is obtained by gauging $\bZ\times \bZ$.  This gauging corresponds to imposing $\bZ\times \bZ$ Gauss's law
\ie\label{untildeconst0}
&{\cal C}_I^{(\mathrm{L})}= e^{2\pi i \pi_I}e^{-i \Lambda_I^{(\mathrm{L})} } =1\,, \qquad {\cal C}_I^{(\mathrm{S})}= e^{2\pi i \pi_I}e^{-i \Lambda_I^{(\mathrm{S})}} =1\,.
\fe
This restricts the large Hilbert space $\mathscr H$ to $\mathcal{H}^{(\mathrm{L})}, \mathcal{H}^{(\mathrm{S})}\subset \mathscr H$ of \eqref{eq:wf}. (See a pedagogical discussion in Appendix A of \cite{Seiberg:2024yig}.) Note that the correction term $iK\widetilde{Q}_{\tau}^1 \varphi^2 (\tau_0^ + )$ in the Euclidean action \eqref{eq:S1} leads to the term $-i \Lambda_1^{(\mathrm{L})} = -iK\varphi^2$ in the first constraint ${\cal C}_1^{(\mathrm{L})}$ in \eqref{untildeconst0}. For the gauge field $\mathcal{A}^{(\mathrm{S})}$ and the transition functions $\Lambda^{(\mathrm{S})}_I$, \eqref{eq:wf} and \eqref{untildeconst0} are the zero-mode versions of \eqref{eq:wfn1} and \eqref{eq:C1}, as well as \eqref{eq:wfn2} and  \eqref{eq:C2}.  

\subsection{Relation between the two Hamiltonian formulations}

The two actions \eqref{eq:S1} and \eqref{eq:S2} are identical, and therefore they lead to the same physics. 
Yet, their Hamiltonian formulations appear to differ.

To understand it, we note that the two descriptions involve different Hilbert spaces, ${\cal H}^{(\mathrm{L})}$ and ${\cal H}^{(\mathrm{S})}$ of \eqref{eq:wf}. We can relate them using the unitary operator $U: \mathscr H \to \mathscr H$ in \eqref{Uchanget}. Indeed, it is straightforward to see that, when we act by $U$ on a wavefunction $\Psi^{(\mathrm{L})} \in {\cal H}^{(\mathrm{L})}$, we obtain a new wavefunction $\Psi^{(\mathrm{S})} = U \Psi^{(\mathrm{L})} \in {\cal H}^{(\mathrm{S})}$, and therefore $U$ maps the subspace ${\cal H}^{(\mathrm{L})} \subset \mathscr H$ to ${\cal H}^{(\mathrm{S})} \subset \mathscr H$.\footnote{As mentioned before, $U$ is not a globally well-defined gauge transformation. In the quantum theory, this corresponds to the statement that $U$ is a unitary operator acting on the larger Hilbert space $\mathscr{H}$, but it does not act on $\mathcal{H}^{(\mathrm{L})}$ or $\mathcal{H}^{(\mathrm{S})}$.}

The two Hamiltonians in \eqref{eq:Hz}, and the $\mathbb{Z} \times \mathbb{Z}$ Gauss's laws in \eqref{untildeconst0}, are related to each other by\footnote{When $\varphi^I$ in \eqref{eq:action1Euc} and \eqref{eq:action1Euc2} are independent of $x$, the Hamiltonians become $H^{(\mathrm{L})}$ and $H^{(\mathrm{S})}$ in \eqref{eq:Hz}, respectively. And they are related by $U$.  This is to be contrasted with the unitary operator in \eqref{eq:U_cont}.
The latter relates the Hamiltonian formulation of \eqref{eq:action1Euc2} with that of \eqref{eq:action2Euc}, and becomes trivial when $\varphi^I$ are $x$-independent.}
\begin{equation}
    H^{(\mathrm{S})} = U H^{(\mathrm{L})} U^{-1} \,, \quad {\cal C}^{(\mathrm{S})}_I = U {\cal C}_I^{(\mathrm{L})} U^{-1} \,.
\end{equation}

\section{More on the bulk action} \label{app:bulk}

Here, we will discuss in more detail the three-dimensional bulk action in Section \ref{sec:higher_group} of the continuum $c=2$ theory.
We will take the bulk to have Euclidean signature, and as before, we will suppress terms in the action that are valued in $2\pi i \mathbb{Z}$.

We will explicitly demonstrate that the bulk Lagrangian \eqref{Btotald}, repeated here in Euclidean signature,
\ie \label{eq:bulk_action_appendix}
&\mathcal{L}_{bulk} =  d \left[ -\frac{i}{2\pi}
            B (d\varphi^1 - A^1 ) \wedge (d\varphi^2 - A^2 ) - \frac{i}{2\pi}  \widetilde A_I \wedge (d\varphi^I - A^I )
      \right] +\mathcal{L}_{anomaly}\,,\\
&\mathcal{L}_{anomaly} = \frac{i}{2\pi} \widetilde{A}_I \wedge dA^I  \,
\fe
does not depend on the bulk extensions of the bosons $\varphi^I$. (Recall that, as we stated above, if $A^I$ are topologically nontrivial, then an insertion of a winding operator is needed.)

We take our two-dimensional spacetime to be a torus $M_2 = T^2$ with $2\pi$-periodic coordinates $x$ and $\tau$, where $\tau$ is the Euclidean time.
The bulk manifold is then taken to be $M_3 = \mathbb{R}_z^- \times M_2$, similar to Section \ref{sec:higher_group}.\footnote{The orientation on $M_3$ is such that the volume form is $dz \wedge d\tau \wedge dx$, and the induced orientation on $M_2$ is such that its volume form is $d\tau \wedge dx$.}

When we integrate $\mathcal{L}_{bulk}$ over $M_3$, the terms inside the square brackets in \eqref{eq:bulk_action_appendix} give rise to boundary terms.
To analyze this, we first trivialize various bundles.
We pick reference points $x_0$ and $\tau_0$, and let all the fields be real and single-valued for $x \neq x_0$, $\tau \neq \tau_0$.
We impose the following transition functions (and transition functions of transition functions) at $x_0$ and $\tau_0$:
\begin{align} \label{eq:transition_functions}
\begin{split}
    \varphi^I (\tau ,x_0^-  ,z ) &= \varphi^I ( \tau , x_0^+ , z)  + 2\pi \widetilde{Q}^I + \Lambda_x^I (\tau , x_0 , z) \,,\\
    \varphi^I ( \tau_0^- , x , z) &= \varphi^I ( \tau_0^+ , x , z) + 2\pi \widetilde{Q}_{\tau}^I + \Lambda_\tau^I (\tau_0 , x ,z )  \,,\\
    B ( \tau , x_0^- , z) &= B (\tau , x_0^+ , z) + K \,,\\
    B (\tau_0^- , x, z ) &= B (\tau_0^+ , x, z) + K_{\tau} \,,\\
    A^I (\tau , x_0^- , z) &= A^I (\tau , x_0^+ , z) + d\Lambda_x^I (\tau , x_0 , z ) \,,\\
    A^I (\tau_0^- , x , z ) &= A^I (\tau_0^+ , x , z ) + d\Lambda_\tau^I (\tau_0 , x , z) \,,\\
    \widetilde{A}_I ( \tau , x_0^- , z) &= \widetilde{A}_I (\tau , x_0^+ , z) + d \widetilde{\Lambda}_{I,x} ( \tau , x_0 , z) - K\epsilon_{IJ} A^J ( \tau, x_0^+ , z) \,,\\
     \widetilde{A}_I (\tau_0^- , x , z ) &= \widetilde{A}_I (\tau_0^+ , x , z) + d \widetilde{\Lambda}_{I,\tau} (\tau_0 , x ,z) - K_{\tau}\epsilon_{IJ} A^J ( \tau_0^+ ,x ,z) \,,\\
     \Lambda_x^I (\tau_0^- , x_0 , z ) &= \Lambda_x^I (\tau_0^+ , x_0 , z ) + 2\pi M_{x,\tau}^I \,,\\
     \Lambda_\tau^I (\tau_0 , x_0^- ,z  ) &= \Lambda_\tau^I (\tau_0 , x_0^+  ,z) + 2\pi M^I_{\tau,x} \,,\\
     \widetilde{\Lambda}_{I,x} (\tau_0^- , x_0 , z ) &= \widetilde{\Lambda}_{I,x} (\tau_0^+ , x_0 , z ) + 2\pi \widetilde{M}_{I,x,\tau} - K_{\tau}\epsilon_{IJ} \Lambda^J_x (\tau_0^+ , x_0 , z ) \,,\\
     \widetilde{\Lambda}_{I,\tau} (\tau_0 , x_0^-  ,z ) &= \widetilde{\Lambda}_{I,\tau} (\tau_0 , x_0^+,z ) + 2\pi \widetilde{M}_{I,\tau,x} - K\epsilon_{IJ} \Lambda^J_\tau (\tau_0 , x_0^+ ,z) \, .
\end{split}
\end{align}
Here, $\widetilde{Q}^I$, $\widetilde{Q}_{\tau}^I$, $K$, $K_{\tau}$, $M_{x,\tau}^I$, $M^I_{\tau,x}$, $\widetilde{M}_{I,x,\tau}$, and $\widetilde{M}_{I,\tau,x}$ are arbitrary integers, and $\Lambda_x^I$, $\Lambda_\tau^I$, $\widetilde{\Lambda}_{I,x}$, and $\widetilde{\Lambda}_{I,\tau}$ are arbitrary real functions for $x \neq x_0$, $\tau \neq \tau_0$.
The transition functions of $\widetilde{A}_I$, $\widetilde{\Lambda}_{I,x}$, and $\widetilde{\Lambda}_{I,\tau}$ are determined such that the modified field strengths $\widetilde{F}_I$ in \eqref{eq:tildeF} are single-valued across $x_0$, $\tau_0$, and also such that the set of transition functions are consistent with each other.

In the three-dimensional integral
\ie
    \int_{\mathbb{R}_z^- \times (\tau_0 , \tau_0 + 2\pi ) \times (x_0 , x_0 + 2\pi )} \mathcal{L}_{bulk} \,,
\fe
the terms that are localized at $z = 0$ are
\begin{align} \label{eq:app_bdy}
\begin{split}
    &~~~~-\int_{ (\tau_0 , \tau_0 + 2\pi ) \times (x_0 , x_0 + 2\pi ) } \left[ \frac{i}{2\pi}
            B (d\varphi^1 - A^1 ) \wedge (d\varphi^2 - A^2 ) + \frac{i}{2\pi}  \widetilde{A}_I \wedge (d\varphi^I - A^I )
      \right] \\
      &- \int_{(\tau_0 , \tau_0 + 2\pi )} \left(
        \frac{i}{2\pi} K \varphi^1 d\varphi^2 + \frac{i}{2\pi} \widetilde{\Lambda}_{I,x} d \varphi^I
      \right)\Big|_{x=x_0^+} \\
      &+\int_{(x_0 , x_0 + 2\pi )} \left(
        \frac{i}{2\pi} K_{\tau} \varphi^1 d\varphi^2 + \frac{i}{2\pi} \widetilde{\Lambda}_{I,\tau} d \varphi^I
      \right)\Big|_{\tau=\tau_0^+} \\
      & +\Bigg[ i K \widetilde{Q}_{\tau}^1 \left(\varphi^2 + \Lambda_\tau^2 \right)
        + i \widetilde{M}_{I,x,\tau} 
        \left(\varphi^I + \Lambda_\tau^I \right)
        - i K_\tau \widetilde{Q}^1  \left(\varphi^2 + \Lambda_x^2 \right)
        -i
            \widetilde{M}_{I,\tau,x} 
        \left(\varphi^I + \Lambda_x^I \right) \\&~~~~~~~~~~
        ~~~~~~~~+ \frac{i}{2\pi} \varphi^1 \left(
            K \Lambda_\tau^2 - K_{\tau} \Lambda^2_x
      \right)
      \Bigg]\Big|_{x= x_0^+, \tau= \tau_0^+} \, .
\end{split}
\end{align}
This gives the desired two-dimensional couplings, including the correction terms. 
We drop the contribution from $z= -\infty$. (See footnote \ref{nobulk}.)

The terms that contain the integral along the bulk $z$-direction are
\begin{align} \label{eq:app_z}
\begin{split}
    &~~~~~~~~~\int_{\mathbb{R}_z^- \times (\tau_0 , \tau_0 + 2\pi ) \times (x_0 , x_0 + 2\pi )  } \mathcal{L}_{anomaly} \\
    &- \int_{\mathbb{R}_z^- \times (\tau_0 , \tau_0 + 2\pi)} \left(
        \frac{i}{2\pi} K A^1 \wedge A^2 - \frac{i}{2\pi} d\widetilde{\Lambda}_{I,x} \wedge A^I
    \right)\Big|_{x= x_0^+} \\
    &+\int_{\mathbb{R}_z^- \times (x_0 , x_0 + 2\pi)} \left(
        \frac{i}{2\pi} K_{\tau} A^1 \wedge A^2 - \frac{i}{2\pi} d\widetilde{\Lambda}_{I,\tau} \wedge A^I
    \right)\Big|_{\tau= \tau_0^+} \\
    &+ \int_{\mathbb{R}_z^-} \Bigg[
        \frac{i}{2\pi} K \Lambda^1_\tau d\Lambda^2_\tau  + \frac{i}{2\pi} \left(
            \widetilde{\Lambda}_{I,x} - K_{\tau}\epsilon_{IJ}\Lambda^J_x
        \right) d\Lambda^I_\tau \\
        &~~~~~~~~~~-\frac{i}{2\pi} K_{\tau} \Lambda^1_x d\Lambda^2_x  - \frac{i}{2\pi} \left(
            \widetilde{\Lambda}_{I,\tau} - K \epsilon_{IJ}\Lambda^J_\tau
        \right) d\Lambda^I_x
    \Bigg]\Big|_{x=x_0^+,\tau=\tau_0^+} \, .
\end{split}
\end{align}
This gives the bulk anomaly action plus the correction terms.
Importantly, there is no dependence on $\varphi^I$ or their winding numbers $\widetilde{Q}^I$, $\widetilde{Q}_{\tau}^I$.
We conclude that the bulk Lagrangian \eqref{eq:bulk_action_appendix} indeed defines a two-dimensional theory.

If we turn off the background gauge fields $A^I$, $\widetilde{A}_I$, as well as their transition functions, \eqref{eq:app_z} vanishes.
Furthermore, as in the main text, suppose $B$ depends only on the spatial coordinate $x$, and therefore $K_\tau = 0$.
Then, \eqref{eq:app_bdy} becomes
\begin{align} \label{eq:appEuc}
\begin{split}
    &~-\int_{ (\tau_0 , \tau_0 + 2\pi ) \times (x_0 , x_0 + 2\pi ) }  \frac{i}{2\pi}
            B d\varphi^1 \wedge d\varphi^2 \\
    &~~~~~~~  - \int_{(\tau_0 , \tau_0 + 2\pi )} 
        \frac{i}{2\pi} K \varphi^1 d\varphi^2 \Big|_{x=x_0^+} + iK \widetilde{Q}_{\tau}^1 \varphi^2 \Big|_{x=x_0^+ , \tau=\tau_0^+} \,.
\end{split}
\end{align}
Combined with the kinetic term $\mathcal{L}_G$ in \eqref{eq:lag2}, this gives the Euclidean action in \eqref{eq:action1Euc}.

\section{Spectrum with linear $B$}
\label{app:spectrum}

In this appendix, we compute the spectra of the continuum and lattice theories when the $B$ field has a linear profile.

\subsection{Continuum}\label{contspec}

We begin by analyzing the spectrum of the continuum theory with linear $B(x)$ as in \eqref{eq:linear_B}. For simplicity, we place the trivialization point at $x_0 = 0$ and omit $x_0$ in the expressions below. We start with \eqref{LagrsumwithF} and \eqref{mode.decoupling}, and drop the total derivative term $dF/dt$.

As in Appendix \ref{particleontorus}, the conjugate momenta are ill-defined.   Instead, as we did there, we use
\ie
\mathsf{\Pi}_{n,I} = G_{IJ} \partial_t \varphi_n^J\,,
\fe
with the commutation relations
\ie\label{ftcommr}
    [\varphi_n^I, \mathsf{\Pi}_{m,J}] = i \delta^I_J \delta_{n,-m}\,, \qquad [\mathsf{\Pi}_{n,I}, \mathsf{\Pi}_{m,J}] = i\frac{K}{2\pi} \epsilon_{IJ} \delta_{n,-m} \, .
\fe

Following \eqref{mode.decoupling}, the Hamiltonian \eqref{eq:Happ2} in terms of these modes is
\ie \label{eq:tildeH_momentum}
   H' = \frac{1}{2} G_{IJ} \widetilde{Q}^{I} \widetilde{Q}^{J} + \frac{1}{2} G^{IJ} \mathsf{\Pi}_{0,I} \mathsf{\Pi}_{0,J}+\frac{1}{2} \sum_{n \neq 0} \left( G^{IJ} \mathsf{\Pi}_{n,I}\mathsf{\Pi}_{-n,J}+n^2G_{IJ} \varphi^I_n \varphi^J_{-n} \right) \, .
\fe
As discussed at the end of Section \ref{sec:trasnlation_continuum}, different Fourier modes are decoupled due to the non-invertible continuous translation symmetry and the locality of the quadratic Hamiltonian.

Let us first consider the zero modes $\mathsf{\Pi}_{0,I}$. Their commutation relations \eqref{ftcommr} and their Hamiltonian \eqref{eq:tildeH_momentum} are exactly as in Appendix \ref{app:HS}. We use
\ie \label{eq:a0cont}
    & a_0 = \frac{1}{\sqrt{2|\omega_B|}}\Big( E_1^I \mathsf{\Pi}_{0,I} + i \, \text{sign}(\omega_B) E_2^I \mathsf{\Pi}_{0,I} \Big) \,, \\
    & \omega_B = \frac{K}{2\pi} \epsilon_{IJ} E_1^I E_2^J = \frac{K}{2\pi \sqrt{\det G}} \,,
\fe
which satisfies
\ie
    \frac{1}{2} G^{IJ} \mathsf{\Pi}_{0,I} \mathsf{\Pi}_{0,J} = |\omega_B| \left( a_0^\dagger a_0 + \frac{1}{2} \right) \,, \qquad [a_0 , a_0^\dagger] = 1 \, .
\fe

This is the same as a particle on a torus with a constant magnetic field.  The spectrum consists of Landau levels with energies $|\omega_B|(N_0+\frac{1}{2})$ with $N_0 =\langle a_0^\dagger a_0\rangle \in \bZ_{\geq 0}$. Each level is $|K|$-fold degenerate, and it is in a projective representation of the $\bZ_{|K|} \times \bZ_{|K|}$ momentum symmetry \eqref{momentumsymmmetryb}.  

Using the second expression in \eqref{eq:mosymm_ops} with $x_0=0$ and the mode expansion \eqref{modeexpa}, this symmetry is generated by 
\ie \label{eq:Umodes}
    & U_I = \exp{\left(\frac{2\pi i}{K} \mathcal{M}_I \right)} \,, \qquad [\mathcal{M}_1, \mathcal{M}_2] = - i\frac{K}{2\pi} \,, \\
    & \mathcal{M}_I = \mathsf{\Pi}_{0,I} - \frac{K}{2\pi} \epsilon_{IJ} \varphi_0^J -\left( \frac{K}{2}-B(0^+) \right) \epsilon_{IJ} \widetilde{Q}^J \,.
\fe
Note that $U_I$ are well-defined, even though ${\cal M}_I$ are not. Also, $U_I$ are independent of the nonzero modes.

The non-zero modes have similar commutation relations to the zero modes \eqref{ftcommr}.  But instead of their coordinates $\varphi^I_n$ being compact, they are non-compact and subject to a harmonic force \eqref{eq:tildeH_momentum}. Here, we define the creation and annihilation operators
\ie \label{eq:ancont}
    &a_{n,\pm} = \frac{\left(4 n^2+\omega_B^2\right)^{-1 / 4}}{\sqrt{2}} \bigg( E_1^I \mathsf{\Pi}_{n,I} \pm iE_2^I\mathsf{\Pi}_{n,I} - i\omega_{n,\mp}\Big(E_I^1 \varphi_n^I \pm i E_I^2 \varphi_n^I\Big) \bigg) \,, \\
    & \omega_{n,\pm} = \sqrt{n^2+\frac{\omega_B^2}{4}} \pm \frac{\omega_B}{2} \, ,
\fe
satisfying $[a_{n,\sigma},a_{m,\sigma'}^\dagger] = \delta_{n,m} \delta_{\sigma,\sigma'}$ and $[a_{n,\sigma},a_{m,\sigma'}]=0$.

In terms of these, the total Hamiltonian \eqref{eq:tildeH_momentum}, including the contribution from the zero modes, is, up to a normal-ordering constant,
\ie
    H' = \frac{1}{2} G_{IJ} \widetilde{Q}^{I} \widetilde{Q}^{J} + |\omega_B|  a_0^\dagger a_0  + \sum_{n \neq 0} \sum_{\sigma = \pm} \omega_{n,\sigma} a^\dagger_{n,\sigma} a_{n,\sigma} \, .
\fe

Using the mode expansion \eqref{modeexpa},
\ie \label{momentum.with.B}
    \mathscr{P} &= \int_0^{2\pi} dx \, \Theta_{tx} - \frac{K}{2\pi} \widetilde{Q}^1 \varphi^2(0^+) \\ &= \widetilde{Q}^I \left( \mathsf{\Pi}_{0,I} - \frac{K}{2\pi} \epsilon_{IJ} \varphi_0^J \right) + \frac{K}{2}\widetilde{Q}^1 \widetilde{Q}^2 - i \sum_{n \neq 0}n \left( \mathsf{\Pi}_{n,I} \varphi^I_{-n} + \frac{K}{2\pi} \varphi_n^1 \varphi_{-n}^2 \right) \\&
    = \widetilde{Q}^I \mathcal{M}_I + \frac{K}{2} \widetilde{Q}^1 \widetilde{Q}^2 - \sum_{n \neq 0} \sum_{\sigma = \pm} n a_{n,\sigma}^\dagger a_{n,\sigma} \,.
\fe
Importantly, $\mathscr P$ is not well-defined, but ${\cal P}=e^{{2\pi i\over K} \mathscr{P}}$ is well-defined.  It satisfies
\ie
    {\cal P}^K=e^{2\pi i \mathscr{P}}=e^{2\pi i (\widetilde{Q}^I \mathcal{M}_I + \frac{K}{2} \widetilde{Q}^1 \widetilde{Q}^2)} = e^{2\pi i \widetilde{Q}^1 \mathcal{M}_1} e^{2\pi i \widetilde{Q}^2 \mathcal{M}_2} = (U_1^{K})^{\widetilde{Q}^1} (U_2^{K})^{\widetilde{Q}^2} = 1\,.
\fe
The third equality above follows from the Baker–Campbell–Hausdorff formula, and the last equality follows from $(U_I)^K=1$ shown in \eqref{eq:tCU}.  This proves \eqref{twopishif}.

To summarize, the energy eigenstates
\ie \label{eq:conteigenlabel}
    \ket{w^I, N_0, N_{n,\sigma},\chi}
\fe
are labeled by the winding charges $w^I=\langle \widetilde{Q}^I \rangle \in \bZ$, the Landau level $N_0 =\langle a_0^\dagger a_0\rangle \in \bZ_{\geq 0}$, the oscillators $N_{n,\sigma} =\langle a_{n,\sigma}^\dagger a_{n,\sigma}\rangle \in \bZ_{\geq 0} $, and $\chi \in \bZ_{|K|}$ labels the $|K|$-fold degeneracy due to the projective representation of $\bZ_{|K|} \times \bZ_{|K|}$ momentum symmetry. In more detail, following the algebra \eqref{momentumsymmmetryb}, we can choose a clock-and-shift basis, such that the action of $U_I$ on the energy eigenstates is
\ie \label{eq:Uonstates}
    U_1 \ket{w^I, N_0, N_{n,\sigma},\chi} &= \ket{w^I, N_0, N_{n,\sigma},\chi-1} \,, \\
    U_2 \ket{w^I, N_0, N_{n,\sigma},\chi} &= e^{\frac{2\pi i \chi}{K}} \ket{w^I, N_0, N_{n,\sigma},\chi} \, .
\fe
The energy eigenvalues
\ie
        E(w^I,N_0,N_{n,\sigma},\chi) = \frac{1}{2} G_{IJ} w^Iw^J + |\omega_B|  N_0  + \sum_{n \neq 0} \sum_{\sigma = \pm} \omega_{n,\sigma}  N_{n,\sigma}
\fe
are indeed independent of $\chi \in \bZ_{|K|}$.  Because of the broken momentum symmetry, unlike the case of constant $B$ (i.e., $K=0$), now, $Q_{ I}$ are well-defined only modulo $K$ and do not label the states.  Instead, the states are organized into Landau levels labeled by $N_0$ and $\chi$.

Let us now discuss the spatial translation symmetry. On the modes, it acts as
\ie \label{eq:Paction}
    &\mathcal{P}\, \widetilde{Q}^I \,\mathcal{P}^{-1} = \widetilde{Q}^I \,, \\
    & \mathcal{P}\, \varphi_0^I \,\mathcal{P}^{-1} = \varphi_0^I + \frac{2\pi}{K} \widetilde{Q}^I \,, \qquad
    \mathcal{P}\, \mathsf{\Pi}_{0,I} \,\mathcal{P}^{-1} = \mathsf{\Pi}_{0,I} \,,\\
    &\mathcal{P}\, \varphi_n^I \,\mathcal{P}^{-1} = e^{\frac{2\pi i n}{K}} \varphi_n^I \,, \qquad
    \mathcal{P}\, \mathsf{\Pi}_{n,I} \,\mathcal{P}^{-1} = e^{\frac{2\pi i n}{K}} \mathsf{\Pi}_{n,I} \,, \qquad (n \neq 0)\, .
\fe
The shift of $\varphi_0^I$ reflects the winding of $\varphi^I$ across $x=x_0=0$.  Using \eqref{eq:Umodes}, these transformations reproduce the algebra \eqref{eq:cont-translation}. Then the operator $\mathcal{P}$ preserves $w^I$, $N_0$, and $N_{n,\sigma}$, and acts only on the degeneracy label $\chi$. This action is fixed by the algebra \eqref{eq:cont-translation},
\ie \label{eq:Ponstates}
      & \mathcal{P} \ket{w^I, N_0, N_{n,\sigma},\chi} = e^{-\frac{2\pi i}{K} \sum n N_{n,\sigma}}\, e^{\frac{2\pi i}{K} w^2 \left( \chi + \frac{K-1}{2} w^1 \right)} \ket{w^I, N_0, N_{n,\sigma},\chi - w^1} \, .
\fe
This means, unless $w^1 = 0 \bmod K$, the energy eigenstates are not eigenstates of $\mathcal{P}$. Translation permutes the $|K|$ states of the Landau level. This is the manifestation of the Witten effect \cite{Witten:1979ey}.

The non-invertible operators \eqref{eq:noninv_proj} act similarly.  The projectors \eqref{eq:projectors_cont} annihilate the states with $w^I \neq 0 \bmod r$, and on the rest, the operator $\mathcal{D}_{s\over rK}$ acts as $r$ times the translation by $2\pi {s\over rK}$, shifting $\chi \to \chi - {s w^1 \over r}$.  This shift is an integer only because of the projection, which is why the exponential in \eqref{eq:noninv_proj} is not well-defined by itself.

\subsection{Lattice}

Let us now analyze the spectrum of the lattice Hamiltonian, with a linearly varying $B_{j,j+1}$ as in \eqref{constantbeta}.
We define, as in \eqref{eq:BigPiLat},
\ie
\Pi_{j,I} = p_{j, I}+\frac{K}{4 \pi L} \epsilon_{I J} \phi_j^{J}-\frac{B_{0,1}}{2} \epsilon_{I J}\left(\Delta_{j, j+1}^{J}+\Delta_{j-1, j}^{J}\right)\,,
\fe
which is the lattice counterpart of the continuum operator $\Pi_I (x)$ in \eqref{eq:BigPicont}.  In terms of it, the Hamiltonian \eqref{eq:H2Linear} takes the form 
\ie
H' = \frac{1}{2}\sum_{j=0}^{L-1}G^{IJ}\,\Pi_{j,I}\Pi_{j,J}+\frac{1}{2}\sum_{j=0}^{L-1}G_{IJ}\,\Delta_{j,j+1}^{I}\Delta_{j,j+1}^{J}\, .
\fe
(Compare with the continuum expression \eqref{eq:Happ2}.) These operators also follow the algebra described in \eqref{eq:BigPiLatCom},
\ie\label{Latticemomcom} 
\left[\Pi_{j, I}, \Delta_{j',j'+1}^J\right]=\frac{i}{2 \pi}\left(\delta_{j, {j'}}-\delta_{j, {j'}+1}\right) \delta_I^J, \quad\left[\Delta_{j,j+1}^I, \Delta_{j',j'+1}^J\right]=0, \quad\left[\Pi_{j, I}, \Pi_{{j'}, J}\right]=\frac{i K}{2 \pi L} \epsilon_{I J} \delta_{j {j'}} \,.
\fe 

As in the continuum discussion, we Fourier transform the gauge-invariant variables
\ie
\mathsf{\Pi}_{n,I} = \frac{1}{\sqrt L}\sum_{j=0}^{L-1} e^{-2\pi i n j/L}\Pi_{j,I}\,,
\qquad
\Delta_{n}^{I} = \frac{1}{\sqrt L}\sum_{j=0}^{L-1} e^{-2\pi i n(j+\frac{1}{2})/L}\Delta_{j,j+1}^{I}\, .
\fe
As in the continuum, we use $\mathsf{\Pi}_{n, I}$ for the Fourier modes of $\Pi_{j, I}$. The position space commutators \eqref{Latticemomcom} become
\ie
[\mathsf{\Pi}_{n,I},\Delta_m^J] &= -\frac{\sin (\pi n / L)}{\pi}\delta_{n,-m}\delta_I^J, \\
[\mathsf{\Pi}_{n,I},\mathsf{\Pi}_{m,J}] &= \frac{iK}{2\pi L}\epsilon_{IJ}\delta_{n,-m}\,,
\qquad
[\Delta_n^I,\Delta_m^J]=0\, .
\fe
We define
\ie
\omega_B = \frac{K}{2\pi L\sqrt{\det G}},
\qquad
\omega_n = \frac{\sin(\pi n/L)}{\pi},
\qquad
\omega_{n,\pm} = \sqrt{\omega_n^2+\frac{\omega_B^2}{4}} \pm\frac{\omega_B}{2}.
\fe
For $n=0$, we have ${\Delta}_{0}^I=\widetilde{Q}^I/\sqrt{L}$ and the oscillator,
\ie
a_0 = \frac{1}{\sqrt{2|\omega_B|}} \left[ E_{1}^{I}\,\mathsf{\Pi}_{0,I} +i\,\operatorname{sign}(\omega_B)\, E_{2}^{I}\,\mathsf{\Pi}_{0,I} \right]\,,
\fe
which satisfies $[a_0,a_0^\dagger]=1$, where $a_0^\dagger$ is the Hermitian conjugate and $E_I^r$ is the vielbein \eqref{eq:vielbein}.

For the nonzero modes, the annihilation operators are
\ie
a_{n,\pm}=\frac{\left(4 \omega_n^2+\omega_B^2\right)^{-1 / 4}}{\sqrt{2}}\left[\left(E_{1}^{I}\pm i\,E_{2}^{I}\right)\mathsf{\Pi}_{n,I}-\frac{\omega_{n,\mp}}{\omega_n}\left(E_I^1\pm i\,E_I^2\right)\Delta_n^I\right]\, ,
\fe
and the creation operators $a_{n,\pm}^{\dagger}$ are their Hermitian conjugates.

Collecting all these facts, the Hamiltonian, up to a normal-ordering constant, becomes
\ie
H'=\frac{1}{2 L} G_{IJ}\,\widetilde{Q}^{I} \widetilde{Q}^{J}+\left|\omega_B\right|a_0^{\dagger} a_0+\sum_{n=1}^{L-1}\left(\omega_{n,+} a_{n,+}^{\dagger} a_{n,+}+\omega_{n,-} a_{n,-}^{\dagger} a_{n,-}\right)\, .
\fe
Interestingly, as in the continuum, even though we have only a subgroup of the $\bZ_L$ lattice translation symmetry, the Fourier modes with different values of $n$ above do not mix.  This follows from the non-invertible translation symmetry generated by $\mathsf{D}$, which includes a translation by one site.

As in the continuum, the energy eigenstates are
\ie
\ket{w^I,N_0,N_{n,\sigma},\chi}
\fe
with the same set of labels as in  \eqref{eq:conteigenlabel}. The energy eigenvalues are
\ie
E(w^I,N_0,N_{n,\sigma},\chi) = \frac{1}{2L}G_{IJ}\,w^{I}w^{J}+|\omega_B|N_0+\sum_{n=1}^{L-1}\sum_{\sigma=\pm}\omega_{n,\sigma}N_{n,\sigma}\,,
\fe
up to the normal-ordering constant, and they are independent of $\chi$.

Finally, ${\sf P}$ and $U_I$ obey the algebra \eqref{eq:latalg} with $g=\gcd(L,K)$. Repeating the argument that led to \eqref{eq:Ponstates} and fixing the overall phase such that ${\sf P}^g=1$, we find
\ie
 {\sf P} &\ket{w^I,N_0,N_{n,\sigma},\chi} \\
&=  \,e^{-\frac{2\pi i}{g}\sum_{n}\sum_{\sigma=\pm} n N_{n,\sigma}}\, e^{\frac{2\pi i}{g}w^2 \left(\chi+\frac{K}{g}w^1\frac{g-1}{2}\right)} \ket{w^I,N_0,N_{n,\sigma},\chi-K w^1/g}\, ,
\fe
which matches the continuum result \eqref{eq:Ponstates} with $\mathsf{P} \to \mathcal{P}^{K/g}$. And similarly, the one-site non-invertible translation $\mathsf{D}$ \eqref{eq:noninv1site} annihilates the states with $w^I \neq 0 \bmod \frac{L}{g}$ and shifts $\chi \to \chi - \frac{K}{L}w^1$, which is an integer because of the projection.

\section{More on non-invertible translation symmetries} \label{app:noninv}

\subsection{Continuum} \label{app:noninv.cont}

In this appendix, we start with the expression \eqref{eq:noninvT_corr} for the non-invertible translation symmetry operator of the continuum theory, explicitly integrate out the auxiliary degrees of freedom $u$ and $v$ on the operator, and show that it reduces to the expression \eqref{eq:noninv_proj}, which includes projection operators.

Let us reiterate \eqref{eq:noninvT_corr},
\begin{align}
\begin{split}
    \mathcal{D}_{\frac{s}{rK}} = \int [Du Dv] \exp \Bigg[
        i \int_{x_0}^{x_0 + 2\pi} dx &\Bigg(
            \frac{2\pi s}{rK} \hat{\Theta}_{tx} + \frac{r}{2\pi} u \partial_x v - \frac{s}{2\pi} u \partial_x \varphi^1 + \frac{1}{2\pi} v \partial_x \varphi^2 \Bigg) \\
        & -  i\left( r \widetilde{Q}^{(u)} v  - s \widetilde{Q}^{(u)} \varphi^1  + \widetilde{Q}^{(v)} \varphi^2  \right)\Big|_{x=x_0^+}
    \Bigg] \,.
\end{split}
\end{align}
We will first integrate out $u$, including the infinite discrete sum over the winding number $\widetilde{Q}^{(u)}$.
The path integral over topologically trivial configurations of $u$ sets $r \partial_x v - s\partial_x \varphi^1 = 0$, whereas the sum over the winding number $\widetilde{Q}^{(u)}$ sets $rv - s\varphi^1 \equiv 2\pi\mathcal{N} \in 2\pi\mathbb{Z}$.

From $r \partial_x v - s\partial_x \varphi^1 = 0$, it follows that $r \widetilde{Q}^{(v)} - s \widetilde{Q}^{1} = 0$.
Since $\mathrm{gcd}(s,r) = 1$, this condition cannot be satisfied unless $\widetilde{Q}^1=0\bmod r$ and $\widetilde{Q}^{(v)}=0\bmod s$. 
Therefore, the path integral over $u$ leads to a factor of a projection operator on $\widetilde{Q}^1=0\bmod r$, and the winding number of $v$ is constrained to be $\widetilde{Q}^{(v)} = {s \over r}\widetilde{Q}^1$.

We obtain
\begin{align}
\begin{split}
    \mathcal{D}_{\frac{s}{rK}} &\propto \mathbb{P}_r^1 \sum_{\mathcal{N}\in\mathbb{Z}} \exp \Bigg[
        i \int_{x_0}^{x_0 + 2\pi} dx \Bigg(
            \frac{2\pi s}{rK} \hat{\Theta}_{tx} + \left( 
                \frac{s}{2\pi r}\varphi^1 + \frac{1}{r}\mathcal{N}  
            \right) \partial_x \varphi^2\Bigg)
         -  i \frac{s}{r} \widetilde{Q}^{1} \varphi^2 \Big|_{x=x_0^+}
    \Bigg] \\
    &= \mathbb{P}_r^1 \sum_{\mathcal{N}\in\mathbb{Z}} \exp \Bigg[
        \frac{2\pi i s}{rK}\left(
            \int_{x_0}^{x_0+2\pi} {\Theta}_{tx}\, dx  - \frac{K}{2\pi } \widetilde{Q}^1 \varphi^2 \Big|_{x=x_0^+}
        \right)  + \frac{2\pi i}{r} \mathcal{N} \widetilde{Q}^2
    \Bigg] \,.
\end{split}
\end{align}
The overall normalization of the path integral will be fixed soon.
Note that in the second line, we see the energy-momentum tensor ${\Theta}_{tx} =\hat{\Theta}_{tx} + \frac{K}{(2\pi)^2} \varphi^1 \partial_x \varphi^2$ \eqref{eq:T}.

The sum over the integer $\mathcal{N}$ gives an additional projection operator on $\widetilde{Q}^2 =0$ mod $r$, and the operator becomes
\ie
    \mathcal{D}_{\frac{s}{rK}} = r \,\mathbb{P}_r^1 \mathbb{P}_r^2 \exp \left[
        \frac{2\pi i s}{rK} \left(
            \int_{x_0}^{x_0+2\pi} {\Theta}_{tx}\, dx  - \frac{K}{2\pi } \widetilde{Q}^1 \varphi^2 \Big|_{x=x_0^+}
        \right)
    \right] \,,
\fe
as in \eqref{eq:noninv_proj}.
The overall normalization is chosen such that the quantum dimension of the operator agrees with the partition function of the fuzzy torus quantum mechanics on a circle (which is $r$).

\subsection{Lattice} \label{app:noninv.lattice}

We now present a lattice version of \eqref{eq:noninvT_corr}.  

We start with the continuum expression 
\begin{align}\label{QMEuclideanc}
\begin{split}
        {\cal S}_\text{Continuum}=-{i\over 2\pi} \int_{x_0}^{x_0 + 2\pi} dx &\Bigg(
             {r} u \partial_x v - {s} u \partial_x \varphi^1 + v \partial_x \varphi^2 \Bigg) \\
        & +  i\left( r \widetilde{Q}^{(u)} v  - s \widetilde{Q}^{(u)} \varphi^1  + \widetilde{Q}^{(v)} \varphi^2  \right)\Big|_{x=x_0^+}\,,
\end{split}
\end{align}
and view it as the action of a quantum mechanical system in Euclidean time.  Specifically, this is the action of the fuzzy torus coupled to background fields $\varphi^I$.

Ignoring the background fields, the Euclidean lattice version of the fuzzy torus is constructed by placing real-valued $u_j$ and $v_j$ on the sites and integer-valued $\bZ\times \bZ$ gauge fields $n^u_{j,j+1}$ and $n^v_{j,j+1}$ on the links with the action \cite{Gorantla:2021svj}
\ie\label{Fuzzytorusa}
{\cal S}_\text{Fuzzy torus}=-i{r\over 2\pi}\sum_j\Big( u_j(v_{j+1}-v_j+2\pi n^v_{j,j+1})- 2\pi  v_j n^u_{j-1,j}\Big)\,.
\fe
(Unless stated otherwise, the labels of the fields satisfy $j\sim j+L$.) This action is not $\bZ\times \bZ$ gauge-invariant.  But  $e^{-{\cal S}_\text{Fuzzy torus}}$ is invariant.  

Our lattice dynamical fields $\phi_j^I$ and $n_{j,j+1}^I$ couple to this system as background fields in the quantum mechanical problem:
\ie\label{Fuzzytorusc}
{\cal S}_\text{Lattice}={\cal S}_\text{Fuzzy torus}&+i{s\over 2\pi}\sum_j\Big( u_j(\phi^1_{j+1}-\phi^1_j+2\pi n^1_{j,j+1})- 2\pi \phi^1_j n^u_{j-1,j}\Big)\\
&-i{1\over 2\pi}\sum_j\Big( v_j(\phi^2_{j+1}-\phi^2_j+2\pi n^2_{j,j+1})- 2\pi \phi^2_j n^v_{j-1,j}\Big)\,.
\fe
Again, this action is not  $\bZ\times \bZ\times\bZ\times\bZ$ (with the factors acting on $u$, $v$, $\phi^1$, and $\phi^2$, respectively) gauge-invariant, but its exponential is.  Note that this exponential is manifestly local and does not need any correction terms.

Next, we follow the steps in Appendix \ref{app:noninv.cont}.
Integrating out $u_j$ sets\footnote{This integration out leads to prefactors in the path integral measure.  For the moment, we ignore them.  Below, we will determine them.\label{normalfi}}
\ie\label{vjcon}
 v_{j+1}-v_j+2\pi n^v_{j,j+1} ={s\over r}(\phi^1_{j+1}-\phi^1_j+2\pi n^1_{j,j+1})\,,
\fe
i.e.,
\ie\label{vjinto}
v_{j}=v_0-2\pi {\cal N}^v_{j} +{s\over r}\left(\phi^1_{j}-\phi^1_{0}+2\pi{\cal N}^1_{j}\right) \,, \qquad j=1,\cdots,L\,,
\fe
where we used the notation
\ie
{\cal N}^x_j= \sum_{j'=0}^{j-1}n^x_{j',j'+1} \quad, \quad x=1,2,u,v\quad,\quad j=1,\cdots,L\,.
\fe
As a check, \eqref{vjinto} is gauge covariant.  One consequence of \eqref{vjcon} is
\ie\label{sumjnv}
r {\cal N}^v_L= s {\cal N}^1_L=s\widetilde{Q}^1\,,
\fe
and therefore, $\widetilde{Q}^1$ should be a multiple of $r$.

Then, integrating out $v_0$ sets
\ie\label{sumjnu}
{\cal N}^2_{L} =\widetilde{Q}^{2}=r{\cal N}_{L}^u\,,
\fe
and therefore, $\widetilde{Q}^2$ is also constrained to be multiples of $r$.

As a result, integrating out the added auxiliary fields on the operator maps
\ie\label{subcalSLs}
&\exp\Big(-{\cal S}_\text{Lattice}\Big)\to \\  &\qquad \mathbb{P}_r^1\mathbb{P}_r^2\exp\left(i{s\over 2\pi r}\sum_{j=1}^{L}\left(\phi^1_{j}+2\pi{\cal N}^1_{j}\right)(\phi^2_{j+1}-\phi^2_j+2\pi n^2_{j,j+1}) - \frac{is}{r} \widetilde{Q}^1\phi^2_1 \right)\,.
\fe
Using \eqref{sumjnv} and \eqref{sumjnu}, it is easy to check that this expression is gauge invariant.

The result \eqref{subcalSLs} of integrating out the auxiliary fields on the operator \eqref{Fuzzytorusc} contains one of the factors in ${\cal V}_{\frac{s}{r}}$ of \eqref{calVdef}.  To match them, we should remember that we view $\phi_j^I$ and $n_{j,j+1}^I$ as classical background fields in the quantum mechanics problem on the operator, and that
$\Delta_{j,j+1}^{I} = \frac{\phi_{j+1}^{I}-\phi_{j}^{I}}{2\pi} + n_{j,j+1}^{I}$ of \eqref{DeltaIdef} is well-defined.
Then, redefining ${\cal S}_\text{Lattice} $ by a local, well-defined counterterm in these background fields, 
\ie\label{latticelocact}
&{\cal S}_\text{Lattice} \to {\cal S}_\text{Lattice} + {\cal S}_\text{Lattice counterterm}\,,\\
&{\cal S}_\text{Lattice counterterm}=-\frac{i \pi s}{r} \sum_{j=0}^{L-1} \Delta_{j, j+1}^1 \Delta_{j, j+1}^2\,,
\fe
makes the result of integrating out the auxiliary fields
\ie
\exp\Big(-{\cal S}_\text{Lattice}\Big)&\to   r \mathbb{P}_r^1\mathbb{P}_r^2 {\cal V}_{\frac{s}{r}}\\
\qquad & = r\mathbb{P}_r^1\mathbb{P}_r^2 \exp \left(i \frac{s}{2 \pi r} \sum_{j=1}^L\left(\phi_j^1+2 \pi \mathcal{N}_j^1\right)\left(\phi_{j+1}^2-\phi_j^2+2 \pi n_{j, j+1}^2\right)-\frac{i s}{r} \widetilde{Q}^1 \phi_1^2\right)\\
&\qquad \cdot\exp\left(\frac{i \pi s}{r} \sum_{j=0}^{L-1} \Delta_{j, j+1}^1 \Delta_{j, j+1}^2\right)\,, 
\fe
where we included the normalization factor $r$ as in footnotes \ref{factorofr} and \ref{normalfi}.  Note also that the local counterterm in \eqref{latticelocact} and the overall normalization guarantee that $({\cal V}_{\frac{s}{r}})^r=\B^s$ \eqref{VtorBtos}.

We conclude that, just as in Appendix \ref{app:noninv.cont}, the lattice non-invertible operator $\mathsf{D}$ of \eqref{eq:noninv1site}, which is constructed out of ${\cal V}_{\frac{s}{r}}$, can be given a manifestly local gauge-invariant presentation at the cost of adding auxiliary degrees of freedom on the operator.

\section{More bosons in two dimensions} \label{app:more_bosons}

Here, we will briefly discuss the generalization of some of the results in Sections \ref{sec:continuum} and \ref{sec:trasnlation_continuum} to the case with $N$ compact bosons $\varphi^I$, $I =1, 2, \cdots, N$.
The Lagrangian is now
\begin{equation}
    \frac{1}{4\pi} G_{IJ} d\varphi^I \wedge \star d\varphi^J + \frac{1}{4\pi} B_{IJ} d\varphi^I \wedge d\varphi^J \,,
\end{equation}
where $B_{IJ} = - B_{JI}$ and $B_{IJ} \sim B_{IJ} + 1$.
We make the couplings $B_{IJ}$ position-dependent,
\ie
    B_{IJ} (x + 2\pi )=B_{IJ} (x) +K_{IJ} \,,
\fe
where $K_{IJ}$ is an anti-symmetric matrix of integers.

Using a $GL(N,\mathbb{Z})$ transformation, we can choose a basis for the bosons $\varphi^I$ such that the matrix $K_{IJ}$ is of the form \cite{newman1972integral}
\ie\label{goodbasis}
& \left(\begin{array}{cc}
0 & K^{[1]} \\
-K^{[1]} & 0
\end{array}\right) \oplus\left(\begin{array}{cc}
0 & K^{[2]} \\
-K^{[2]} & 0
\end{array}\right) \oplus \cdots \oplus\left(\begin{array}{cc}
0 & K^{[M]} \\
-K^{[M]} & 0
\end{array}\right) \oplus {\bf 0}\,,\\
&K^{[m]}\in \bZ_{>0}\,, \qquad m= 1, 2, \cdots , M \,,\\
&K^{[1]}\Big| K^{[2]}\Big| K^{[3]}\Big| \cdots \Big| K^{[M]} \, ,
\fe
with ${\bf 0}$ the $(N-2M) \times (N-2M)$ zero matrix. 
Then, we can repeat the discussion for $N=2$, and write the target space momentum symmetry operators
\begin{align}
\begin{split}
    U_{2m-1} &=\exp \left[
        \frac{2\pi i}{K^{[m]}} \left(
            \int_{x_0}^{x_0 +2\pi} J_{t, 2m-1} \,dx - \frac{K^{[m]}}{2\pi} \varphi^{2m} \Big|_{x = x_0^+}
        \right)
    \right] \,,\\
    U_{2m} &=\exp \left[
        \frac{2\pi i}{K^{[m]}} \left(
            \int_{x_0}^{x_0 +2\pi} J_{t, 2m} \,dx + \frac{K^{[m]}}{2\pi} \varphi^{2m-1} \Big|_{x = x_0^+}
        \right)
    \right] \,,
\end{split}
\end{align}
for $m=1,2,\cdots,M$, and $J_{\mu,I} = \frac{1}{2\pi} \left( G_{IJ} \partial_\mu \varphi^J - \epsilon_{\mu \nu} B_{IJ} (x) \partial^\nu \varphi^J \right)$.
For each $m$, they generate a $\mathbb{Z}_{|K^{[m]}|} \times \mathbb{Z}_{|K^{[m]}|}$ symmetry, which acts projectively,
\ie
    U_{2m-1} U_{2m} = e^{\frac{2\pi i}{K^{[m]}}} U_{2m} U_{2m-1} \, .
\fe
The momentum symmetries for the bosons $\varphi^I$ with $I = 2M+1 , \cdots , N$ are unaffected, and they generate a $U(1)^{N - 2M}$ symmetry, which acts linearly.
The total momentum symmetry is therefore
\ie
    \mathbb{Z}_{|K^{[1]}|}^2 \times \cdots \times \mathbb{Z}_{|K^{[M]}|}^2 \times U(1)^{N - 2M} \, .
\fe

Now, suppose that $B_{IJ} (x)$ depends linearly on $x$,
\begin{equation}
    B_{IJ} (x) = B_{0IJ} + K_{IJ} \frac{x}{2\pi} \, .
\end{equation}
As in the $N=2$ case, the (invertible) spatial translation symmetry is broken to a $\mathbb{Z}_{|K|}$ subgroup, where
\ie
    K = \mathrm{gcd} (K^{[1]} , \cdots , K^{[M]} ) =K^{[1]}\, .
\fe
(Even without choosing the basis \eqref{goodbasis}, we can write it as $K=\gcd(K_{IJ})$.) 
It is generated by
\ie
    \mathcal{P} = \exp \left[
        \frac{2\pi i}{K} \left(
            \int_{x_0}^{x_0+2\pi} {\Theta}_{tx} \,dx - \frac{1}{2\pi} \sum_{m=1}^M K^{[m]} \widetilde{Q}^{2m-1} \varphi^{2m} \Big|_{x=x_0^+}
         \right)
    \right] \,,
\fe
where now
\ie
    {\Theta}_{\mu x} = \hat{\Theta}_{\mu x} - \frac{1}{(2\pi)^2} \sum_{m=1}^M K^{[m]} \epsilon_{\mu\nu} \varphi^{2m-1} \partial^\nu \varphi^{2m}\, .
\fe
The operator $\mathcal{P}$ satisfies
\ie
    U_{2m-1} \mathcal{P} = e^{\frac{2\pi i}{K} \widetilde{Q}^{2m}} \mathcal{P} U_{2m-1} \,, \quad U_{2m} \mathcal{P} = e^{ -\frac{2\pi i}{K} \widetilde{Q}^{2m-1}} \mathcal{P} U_{2m} \,,
\fe
for $m=1 ,2, \cdots, M$.
It commutes with the remaining $U(1)^{N-2M}$ momentum symmetry.

The non-invertible translation symmetry operators in Section \ref{sec:nonInvertibleCont} can also be straightforwardly generalized by using auxiliary degrees of freedom or various projection operators. We will not describe them explicitly.

\section{Higher dimensions} \label{app:higherD}
The discussion on higher-group symmetries in Section \ref{sec:higher_group} can be generalized to arbitrary spacetime dimensions.
Specifically, we take a theory of $N$ compact bosons $\varphi^I\sim \varphi^I +2\pi$ with $I=1,\cdots, N$ in $D$ spacetime dimensions and find a higher-group symmetry involving a $-1$-form symmetry, analogous to that in Section \ref{sec:higher_group} for $D=2$.

We begin with the Lagrangian\footnote{For $D=2$, this normalization of $G_{IJ}$, which was used in Section \ref{sec:continuum}, is convenient.  For other values of $D$, $G_{IJ}$ is dimensionful, and the normalization is insignificant.}
\ie
{1\over 4\pi}G_{IJ}d\varphi^I\wedge \star d\varphi^J\, .
\fe
The theory has a $U(1)^N$ zero-form (target space) momentum symmetry, so we couple it to $U(1)^N$ background gauge fields and write\footnote{In this appendix, the superscripts on various Lagrangians and the background gauge fields $\widetilde{A}$'s below, as well as on the corresponding field strengths $\widetilde{F}$'s and the gauge parameters $\widetilde{\lambda}$'s, are used to indicate their differential form degrees.}
\ie
{\cal L}_G^{(D)}={1\over 4\pi}G_{IJ}(d\varphi^I-A^I)\wedge \star(d\varphi^J-A^J)\,,
\fe
with the gauge transformations
\ie\label{Agaugea}
&\varphi^I\to \varphi^I +\lambda^I\,,\\
&A^I\to A^I +d\lambda^I\, .
\fe
Again, if $A^I$ is topologically nontrivial, we need to add insertions of winding operators.\footnote{In $D$ spacetime dimensions with $D>2$, the winding operators are $D-2$-dimensional extended operators.}

The theory also has a $U(1)^N $ $D-2$-form winding symmetry.
We couple it to $D-1$-form background gauge fields $\widetilde A_I^{(D-1)}$ via
\ie\label{AtildeAg}
{1\over 2\pi} \widetilde A_I^{(D-1)}\wedge (d\varphi^I-A^I)\, .
\fe
Here, we included a term quadratic in the background gauge fields, such that the theory is manifestly gauge-invariant under the gauge transformations of $A^I$ \eqref{Agaugea}.

The coupling \eqref{AtildeAg} makes it clear that the theory with the background gauge fields is not invariant under $\widetilde A_I^{(D-1)}$ gauge transformations.  This reflects a mixed 't Hooft anomaly between the momentum zero-form symmetry and the winding $D-2$-form symmetry.  Soon, we will discuss this anomaly in more detail.

The winding symmetry leads to additional $q$-form symmetries with $q=0,\cdots, D-3$.
Their currents are products of the currents of the winding symmetries,
\ie
{1\over (2\pi)^p}(d\varphi^{I_1}-A^{I_1})\wedge \cdots \wedge (d\varphi^{I_p}-A^{I_p})\,, \quad p=D-q-1\, .
\fe
Since these currents are products of winding currents, we can refer to these symmetries as ``derived symmetries.'' Such symmetries that are derived from higher-form symmetries have been discussed by various authors, including \cite{Heidenreich:2020pkc, Brauner:2020rtz, Choi:2021kmx, Roumpedakis:2022aik, Damia:2022seq, Seiberg:2024wgj, Seiberg:2024yig}. In particular, we follow many of the steps in \cite{Brauner:2020rtz}.

These derived symmetries can be extended to $p=D$, leading to a derived $q=-1$-form symmetry.  The corresponding ``currents'' are integrated over all the $D$-dimensional spacetime, and their background gauge fields are $\theta$-parameters.

Turning on background fields for all these symmetries, we study the Lagrangian
\ie\label{spacetimeLg}
 \mathcal{L}^{(D)} &= \mathcal{L}_G^{(D)} + \sum_{p=1}^D \frac{1}{p!} \frac{1}{(2\pi)^p} \widetilde{A}^{(D-p)}_{I_1 \cdots I_p} \wedge (d\varphi^{I_1} - A^{I_1} ) \wedge \cdots \wedge (d\varphi^{I_p} - A^{I_p} )\, .
 \fe
Here, $\widetilde A^{(D-p)}_{I_1\cdots I_p}$ are $D-p$-forms in spacetime and are antisymmetric in the target space indices $I=1,\cdots ,N$. In this notation, ${\widetilde A}^{(0)}_{I_1\cdots I_D}$ are $\theta$-parameters.\footnote{In the main text, we have $D=2$, $N=2$, and $B = \widetilde{A}^{(0)}_{12}/2\pi$.}
The prefactors $\frac{1}{p!} \frac{1}{(2\pi)^p}$ are such that the background gauge fields couple to currents with quantized charges.

The Lagrangian \eqref{spacetimeLg} is manifestly gauge invariant under the gauge transformations of $A^I$, but the gauge invariance of the other background fields is more subtle. Related to that, the Lagrangian is well-defined only for topologically trivial $\widetilde A^{(D-p)}_{I_1\cdots I_p}$.

First, let us study the theory based on \eqref{spacetimeLg} with topologically trivial $\widetilde A^{(D-p)}_{I_1\cdots I_p}$ and explore its variation under topologically trivial gauge transformations.
We could try the gauge transformations $\widetilde A_{I_1\cdots I_p}^{(D-p)}\to \widetilde A_{I_1\cdots I_p}^{(D-p)} + d\widetilde \lambda_{I_1\cdots I_p}^{(D-p-1)}$ with $p=1,\cdots, D-1$.
Clearly, \eqref{spacetimeLg} is not invariant under this, and therefore one might think that there is an anomaly. However, most of this anomaly can be avoided by deforming the gauge transformations for $p=1,\cdots,D-2$ \cite{Brauner:2020rtz},\footnote{This is essentially the Green-Schwarz anomaly canceling mechanism \cite{Green:1984sg}.}
\ie\label{modgaugeo}
&\widetilde A^{(D-p)}_{I_1\cdots I_p}\to \widetilde A^{(D-p)}_{I_1\cdots I_p}+ d\widetilde \lambda^{(D-p-1)}_{I_1\cdots I_p}+{(-1)^{D+1}\over 2\pi}\widetilde \lambda^{(D-p-2)}_{I_1\cdots I_{p+1}}\wedge dA^{I_{p+1}}\,, \quad &p=1,\cdots, D-2\,,\\
&\widetilde A^{(1)}_{I_1\cdots I_{D-1}}\to \widetilde A^{(1)}_{I_1\cdots I_{D-1}}+ d\widetilde \lambda^{(0)}_{I_1\cdots I_{D-1}}\,, \quad & p=D-1\,,\\
&\widetilde A^{(0)}_{I_1\cdots I_{D}}\to \widetilde A^{(0)}_{I_1\cdots I_{D}}\,, \quad & p=D\, .
\fe
Under this transformation,
\ie\label{boundvariio}
{\cal L}^{(D)}\to {\cal L}^{(D)}
+& \sum_{p=1}^{D-1} \frac{1}{p!} \frac{1}{(2\pi)^p} d\widetilde \lambda^{(D-p-1)}_{I_1\cdots I_p}\wedge (d\varphi^{I_1} - A^{I_1} ) \wedge \cdots \wedge (d\varphi^{I_p} - A^{I_p} )\\
+& \sum_{p=1}^{D-2} \frac{(-1)^{D+p}}{(p+1)!} \frac{1}{(2\pi)^{p+1}} \widetilde \lambda^{(D-p-2)}_{I_1\cdots I_{p+1}}\wedge d\left( (d\varphi^{I_1} - A^{I_1} ) \wedge \cdots \wedge (d\varphi^{I_{p+1}} - A^{I_{p+1}} )\right)\\
={\cal L}^{(D)}
+& d\left(\sum_{p=1}^{D-1} \frac{1}{p!} \frac{1}{(2\pi)^p} \widetilde \lambda^{(D-p-1)}_{I_1\cdots I_p}\wedge (d\varphi^{I_1} - A^{I_1} ) \wedge \cdots \wedge (d\varphi^{I_p} - A^{I_p} )\right)\\
 +&\frac{(-1)^D}{2\pi} \widetilde \lambda^{(D-2)}_{I}\wedge  dA^{I}  \, .
\fe
Since we assumed that the background fields and the gauge transformations are topologically trivial, we can neglect the total derivative and find that the modification of the gauge transformation \eqref{modgaugeo} leads to an anomalous gauge transformation only under $\widetilde \lambda_I^{(D-2)}$.
This mixed momentum-winding anomaly corresponds to the $D+1$-dimensional anomaly theory
\ie\label{anomalygena}
\mathcal{L}_{anomaly}^{(D+1)} = \frac{(-1)^{D+1}}{2\pi} \widetilde{A}^{(D-1)}_I \wedge dA^I  \, .
\fe

Next, we consider the $\theta$-parameters $\widetilde A^{(0)}_{I_1\cdots I_D}$.  In the absence of the background fields $A^I$, the theory is invariant under
\ie
\widetilde A^{(0)}_{I_1\cdots I_D}\to \widetilde A^{(0)}_{I_1\cdots I_D}+ 2\pi k_{I_1\cdots I_D}\,, \quad  k_{I_1\cdots I_D}\in \bZ\, .
\fe
The coupling to the other background gauge fields violates this periodicity.  However, we find that the transformation
\ie\label{thetaperiodicity}
&\widetilde A^{(D-p)}_{I_1\cdots I_{p}}\to \widetilde A^{(D-p)}_{I_1\cdots I_{p}}+ {(-1)^{p(D+1)}\over (D-p)!(2\pi)^{D-p-1}}k_{I_1\cdots I_{D}} A^{I_{p+1}}\wedge \cdots \wedge A^{I_D} \,, \quad& p=1,\cdots, D-1\,,\\
&\widetilde A^{(0)}_{I_1\cdots I_{D}}\to \widetilde A^{(0)}_{I_1\cdots I_{D}}+ 2\pi k_{I_1\cdots I_{D}}\,, \quad& p=D\,,
\fe
leaves the theory unchanged.
In more detail, under this transformation,
\ie
    &\mathcal{L}^{(D)} \rightarrow \mathcal{L}^{(D)} + \frac{k_{I_1 \cdots I_D}}{D! (2\pi)^{D-1}} \Big(
        d\varphi^{I_1} \wedge \cdots \wedge d\varphi^{I_D} - A^{I_1}\wedge \cdots \wedge A^{I_D}
    \Big)
    \, .
\fe
The first term integrates to an integer multiple of $2\pi$ and can be dropped.  The second term  cancels the variation of the anomaly
\ie
\mathcal{L}_{anomaly}^{(D+1)} \rightarrow \mathcal{L}_{anomaly}^{(D+1)} + \frac{k_{I_1 \cdots I_D}}{D! (2\pi)^{D-1}} d \Big( A^{I_1} \wedge \cdots \wedge A^{I_D} \Big) \, .
    \fe

The following field strengths are invariant under the gauge transformations \eqref{modgaugeo} and the periodicity \eqref{thetaperiodicity}
\ie\label{fieldstrengths}
 & \widetilde{F}_{I_1 \cdots I_p}^{(D-p+1)} = d\widetilde{A}^{(D-p)}_{I_1 \cdots I_p} + \frac{(-1)^{D}}{2\pi}\widetilde{A}^{(D-p-1)}_{I_1 \cdots I_{p+1}} \wedge dA^{I_{p+1}} \,, \quad p=1,\cdots, D-1  \,,\\
&\widetilde{F}_{I_1 \cdots I_D}^{(1)} = d\widetilde{A}^{(0)}_{I_1 \cdots I_D}\, .
\fe
The modified gauge transformations \eqref{modgaugeo} and the periodicity \eqref{thetaperiodicity} describe a higher-group structure, where $q$-form symmetries for $q= -1 ,0 , 1, \cdots , D-2$ mix with each other.

With this understanding, we can discuss topologically nontrivial fields and gauge transformations.  This means that when we cover our spacetime (or the bulk that we will attach to it) with patches, the transition functions on the overlap regions include both the continuous gauge transformations \eqref{modgaugeo} and the discrete transformations \eqref{thetaperiodicity}.

We extend all the fields to a $D+1$-dimensional bulk, and write the bulk term
\ie\label{bulktermg_app}
&\mathcal{L}_{bulk}^{(D+1)} = \sum_{p=1}^D \frac{1}{p!} \frac{1}{(2\pi)^p} \widetilde{F}_{I_1 \cdots I_p}^{(D-p+1)} \wedge (d\varphi^{I_1} - A^{I_1} ) \wedge \cdots \wedge (d\varphi^{I_p} - A^{I_p} ) \,,
\fe
with the field strengths $\widetilde{F}_{I_1 \cdots I_p}^{(D-p+1)}$ in \eqref{fieldstrengths}.
Importantly, ${\cal L}_{bulk}$ is  well-defined.
It is invariant under \eqref{modgaugeo} and \eqref{thetaperiodicity}.

This bulk term can be written as
\ie\label{bulktalg}
&\mathcal{L}_{bulk}^{(D+1)} = \sum_{p=1}^D \frac{1}{p!} \frac{1}{(2\pi)^p} d\left(\widetilde{A}_{I_1 \cdots I_p}^{(D-p)} \wedge (d\varphi^{I_1} - A^{I_1} ) \wedge \cdots \wedge (d\varphi^{I_p} - A^{I_p} ) \right)+\mathcal{L}_{anomaly}^{(D+1)}\,,\\
&\mathcal{L}_{anomaly}^{(D+1)} = \frac{(-1)^{D+1}}{2\pi} \widetilde{A}^{(D-1)}_I \wedge dA^I  \, .
\fe
The first term in ${\cal L}_{bulk}$ leads to the second term in the spacetime Lagrangian $\cal L$ \eqref{spacetimeLg}. Importantly, ${\cal L}_{bulk}$ does not depend on the extension of  $\varphi^I$ to the bulk, and therefore the theory is still $D$-spacetime-dimensional.\footnote{For $D=2$, we demonstrated this in Appendix \ref{app:bulk}.}
The remaining bulk term, ${\cal L}_{anomaly}$, represents the mixed anomaly between the momentum and winding symmetries.

Given that $\widetilde A_I^{(D-1)}$ are not standard gauge fields, one might ask whether ${\cal L}_{anomaly}$ is well-defined.  The standard way to make this Chern-Simons theory well-defined is to use a Lagrangian in $D+2$-dimensions.  The Lagrangian for this theory,
\ie\label{Dplus2}
{\cal L}_{anomaly}^{'\ (D+2)} ={(-1)^{D-1}\over 2\pi} d\widetilde A_I^{(D-1)}\wedge d A^I={(-1)^{D-1}\over 2\pi} \widetilde F_I^{(D)}\wedge d A^I\,,
\fe
is well-defined and the corresponding action on a closed $D+2$-dimensional manifold is in $2\pi\bZ$.
The fact that it is well-defined is manifest in the second expression in \eqref{Dplus2}.   (The difference between $d\widetilde A_I^{(D-1)}$ and $\widetilde F_I^{(D)}$ includes a factor of $dA^J$, which vanishes upon anti-symmetrization of $I$ and $J$.)  Then, to see that the integral of \eqref{Dplus2} is in $2\pi\bZ$, we use the fact that transition functions associated with \eqref{modgaugeo} and \eqref{thetaperiodicity} (which make $\widetilde A_I^{(D-1)}$ non-standard gauge fields), are such that $d\widetilde A_I^{(D-1)}$ are shifted across overlaps by terms including $dA^J$.  Such terms do not affect ${\cal L}'_{anomaly}$.  So even though the periods of $d\widetilde A_I^{(D-1)}$ or of $\widetilde F_I^{(D)} $ might not be in $2\pi\bZ$, the integral of \eqref{Dplus2} is in $2\pi\bZ$ and therefore, the $D+2$-dimensional theory \eqref{Dplus2} is trivial on a closed $D+2$-dimensional manifold.

To summarize, the bulk term ${\cal L}_{bulk}$ \eqref{bulktermg_app} is manifestly well-defined.  Writing it as in \eqref{bulktalg} makes it clear that it is independent of the extension of the dynamical fields to the bulk, and it reproduces the second term in the spacetime Lagrangian \eqref{spacetimeLg}.

A notable special case is $D=1$, where the winding symmetry is a $-1$-form symmetry, and then there are no derived symmetries.  In that case, the anomaly is between the zero-form and the $-1$-form symmetries:
\ie
&{\cal L}^{(1)} = {\cal L}_G^{(1)} +{1\over 2\pi } \widetilde A_I^{(0)}(d\varphi^I-A^I)\,,\\
 &{\cal L}_{bulk} ^{(2)}=\frac{1}{2\pi} d\widetilde A^{(0)}_I\wedge(d\varphi^I -A^I ) \,,\\
 &{\cal L}_{anomaly}^{(2)}={1\over 2\pi } \widetilde A_I^{(0)} dA^I \,,
\fe
where $\widetilde A_I^{(0)}$ are standard $\theta$-parameters.
In this simple case, there is no higher-group symmetry.

\bibliographystyle{JHEP}
\bibliography{references}
\end{document}